\documentclass[onecolumn,amsmath,amssymb,nofootinbib,12pt]{article}
\usepackage{jheppub}

\usepackage{graphicx}
\usepackage{dcolumn}
\usepackage{bm,psfrag}

\usepackage{subfigure}
\usepackage{float}
\usepackage{tensor}

\newcommand{\ben}{\begin{equation}}
\newcommand{\een}{\end{equation}}
\newcommand{\be}{\begin{equation}}
\newcommand{\ee}{\end{equation}}
\newcommand{\bea}{\begin{eqnarray}}
\newcommand{\eea}{\end{eqnarray}}
\newcommand{\ba}{\begin{eqnarray}}
\newcommand{\ea}{\end{eqnarray}}

\newcommand{\beq}{\begin{equation}}
\newcommand{\eeq}{\end{equation}}
\newcommand{\beqa}{\begin{eqnarray}}
\newcommand{\eeqa}{\end{eqnarray}}
\newcommand{\beqar}{\begin{eqnarray*}}
\newcommand{\eeqar}{\end{eqnarray*}}

\newcommand{\reef}[1]{(\ref{#1})}

\newcommand{\eg}{{\it e.g.,}\ }
\newcommand{\ie}{{\it i.e.,}\ }
\newcommand{\comment}[1]{{\bf [[[#1]]]}}

\newcommand{\labell}[1]{\label{#1}} 

\newcommand{\cO}{{\cal O}}

\def\t6 {T_\mt{D6}}

\newcommand{\mt}[1]{\textrm{\tiny #1}}

\newcommand{\vk}{{\vec{k}}}
\newcommand{\vx}{{\vec{x}}}

\def\cale         {{\cal E}}

\def\calo         {{\cal O}}

\def\del          {\partial}

\def\ee           {{\rm e}}

\def\sqr#1#2{{\vcenter{\vbox{\hrule height.#2pt
 \hbox{\vrule width.#2pt height#1pt \kern#1pt
 \vrule width.#2pt}\hrule height.#2pt}}}}

\def\w{\omega}

\def\ee{\cale}

\def\aa1{\phi}
\def\cc1{\psi}

\def\vev#1{\langle #1 \rangle}

\def\nnn{\nonumber}

\def\comment#1{{\bf [[#1]]}}

\newcommand{\dg}{\delta \lambda}
\newcommand{\dt}{\delta t}

\newcommand{\sech}{\text{sech}}

\begin{document}

\preprint{arXiv:1411.7710 [hep-th]}

\title{Universality in fast quantum quenches}

\author{Sumit R. Das,$^{1}$ Dami\'an A. Galante$^{2,3}$ and Robert C. Myers$^3$}
\affiliation{$^1$\,Department of Physics and Astronomy, University of Kentucky,\\ 
\vphantom{k}\ \ Lexington, KY 40506, USA}
\affiliation{$^2$\,Department of Applied Mathematics, University of Western Ontario,\\ 
\vphantom{k}\ \ London, ON N6A 5B7, Canada}
\affiliation{$^3$\,Perimeter Institute for Theoretical Physics, Waterloo, ON N2L 2Y5, Canada}

\emailAdd{das@pa.uky.edu}
\emailAdd{dgalante@perimeterinstitute.ca}
\emailAdd{rmyers@perimeterinstitute.ca}

\date{\today}

\abstract{
We expand on the investigation of the universal scaling properties in the early time behaviour of fast but smooth quantum quenches in a general $d$-dimensional conformal field theory deformed by a relevant operator of dimension $\Delta$ with a time-dependent coupling. The quench consists of changing the coupling from an initial constant value $\lambda_1$ by an amount of the order of $\delta \lambda$ to some other final value $\lambda_2$, over a time scale $\delta t$. In the fast quench limit where $\delta t$ is smaller than all other length scales in the problem, $ \delta t \ll \lambda_1^{1/(\Delta-d)}, \lambda_2^{1/(\Delta-d)}, \delta \lambda^{1/(\Delta-d)}$, the energy (density) injected into the system scales as $\delta{\cal E} \sim (\delta \lambda)^2 (\delta t)^{d-2\Delta}$.  Similarly, the change in the expectation value of the quenched operator at times earlier than the endpoint of the quench scales as $\langle {\cal O}_\Delta\rangle \sim \delta \lambda\, (\delta t)^{d-2\Delta}$, with further logarithmic enhancements in certain cases. While these results were first found in holographic studies, we recently demonstrated that precisely the same scaling appears in fast mass quenches of free scalar and free fermionic field theories. As we describe in detail, the universal scaling refers to renormalized quantities, in which the UV divergent pieces are consistently renormalized away by subtracting counterterms derived with an adiabatic expansion. We argue that this scaling law is a property of the conformal field theory at the UV fixed point, valid for arbitrary relevant deformations and insensitive to the details of the quench protocol. Our results highlight the difference between smooth fast quenches and instantaneous quenches where the Hamiltonian abruptly changes at some time.
}

\maketitle


\newpage

\section{Introduction}

In recent years, there has been a great deal of interest in studying quantum quenches \cite{more}, \ie studying of the quantum evolution of an isolated system in the presence of a time-dependent parameter in the Hamiltonian. Amongst other things, these processes are theoretically interesting as probes of two related issues: thermalization and critical points. Considering the first of these, suppose we start with a system in its ground state. If a parameter in the Hamiltonian, \eg an external field, undergoes a rapid change, the system driven to some highly excited state but one would expect that after sufficient time the system will approach a steady state which {\em resembles} a thermal state. The question then is to understand the sense in which the final pure state is close to a thermal state, and to understand the approach to such a state. 
Similar questions can be studied in a thermal quench, where the initial state is a thermal state. 
Of course, these questions lie at the heart of the foundations of statistical mechanics and they are typically difficult to investigate, especially when the system is strongly coupled. Recent experiments with cold atom systems and heavy ion collisions are beginning to yield valuable experimental insights into such processes, which pose both greater motivation and interesting challenges for the theoretical community. 

A second class of interesting quenches are those which cross a critical point. That is, suppose the time-dependent parameter passes through a value which would correspond to a critical point in equilibrium.  One would then expect that the subsequent evolution of the system will carry universal signatures of the critical point. An early example of such a signal is Kibble-Zurek scaling \cite{kibble,zurek}. Suppose one starts in a gapped phase of the system, with the quench rate slow compared to the scale set by the gap. Initially the evolution of the system would be adiabatic. However, as the parameter approaches the critical point, the instantaneous gap vanishes and adiabaticity is lost, producing an excited state. Kibble \cite{kibble}, and subsequently Zurek \cite{zurek}, argued that the density of defects at late times scales as a universal power of the quench rate with the exponent determined by the equilibrium and near-equilibrium critical exponents. In recent years, this argument has been extended to quantum phase transitions and the same arguments have been shown to lead to scaling of other one point functions and correlation functions \cite{qcritkz}. The arguments which lead to Kibble-Zurek scaling are based on rather drastic approximations; nevertheless, there are several model systems where such scaling appears to hold. There is no theoretical framework analogous to the renormalization group which justifies such scaling, and strongly coupled systems remain beyond the reach of current theoretical tools. At the other extreme, Cardy, Calabrese and Sotiriadis \cite{cc2,cc3} derived a set of exact universal results for {\em instantaneous} quenches in two-dimensional field theories from a gapped phase to a critical point, using powerful methods of boundary conformal field theory. Yet another set of scaling relations can be derived from time-dependent perturbation theory when the amplitude of an instantaneous quench to a critical point is small \cite{gritsev}.

In the past few years, the AdS/CFT correspondence has been used to study both quantum and thermal quenches in strongly coupled quantum field theories which possess a gravity dual. In this approach, the couplings in the field theory are related to boundary conditions for the metric and other fields in the dual gravity theory. Therefore studying a quench process reduces to solving of a set of partial differential equations with specified initial conditions and time-dependent boundary conditions --- a problem which is much easier to tackle than the original quantum problem in a strongly coupled field theory. The dual description of thermalization becomes the collapse of an incoming shell leading to the formation of a black hole horizon \cite{holo-therm1, holo-therm2, apparent}. One of the interesting results which emerged from these studies is that few body correlation functions thermalize rapidly --- a phenomenon which is indeed observed in heavy ion collisions. For quenches across critical points, holographic studies point towards a mechanism for emergence of scaling solutions in the critical region \cite{holo-slow} and has led to novel dynamical phases \cite{holo-bhaseen}. Further, progress has been made towards observing Kibble-Zurek scaling of defect densities in symmetry breaking phase transitions \cite{julian}.

Recently, holographic studies also revealed a new set of scaling relations in the early time behaviour of fast but smooth quenches in a critical theory deformed by a relevant operator $\calo_\Delta$ with conformal dimension $\Delta$ \cite{numer,fastQ}. The quenches in question involve introducing a time-dependent coupling $\lambda (t)$ for the latter operator. If the coupling varies by an amount $\delta \lambda$ in a time $\delta t$, a fast quench means 
\ben
\delta \lambda\, (\delta t)^{d-\Delta} \ll 1\,.
\label{1-1}
\een
In this fast regime, studying quenches where the relevant coupling goes from being zero initially to $\delta\lambda$ at late times, it was found that the change of the holographically renormalized energy density $\delta \cale$ scales as 
\ben
\delta \cale \sim \delta \lambda^2 (\delta t)^{d-2\Delta}\,.
\label{1-2}
\een
Similarly, the peak of the renormalized expectation value of the quenched operator 
was found to scale as 
\ben
\langle \calo_\Delta \rangle_{ren} \sim \delta \lambda\, (\delta t)^{d-2\Delta},
\label{1-3}
\een
consistent with certain Ward identities. These same results also hold for reverse quenches where the relevant coupling goes from $\delta\lambda$ at early times to zero at late times.
For $\Delta > d/2$, this implies that $\delta \cale$ and $\langle \calo_\Delta\rangle$ grow with the quench rate, \ie as $\delta t$ shrinks. In fact, the growth in $\langle \calo_\Delta\rangle$ is enhanced by a logarithmic factor for even $d$ and integer $\Delta$ and for odd $d$ and half-integer $\Delta$. 

Implicitly, eqs.~\reef{1-2} and \reef{1-3} indicate that for $\Delta>d/2$, these quantities diverge in the limit of an infinitely fast quench, \ie with $\delta t\to0$.
Hence these results seem to be at odds with known results for {\it instantaneous} quenches, \eg \cite{cc2} -- \cite{gritsev}. In these works, a parameter in the Hamiltonian is taken to change instantaneously from one constant value to another value at some time $t_0$, and the dynamics is treated in the sudden approximation. This means that in the Schroedinger picture, the state at $t=t_0$ is treated as an initial condition for standard evolution by the new time independent Hamiltonian. Na\"ively, one may think that such an instantaneous quench should correspond to the $\dt\to0$ limit of a smooth quench but this is clearly not the case since in the setup just described, the renormalized expectation values are certainly not divergent.

Of course, the holographic studies \cite{numer,fastQ} were implicitly considering strongly coupled quantum field theories whereas the work on instantaneous quenches typically considered free (or weakly coupled) field theories, \eg \cite{cc2,cc3}, except in two space-time dimensions. Hence, one possibility is that the new divergences appearing as $\dt\to0$ are only a feature of the special class of strongly coupled theories which have gravity duals. However, we recently showed that this is not the case \cite{dgm}. In fact, precisely the same scaling as in eqs.~\reef{1-2} and \reef{1-3} was found to be exhibited in mass quenches of {\em free} field quantum theories. Further, we argued that this behaviour is rather generic.
In the present paper, we provide more details of the calculations presented in \cite{dgm} and report several new results. We also provide a new argument that the universal scaling in the early time response shown in eqs.~\reef{1-2} and \reef{1-3} holds for a quench from any constant value of the relevant coupling $\lambda_1$  to any other value $\lambda_2$ as long as the time scale $\delta t$ is small compared to all other physical length scales in the problem,
\ben
\delta t \ll \lambda_1^{1/(\Delta-d)}, \lambda_2^{1/(\Delta-d)}, \delta \lambda^{1/(\Delta-d)}
\label{regime}
\een

In the following, we first consider free bosonic and fermionic field theories in arbitrary dimensions, with a time-dependent mass which evolves smoothly in some time interval $\delta t$. We consider a variety of different protocols, \ie different profiles for $m(t)$, which allow us to solve the problem exactly for arbitrary $\dt$. Hence, we are able to calculate $\langle \calo_\Delta \rangle$ for finite $\dt$ and then examine the result in the fast regime where $\delta \lambda\, (\delta t)^{d-\Delta} \ll 1$. We find that the (renormalized) expectation value indeed obeys the universal scaling law \reef{1-3}, originally found in the holographic models studied in \cite{numer,fastQ}. 

Our analysis clearly exposes the difference between fast but smooth quenches arising in the limit $\dt\to0$, and instantaneous quenches where one works with the sudden approximation. Since we are considering a quantum field theory, the quench rate $1/\dt$ and the quench amplitude $m$ (\eg the initial mass for the quenches studied in section \ref{responseq}) are not the only scales in the problem. There is, in addition, the UV (momentum space) cutoff $\Lambda$. Implicitly, our fast quench limit involves a quench rate which is large compared to the initial mass but still small compared to cutoff, \ie 
\ben
\Lambda \gg 1/\delta t \gg m\ .
\label{1-4}
\een
Although the quench rate never appears in the discussion of instantaneous quenches, they can be considered as having $1/\dt \sim \Lambda$. However, local quantities like $\langle \calo_\Delta \rangle$ receive contributions from all scales, and are therefore sensitive to whether or not the quench rate is comparable to the cutoff scale. Indeed we show explicitly that the correlation functions of individual momentum modes for fast and smooth quenches reduce to those in the instantaneous quenches (as reported in \cite{cc2,cc3}) only when the quench rate is large compared to the momenta --- hence matching local quantities would require rates comparable to the cutoff scale. The regime of our interest is quite distinct from the latter and arguably more physical. Nevertheless, we expect that for certain quantities, \eg correlation functions at finite distances, the results for both types of quenches will agree when the distance is large compared to $\delta t$ since one expects that only small momenta contribute to the result. 
Our calculations, which are contained in a forthcoming publication \cite{dgm2}, show that this is indeed the case. We expect a similar result for other quantities which are not UV sensitive.
Similarly, one might expect that the results of smooth fast quench should agree with those of instantaneous quench at late times, $t \gg \delta t$. For free fields we will find that the late time behavior indeed agrees for $d=3$. However, in higher dimensions the late time results for smooth and instantaneous quenches differ \cite{dgm2}. While we trace the technical origin of the difference, we can not provide any good physical intuition as to why this should be the case. 

A key ingredient in our work\footnote{The same is true of the corresponding holographic studies \cite{numer,fastQ}.} is the renormalization of the underlying quantum field theory. The bare quantity $\langle \calo_\Delta \rangle$ is, of course, UV divergent and we need to add suitable counterterms to extract physical quantities at resolutions much coarser than the cutoff scale. Our problem is quite similar in spirit to quantum field theories in curved space-times, \eg see \cite{BD2,BD,Duncan}. In that case, the required counterterms involve operators made out of quantum fields, as well as curvature tensors of the background space-time. Further, in this context, diffeomorphism invariance provides an important guide restricting the form of the counterterms, which may appear. In the present case with a time-dependent mass, we find that we need to add counterterms which involve time derivatives of the mass function, in higher dimensions ($d \geq 6$) where stronger divergences appear. Further, the underlying theory is  invariant under coordinate transformations if we treat the mass as a background scalar field. Hence diffeomorphism invariance is again a useful guide in restricting the form of the required counterterms. 

However, we are still left with the problem of determining the precise coefficients of the counterterms which render the renormalized observables finite.  We find that these coefficients can be determined by examining the quenches in an adiabatic limit.\footnote{Note that similar calculations appear in the context of inflationary cosmology where it was found that the leading adiabatic contribution is sufficient to cancel the UV divergence \cite{brandenberger}. These calculations are, of course, in $d=4$ where counterterms with time derivatives are not required.} That is, the counterterm coefficients determined for an adiabatic quench still remove all of the UV divergences in $\langle\calo_\Delta\rangle$ for fast (but smooth) quenches. We argue that this result can be anticipated as follows: in renormalizing the theory, we are always considering quench rates $1/\dt$ which are much smaller than the UV cutoff $\Lambda$. In this situation, we expect the high momentum modes, near the cutoff scale, do not  care if the quench rate is large or small compared to the mass. Hence any UV divergences should be the same in fast quenches with $1/(\delta t)\gg m$ and in adiabatic quenches where $1/\dt\ll m$. Of course, the latter adiabatic limit is relatively straightforward to analyze since one is performing an expansion in derivatives with respect to time. 

It is worthwhile emphasizing that the cutoff which we use in our calculations is on the spatial momenta. If the microscopic theory were to live on a lattice, we would think of a Hamiltonian lattice theory with continuous time and a spatial lattice. The renormalization procedure described above means that we only need to adjust a finite number of parameters in the microscopic theory to get finite results for composite operators like the energy density. 

To conclude the introduction, we outline our key results and provide their locations throughout the paper.

\begin{enumerate}

\item{} We show in detail that the adiabatic expansion provides the correct counterterms which renormalize one point functions for free bosonic and fermionic theories for time-dependent masses. The bosonic case is discussed in section \ref{sect21} while the fermionic case is contained in section \ref{fermi}.  

\item{} We obtain numerical results for the renormalized one-point function of the mass operator and therefore of the energy production as well. In the limit of fast quenches
\reef{1-1}, our results clearly display the scaling behavior shown in eqs.~\reef{1-2} and \reef{1-3}. We also find explicit analytic expressions for the leading order response at early times, which again confirm this scaling. The bosonic case is described in sections \ref{responseq}, \ref{cft_cft} and \ref{massive_massive} while the fermionic case is contained in Section \ref{fermi}.

\item{} In section \ref{higher_scaling} (and appendix \ref{appendix}), we construct higher spin currents for free massive scalars. We argue that these also obey a set of universal scaling relations. The latter is explicitly shown for the spin-2 and spin-4 currents.

\item{} In section \ref{comparin}, we briefly discuss the relationship between fast smooth quenches and instantaneous quenches. We explain why the response is clearly different for these two protocols: this stems from the fact that the renormalized quantity deals with quench rates which are fast compared to the physical mass scale, while instantaneous quenches involve quench rates fast compared to all scales, including the UV cutoff scale.  The comparison between the fast smooth quenches and the instantaneous quenches will be discussed in much greater detail in \cite{dgm2}.

\item{} In section \ref{late2}, we compare the late-time response (\ie $t \gg \dt$) of a smooth fast quench with that of an abrupt quench for free bosonic field theory. In particular, we explicitly show that for $d=3$,  that the response is independent of $\dt$ at late times and leads to a logarithmic growth of $\langle\phi^2\rangle$ with time, in exact agreement  with the abrupt quench result. For $d=5$, we show once again that at late times, the $\dt \rightarrow 0$ limit is smooth.

\item{} In section \ref{interacting}, we argue that the universal scaling discussed in this paper is a property of {\em any} quantum field theory whose UV limit is a conformal field theory, \eg a conformal field theory in any number of dimensions deformed by a relevant operator. For quenches which take the system from any nonzero value of the corresponding coupling to some other value of the coupling this universal scaling holds for early time response --- so long as the time scale of quench is the smallest physical scale in the problem, as in eq.~(\ref{regime}). The scaling is purely a property of the UV conformal field theory.

\end{enumerate}

\section{Quenching a free scalar field}
\label{free}

We start by analyzing mass quenches for the simple case of a free scalar field $\phi$ in $d$ spacetime dimensions, \ie $d-1$ spatial dimensions. In particular, we focus on varying the mass with the following profile:
\beq
m^2(t) = m^2\, (A + B\, \tanh (t/\dt))\,.
\labell{massprofile}
\eeq
Hence the mass goes smoothly from the value $m^2 (A-B)$ in the infinite past to $m^2 (A+B)$ in the infinite future but the transition occurs essentially in a time period of duration $\dt$ centered around $t=0$.\footnote{Note that here and throughout the paper, we are only considering global quenches. That is, the mass is only a function of time and varies in the same way throughout all of the spatial directions.} While much of our discussion does not depend on specific values of $A$ and $B$, we will begin with a discussion of the case $A=-B=1/2$, with which the theory is massive with mass $m^2$ in the past and becomes massless in the future.\footnote{With this choice and taking the limit $\dt\to0$, we will be able to compare our results directly to the previous results for instantaneous quenches in \cite{cc2,cc3} . We might also comment that a `tanh' profile similar to eq.~\reef{massprofile} appeared in the holographic studies of \cite{numer}.} As we will show that the scaling behaviour in eqs.~\reef{1-2} and \reef{1-3} is recovered with this particular choice. 

In section \ref{cft_cft}, we also examine quenches with the mass profile
\beq
m^2(t) = m^2 \, \sech^2 (t/\delta t)
\labell{pulse}
\eeq
where the mass vanishes both the infinite past and the infinite future. We again find that the renormalized expectation values show the same scaling as in eqs.~\reef{1-2} and \reef{1-3}. 

Finally in section \ref{massive_massive} we show that the analysis easily extends to general $A$ and $B$ and we again find the same scaling as long as the coefficients satisfy
\beq
\delta (m^2)\,\dt^2=|m^2(-\infty)-m^2(+\infty)|\,\dt^2 = 2 |B| m^2\,\dt^2\ll1\,,
\label{accord}
\eeq
in accord with eq.~\reef{1-1}, and also 
\beq
\frac{m^2(-\infty)+m^2(+\infty)}{2}\,\dt^2=A\,m^2\,\dt^2\ll1\,.
\labell{extra}
\eeq
 
The particular protocols or mass profiles in eqs.~\reef{massprofile} and \reef{pulse} were chosen because they allow us to completely solve the corresponding quantum field theory. That is, the mode functions for the scalar field can be written in closed form, as we will show below. In fact, for the profile \reef{massprofile}, we can use results first derived in studying quantum fields propagating in curved spacetimes \cite{BD,BD2}. Specifically, in that case, scalar field was examined in an expanding flat Freedman-Robertson-Walker cosmology, which corresponds to a conformally flat geometry described by metric
\beq
ds^2 = a^2(t)\, (-dt^2 + d\vec{x}^2)\,.
\labell{frw}
\eeq
A (minimally coupled) free massive scalar field $\phi$, with a constant mass $m$, propagating in this cosmological background obeys the equation of motion
\beq
(\Box - m^2 a^2(t) )\, \phi = 0\,,
\labell{eom0}
\eeq
where $\Box$ denotes the ordinary flat space d'Alembertian. That is, the scalar field equation in this curved geometry is identical to that of a scalar field in flat space but with a time-varying mass $m^2(t)= m^2 a^2(t)$. Further, it was noted in \cite{BD,BD2}, that with $a^2(t)=(A + B \tanh t/\dt)$, \ie with the mass profile \reef{massprofile}, the corresponding mode functions are given in terms of hypergeometric functions. Hence we may use these results but now interpret the theory as a scalar field undergoing a mass quench. It is also important to mention that with these closed form solutions, we are able to study the behaviour of the theory for arbitrary quenches rates $1/\dt$ and hence we can take the limit $\dt \rightarrow 0$ to approach an \textit{instantaneous} quench. 

Let us begin with analyzing the theory with the mass profile \reef{massprofile}. We start by decomposing our field in mode functions
\beq
\phi= \int\!\! \frac{d^{d-1}k}{(2\pi)^{(d-1)/2}} \ \left( a_\vk\, u_\vk + a^\dagger_\vk\, u^*_\vk\right)\,,\qquad
{\rm where}\ \ \ [a_\vk , a^\dagger_{\vk^\prime} ] = \delta^{d-1}(\vk - \vk^\prime)\,.
\labell{fieldx}
\eeq
As a boundary condition, we will choose the $u_\vk$ to be the in-modes which behave as plane waves in the infinite past. Similarly there will be a corresponding set of out-modes which become plane waves in the infinite future. The operators $a_\vk$ above are then defined to annihilate the in-vacuum, \ie $a_\vk |in,0\rangle=0$. Exact solutions for these in-modes are \cite{BD,BD2}
\begin{eqnarray}
u_\vk & = & \frac{1}{\sqrt{2 \omega_{in}}} \exp(i\vk\cdot\vec{x}-i\omega_+ t - i\omega_- \dt \log (2 \cosh t/\dt)) \times \nonumber \\
& &\qquad_2F_1 \left( 1+ i \omega_- \dt, i \omega_- \dt; 1 - i \omega_{in} \dt; \frac{1+\tanh(t/\dt)}{2} \right)\,,
\label{modes}
\end{eqnarray}
where $_2F_1$ is the usual hypergeometric function and
\begin{eqnarray}
\omega_{in}  & = &  \sqrt{\vk^2+m^2 (A-B)}\,, \nonumber\\
\omega_{out}  & = & \sqrt{\vk^2+m^2 (A+B)}\,, \label{omegadef}\\
\omega_{\pm}  & = & (\omega_{out}\pm\omega_{in})/2\,.
\nonumber
\end{eqnarray}

\subsection{Regularization and Renormalization} \label{sect21}

The quantities we are interested in involve a sum over all modes and are typically UV divergent and need to be renormalized by adding suitable counterterms. In this subsection we show how this can be done. The discussion is valid for generic $m(t)$ - in fact we will find the counterterm in terms of the function $m(t)$ and its derivatives. However it is useful to begin the discussion with the mass profile (\ref{massprofile}).

First focus on the case where $A=-B=1/2$, in which case we have $\omega_{in} = \sqrt{\vk^2+m^2}$ and $\omega_{out} = |\vk|$. Now we adopt the perspective presented in the holographic analysis of \cite{numer, fastQ} in the following. In particular, we think of the scalar field theory as a CFT deformed by the operator $\cO_\Delta \sim \phi^2$, with conformal dimension $\Delta=d-2$. Further, the quenches are made by varying the corresponding coupling in time, \ie $\lambda(t) = m^2(t)$. Our first calculation will be to determine the expectation value of $\langle \phi^2 \rangle$, which is straightforward given the mode decomposition above
\beq
\langle\phi^2\rangle\equiv
\langle in,0|\phi^2|in,0\rangle = \frac{1}{2(2\pi)^{d-1}}\int \frac{d^{d-1}k}{\omega_{in}}\, |_2F_1|^2\,.
\label{phi_squared}
\eeq

Of course, this expectation value \reef{phi_squared} contains UV divergences associated with the integration of
$k=|\vk|\to\infty$. The standard approach to deal with these UV divergences is to add suitable counterterms involving
the time-dependent mass to the effective action, as in the holographic renormalization of \cite{numer}. We turn to the
determination of the counterterms in section \ref{effaction}. However, as described in \cite{dgm}, it is straightforward to 
find the counterterms which render the expectation value \reef{phi_squared} finite. Hence let us write the renormalized
expectation value as
\beq
\langle\phi^2\rangle_{ren} \equiv \frac{\Omega_{d-2}}{2(2\pi)^{d-1}} \int dk \left( \frac{k^{d-2}}{\w_{in}}\, |_2F_1|^2 - f_{ct}(k,m(t)) \right)\,. \label{renorm}
\eeq
where $f_{ct}(k,m(t))$ designates the counterterm contribution and $\Omega_{d-2}$ denotes the angular volume of a unit
($d$--2)-dimensional sphere, \ie  
\beq
\Omega_{d-2}\equiv\frac{2\,\pi^{(d-1)/2}}{\Gamma\left((d-1)/2\right)}\,.
\label{volS}
\eeq

As a first attempt to evaluate  $f_{ct}(k,m(t))$, we might na\"ively think that the counterterm contributions needed to regulate $\langle\phi^2\rangle$ are those related to the divergences in the constant mass case.  That is, with a constant mass, we can identify the UV divergences by expanding
\beqa
\langle\phi^2\rangle &= &\frac{\Omega_{d-2}}{2(2\pi)^{d-1}} \int dk \frac{k^{d-2}}{\sqrt{k^2+m^2}} \label{naive}\\
 &=& \frac{\Omega_{d-2}}{2(2\pi)^{d-1}} \int dk \  k^{d-3} \left(1-\frac{m^2}{2 k^2}+\frac{3 m^4}{8 k^4} 
-\frac{5 m^6}{16 k^6} + O(m^8/k^{8}) \right) \,.  \nonumber
\eeqa
With the simple substitution $m^2\rightarrow m^2(t)$, we might then conjecture that eq.~\reef{renorm} becomes finite with
\beq
f_{ct}(k,m(t)) = k^{d-3} \left(1-\frac{m(t)^2}{2 k^2}+\frac{3 m(t)^4}{8 k^4} -\frac{5 m(t)^6}{16 k^6} + O(m(t)^8/k^{8}) \right) \,, 
\label{naive2}
\eeq
where we would only include the terms proportional $k^n$ with $n\ge-1$. As we will see below, this conjecture is only correct for $d \leq 5$.
For higher spacetime dimensions (\ie $d\ge6$ in the scalar case), new counterterms involving time derivatives of the mass are allowed by dimensional counting. For example, in $d=6$, the term proportional to  $m(t)^4/k$ is associated with a logarithmic divergence in $\langle\phi^2\rangle$. However, by dimensional analysis, $f_{ct}(k,m(t))$ could also contain a term of the form $\partial_t^2 m(t)^2/k$, which might cancel a new logarithmic divergence proportional to $\partial_t^2 m(t)^2$ in $d=6$. 
Of course, in the case of a constant mass \reef{naive}, no such divergence appears but in the present case of a mass quench, a new UV divergence of this form will be found. As we go to higher and higher dimensions, the set of dimensionally allowed terms involving time derivatives of the mass quickly grows and in fact, the corresponding divergences (generically) do appear, as we will see below. However, let us note that the same dimensional arguments would have identified a potential contribution of the form $\partial_t m(t)^2/k$ in $d=5$ but no corresponding divergence is found. Hence this makes evident that these terms are subject to constraints beyond simple dimensional analysis. In particular, we will show that this single-derivative contribution can be ruled out by diffeomorphism invariance.

Finally, let us comment that in holographic calculations \cite{numer,fastQ}, these kind of terms naturally appear since couplings are not just constants but boundary values of spacetime-dependent  bulk fields. Holographic renormalization then requires introducing counterterms in the gravitational action constructed out of derivatives of the boundary values. 

\subsubsection{Regulating the theory using an adiabatic expansion}
\label{adiabat}

An elegant way to find the necessary counterterm contributions is to look at the divergences appearing in eq.~\reef{phi_squared} for an adiabatic quench, \ie an infinitely slow quench. In that way, one can organize all contributions with an adiabatic expansion and exactly find the divergent pieces. The discussion below is for a general function $m(t)$.

 
The adiabatic expansion is an expansion in time derivatives, more precisely in powers of $\partial_t^n m/m^{n+1} \ll1$. These ratios are, of course, small if the time variation of the mass is infinitely slow. In a generic quantum mechanical system, this expansion is achieved by expanding the state as a linear superposition of {\em instantaneous eigenstates} and solving the resulting differential equations for the coefficients in a derivative expansion. For a free field theory, the procedure is easier --- one can obtain mode solutions of the equations of motion for each momentum mode,
\beq
\frac{d^2u_\vk}{dt^2} + (k^2 + m^2(t))\, u_\vk =0\,,
\label{adiabatic_eom}
\eeq
 in a WKB type approximation.
That is, we wish to find solutions of this equation which are of the form
\beq
u_\vk = \frac{1}{\sqrt{2\, \Omega_k(t)}} \exp \left(i \, \vk\cdot\vec{x} -i \int^t \Omega_k(t') dt' \right)\,.
\label{outbk}
\eeq
Demanding that this ansatz solves eq.~(\ref{adiabatic_eom}) requires that $\Omega_k$ satisfies
\beq
\Omega_k^2 = \omega_k^2 - \frac{1}{2}  \frac{\partial_t^2{\Omega}_k}{\ \Omega_k} + \frac{3}{4} \left(\frac{\partial_t{\Omega}_k}{\Omega_k} \right)^2 \,,
\qquad{\rm with}\ \ \omega_k^2= k^2 + m^2(t)\,. \label{Omega}
\eeq
The adiabatic expansion is then obtained in eq.~(\ref{Omega}) by expanding the solution as  
\beq
\Omega_k = \Omega_k^{(0)} + \Omega_k^{(1)} + \Omega_k^{(2)} + \cdots,
\label{o_expansion}
\eeq
where $\Omega_k^{(n)}$ is $n^{th}$ order in time derivatives. 
We can now substitute this expansion into eq.~(\ref{Omega}) and solve it order by order. The first two orders are trivial, yielding
\begin{eqnarray}
\Omega_k^{(0)} & = & \omega_k\,, \label{onnne}\\
2 \Omega_k^{(0)} \Omega_k^{(1)} & = & 0\,,
\nonumber
\end{eqnarray}
where the latter yields $\Omega_k^{(1)} = 0$.
The next two orders produce
\begin{eqnarray}
\left( \Omega_k^{(1)} \right)^2 + 2 \Omega_k^{(0)} \Omega_k^{(2)} & = & -\frac{1}{2} \left(\frac{\ddot{\omega}_k}{\omega_k} -\frac{3 \dot{\omega}_k^2}{2\omega_k^2} \right)\,, \label{argent} \\
2 \Omega_k^{(1)} \Omega_k^{(2)} + 2 \Omega_k^{(3)} \Omega_k^{(1)} & = & 0\,, \nonumber
\end{eqnarray}
which are solved by
\beq
\Omega_k^{(2)} = -\frac{1}{4 \omega_k} \left(\frac{\ddot{\omega}_k}{\omega_k} -\frac{3 \dot{\omega}_k^2}{2\omega_k^2} \right)
\qquad {\rm and}\qquad
\Omega_k^{(3)} = 0\,.
\labell{solargent}
\eeq
Again substituting these results into eq.~(\ref{Omega}), we find the next order equation
\beq
\left(\Omega_k^{(2)}\right)^2+2 \Omega_k^{(1)} \Omega_k^{(3)}+2 \Omega_k^{(0)} \Omega_k^{(4)} = \frac{\left(-36 \dot{\omega}_k^4+48 \omega_k \dot{\omega}_k^2 \ddot{\omega}_k-6 \omega_k^2 \ddot{\omega}_k^2-10 \omega_k^2 \dot{\omega}_k \dddot{\omega_k}+\omega_k^3 \ddddot{\omega}_k\right)}{8 \omega_k^6}\, , 
\label{chill}
\eeq
which gives
\beq
\Omega_k^{(4)} = \frac{-297 \dot{\omega}_k^4+396 \omega_k \dot{\omega}_k^2 \ddot{\omega}_k-52 \omega_k^2 \ddot{\omega}_k^2-80 \omega_k^2 \dot{\omega}_k \dddot{\omega}_k+8 \omega_k^3 \ddddot{\omega}_k}{128 \omega_k^7}\,.
\label{solchill}
\eeq

As we will see, it is enough to expand up to this order to get all the necessary counterterm contributions to regulate present theories up to $d=9$. Now, we want to extract the large-$k$ behaviour of
\beq
\langle \phi^2 \rangle = \frac{\Omega_{d-2}}{2(2\pi)^{d-1}} \int dk\  \frac{k^{d-2}}{\Omega_k}\,,
\labell{2-23}
\eeq
and so we will need to expand $1/\Omega_k$ for large $k$, as well as in time-derivatives. Using $\omega_k^2 = k^2 + m^2(t)$, we find
\begin{eqnarray}
\frac{1}{\Omega_k} & \simeq & \frac{1}{k}\left[1-\frac{m^2(t)}{2 k^2}+\frac{3 m^4(t)}{8 k^4}-\frac{5 m^6(t)}{16 k^6} + \cdots\right.\nonumber \\
& & \qquad +\, \frac{\partial_t^2 m^2(t)}{8 k^4}-\frac{5}{32k^6}\left((\partial_t m^2(t))^2+2 m^2(t) \partial_t^2 m^2(t)\right) + \cdots \nonumber \\
& & \qquad\left. -\,\frac{\partial_t^4 m^2(t)}{32 k^6} + \cdots\ \ \right]\,,
\end{eqnarray}
where each line in the last expression corresponds to a particular order in time derivatives, \eg the first line is zeroth order; the second line, second order; etcetera.  The ellipsis at the end of each line indicates terms that are higher order in $1/k$, \ie 1/$k^{8}$ and higher. Multiplying by $k^{d-2}$, those are all the divergent terms in spacetime dimensions less or equal to $d=9$. We can see that the first line corresponds to the terms discussed in eq.~(\ref{naive}). But this is only the zeroth-order adiabatic approximation and there are additional divergent terms at higher orders in the expansion in time derivatives.\footnote{We might also note here that all of the terms appearing in this expansion involve an even number of time derivatives.} Of course, the results match those reported in \cite{dgm}, where we found
\begin{eqnarray}
f_{ct} (k,m(t)) & = &  k^{d-3} - \frac{k^{d-5}}{2} m^2(t) + 
\frac{k^{d-7}}{8} \left( 3m^4(t)+ \partial^2_t m^2(t)\right) \label{ict} \\
& & - \frac{k^{d-9}}{32} \left(10 m^6(t) + \partial^4_t m^2(t) + 10 m^2(t)\, \partial^2_t m^2(t)+5\partial_t m^2(t)\, \partial_t m^2(t)\right)  + \cdots \,. 
\nonumber
\end{eqnarray}
For a fixed spacetime dimension $d$, we would only keep the terms up to the power $k^{-1}$ and drop any terms with more negative powers of $k$. Again, the contributions
explicitly written above are sufficient to regulate theories up to and including $d=9$.

The above discussion applies for a general space-time dimensions, however, we should distinguish between odd and even dimensions. For odd $d$, all of the powers of $k$ appearing in eq.~\reef{ict} are even (or zero) and $f_{ct}$ essentially subtracts a series of power-law divergences $\Lambda^n$, where $\Lambda$ is the UV cutoff scale. When $d$ is even, the powers of $k$ are now odd, and similar power-law divergences are appearing for the positive powers of $k$. However, apart from these divergences, we may also find a logarithmic divergence when eq.~\reef{ict} contains a $1/k$ term. If we considered this term alone, the $k$ integral in eq.~\reef{renorm} is divergent both in the UV and in the IR. Hence, we also need to introduce a lower bound $\mu_i$ for each such integral, which then yields $\log(\Lambda/\mu_i)$. Hence we see this amounts to introducing an extra renormalization scale in defining the renormalized expectation value \reef{renorm} for even $d$. The appearance of these new scales reflects certain scheme-dependent ambiguities in defining the renormalized theory, and in particular, as observed in previous holographic studies \cite{numer}, new ambiguities can arise with time-dependent couplings. Of course, the potential $\Lambda$ divergences are all eliminated in eq.~\reef{renorm} and we take the limit $\Lambda\to\infty$ in evaluating the renormalized expectation value. Hence in the final result, an infrared scale must replace the UV cutoff in the logarithmic dependence on the renormalization scale, \eg $\log(\dt\,\mu_i)$ --- see sections \ref{responseq} and \ref{analytical_cont} for further discussion. 

With the subscript on $\mu_i$, we are emphasizing that in principle one can introduce a separate renormalization scale for each such integral corresponding to a separate counterterm. For example, with $d=6$ in eq.~\reef{ict}, there can be a separate renormalization scale associated with the integrals proportional to $m^4(t)$ and $\partial^2_t m^2(t)$, since they correspond to contributions coming from distinct counterterms --- see section \ref{effaction} for further discussion. However, in our explicit calculations in the following, we will set all of these scales to be equal, \ie $\mu_i=\mu$. The effect in the computation is to divide the integral in the expectation value \reef{renorm} into two
parts. The first, from $k=0$ to $k=\mu$ does not include the $1/k$ contribution in $f_{ct}$ while in the second, from $k=\mu$ to $k=\infty$, we use the full expression for $f_{ct}$ including the $1/k$ term. 

Now we claim that the large-$k$ terms appearing in the adiabatic expansion provide the correct counterterm contributions to regulate $\langle\phi^2\rangle$ for general quenches. This claim may seem surprising since the adiabatic expansion should be only valid for {\em slow} quenches. However, one can easily verify numerically that with eq.~\reef{ict}, the renormalized expectation value \reef{renorm} is finite, \eg with the mass profile \reef{massprofile} even outside of the adiabatic regime. The point is that we are considering  a quench rate $1/\dt$ which is always slow compared to the UV cutoff scale, though it may be fast compared to $m$, \eg as described by eq.~\reef{1-4}. The condition for validity of the adiabatic expansion is ${\dot \omega_k} \ll \omega_k^2$. For this condition to hold for all $k$, we must have $m\, \delta t \gg 1$. However, for high momenta $k \gg m$, this condition still holds as long as $k \,\delta t \gg (m/k)^2 $, which is always satisfied for sufficiently large $k$. Hence the fact that we are interested in studying fast quenches where $m\, \delta t \ll 1$ does not matter for the very high momentum modes, whose contributions are producing the UV divergences. This explains why the adiabatic expansion provides a consistent and convenient framework to find the divergent pieces of the expectation value. In fact, the counterterms are universal and, in particular, independent of the rate at which the mass varies.

\subsubsection{Explicit verification for tanh profile}
\label{aminusbhalf}

We now show explicitly that eq.~\reef{ict} provides the correct counterterm contributions for general quenches, we return to the tanh profile in eq.~\reef{massprofile} with $A=  -B = 1/2$. Recall that in this case, the bare expectation value is given in eq.~\reef{phi_squared} where the details of hypergeometric functions appear in eqs.~\reef{modes} and \reef{omegadef}. Now we proceed to expand these hypergeometric functions for large momentum. In the series representation, the hypergeometric function is defined as
\beq
_2F_1 (a,b;c;z)  = \sum_{n=0}^{\infty} \frac{(a)_n (b)_n}{(c)_n} \frac{z^n}{n!},
\label{hyper}
\eeq
where $(x)_n \equiv x (x+1) \cdots (x+n-1)$ and $(x)_0 = 1$. Further $a,b,c,z$ are given in eq. (\ref{modes}). In particular, we recall that the argument $z$ is given by
\beq
 z= \frac12\left(1+\tanh (t/\dt)\,\right)\,,
 \labell{zzz}
\eeq
This variable is the same regardless the choice of $A$ and $B$. Now we can expand $\omega_{in}$ and $\omega_-$ for large $k$ and see how the first few terms of this series behave. Then we have to take the absolute value squared to get the counterterms for the expectation value of eq~ (\ref{phi_squared}). By checking the behaviour of the series, it can be verified that each successive term begins with a lower power of $k$. Hence in order to get all the divergent terms up to $d=9$, it is sufficient to work with \textit{only} the first five terms in eq.~\reef{hyper}.

Focusing again on the mass profile \reef{massprofile} with $A=-B=1/2$ and expanding these terms for large $k$, we find 
\begin{eqnarray}
\frac{k^{d-2}}{\omega_{in}} |_2F_1|^2 & = & k^{d-3}+\frac{1}{2} k^{d-5} \left(-m^2+m^2 z\right)+ \nonumber \\
& & + \frac{1}{8} k^{d-7} \left(3 m^4-6 m^4 z+3 m^4 z^2-\frac{4 m^2 z}{\dt^2}+\frac{12 m^2 z^2}{\dt^2}-\frac{8 m^2 z^3}{\dt^2}\right)+ \nonumber \\
& & + \frac{1}{16} k^{d-9} \Big(-5 m^6+15 m^6 z-15 m^6 z^2+5 m^6 z^3+  \nonumber \\
& & \frac{8 m^2 z}{\dt^4}-\frac{120 m^2 z^2}{\dt^4}+\frac{400 m^2 z^3}{\dt^4}-\frac{480 m^2 z^4}{\dt^4}+\frac{192 m^2 z^{5}}{\dt^4}+ \nonumber \\
& & \left. \frac{20 m^4 z}{\dt^2}-\frac{90 m^4 z^2}{\dt^2}+\frac{120 m^4 z^3}{\dt^2}-\frac{50 m^4 z^4}{\dt^2}\right) + O(k^{d-11})
\label{ict2}
\end{eqnarray}
At first sight this expression does not look similar to eq.~(\ref{ict}), but we will now show that they are both actually the same. To start with, we should notice that we can write $m^2(t)$ as a function of $z$ as $m^2(t)=m^2(1-z)$. Then, for instance the $k^{d-5}$ term in eq. (\ref{ict2}) is just $-m^2(t)/2$, matching the corresponding term in eq.~(\ref{ict}). The same happens with all the terms that are independent of the value of $\dt$; \ie they give $m^4(t)$ and $m^6(t)$, as they should. The appearance of terms which are inversely proportional to $\dt$ reflects the appearance of time-derivatives in those terms. In fact, we can use trigonometric identities to express derivatives of the mass in terms of powers of the same mass function. This is because derivatives of the hyperbolic tangent are formed by terms proportional to $\tanh$ and $\sech$. For instance, the first derivative of $m^2(t)$ gives $\partial_t m^2(t) = -\frac{1}{2}\frac{m^2}{\dt} \sech^2 (t/\dt)$ and the second derivative, $\partial_t^2 m^2(t) = \frac{m^2}{\dt^2} \sech^2(t/\dt) \tanh(t/\dt)$. But now using trigonometric identities, we can write $\sech$ in terms of $\tanh$: $\sech^2 (x) = 1 - \tanh^2(x)$. Moreover, $\tanh (t/\dt) = 2 z - 1$, so we can express every derivative just in powers of $z$. The second derivative will give, for example, $\partial_t^2 m^2(t) =\frac{4m^2}{\dt^2} z  (1 - 3 z + 2 z^2)$ and up to an extra minus sign, this expression matches exactly the last three terms appearing in the $k^{d-7}$ term of eq.~(\ref{ict2}). In the same way, we can translate all the terms of eq.~(\ref{ict2}) to match the universal form that we found in eq.~(\ref{ict}).

We emphasize that the above calculations are valid for any value of $\dt$ and hence this verifies that a single set of counterterms can be chosen to regulate $\langle\phi^2\rangle$ independent of the quench rate. In particular, the same counterterms should be valid in the limit $\dt \rightarrow 0$. We have also performed the same calculations with expanding the hypergeometric function for the \textit{reverse} quench and found the same counterterms, now as functions of the new $m^2(t) = \frac{m^2}{2} (1 + \tanh (t/\dt))$. We will also see that for a pulsed quench \reef{pulse}, as studied in section \ref{cft_cft}, the same counterterm contributions \reef{ict} again regulate the expectation value for any value of $\dt$. Hence all of these examples provide a verification of our claim that studying an adiabatic quench is sufficient to determine the correct counterterm contributions to regulate $\langle\phi^2\rangle$ for general quenches.

\subsubsection{Counterterms in the path integral}
\label{effaction}

Up to this point, we have been interested in finding the necessary contributions which render eq.~\reef{phi_squared} finite and allow us to calculate the renormalized expectation value in eq.~\reef{renorm}. However, we may also be interested in computing other observables, \eg the expectation value of the energy-stress tensor --- see section \ref{stress_energy}. Of course, expectation values of other operators will again generally be UV-divergent and also need regularization. The point we would like to emphasize in this section is that all such divergences should be eliminated by a common set of counterterms regulating the effective action or partition function. Once we have the regulated partition function, we can find the renormalized expectation value of the operators of interest by taking functional derivatives with respect to the appropriate sources.  Suppose the path integral is regulated by a UV cutoff $\Lambda$, then we have 
\beq
Z(m^2,g_{\mu\nu})=\int[{\cal D}\phi]_\Lambda\ \exp\left[-iS_0(\phi,m^2,g_{\mu\nu})-iS_{ct}(m^2,g_{\mu\nu},\Lambda)\right]
\label{part}
\eeq
which includes the free field action
\beq
S_0(\phi,m^2,g_{\mu\nu})=-\frac12\int d^dx\sqrt{-g}\left[g^{\mu\nu}\partial_\mu\phi\partial_\nu\phi + m^2\phi^2\right]
\label{freeact}
\eeq
and the counterterm action\footnote{Note that because we are considering a free field theory, all of the counterterms are pure c-numbers.}
\begin{eqnarray}
&&S_{ct}(m^2,g_{\mu\nu},\Lambda) =-\int d^dx \sqrt{-g} \left[ s_{00} \Lambda^d + s_{10} m^2 \Lambda^{d-2}  + s_{20} m^4 \Lambda^{d-4}\right.
\label{ctact}\\
&&\ \ \ \ \ +  (s_{30} m^6 + s_{31} m^2 \Box m^2) \Lambda^{d-6}    
 + (s_{40} m^8 + s_{41} m^4 \Box m^2 + s_{42} m^2 \Box^2 m^2 ) \Lambda^{d-8}+\cdots  \nonumber \\
& &\ \ \ \ \ + R\, \left[s_{50} \Lambda^{d-2} + s_{51} m^2 \Lambda^{d-4} + (s_{52} m^4 + s_{53} \Box m^2 ) \Lambda^{d-6} +
\right. \nonumber \\
& &\qquad\qquad\qquad \left.\left. + (s_{54} m^6 + s_{55} 
(\partial m^2)^2 + s_{56} m^2 \Box m^2 + s_{57} \Box^2 m^2) \Lambda^{d-8}+\cdots\right]+\cdots \right]\,, \nonumber
\end{eqnarray}
where $R$ is the Ricci scalar of the metric $g_{\mu\nu}$ and $s_{ij}$ are finite numbers. Of course, for a fixed dimension $d$, we only retain the terms above with positive powers of $\Lambda$ and in cases, where the na\"ive power is zero, it should be replaced by a logarithmic divergence $\log(\Lambda/\mu)$ --- as discussed in the previous section. Now the expectation values of the `mass operator' and the stress tensor are given by
\bea
\langle \phi^2 \rangle_{ren}&=&-2i \left[\frac1{\sqrt{-g}}\ \frac{\delta \ \ }{\delta m^2}\log Z\right]_{g_{\mu\nu}=\eta_{\mu\nu};\,\Lambda\to\infty}\,,
\label{gamma1}\\
\langle T_{\mu \nu} \rangle_{ren}& =& -2i \left[\frac1{\sqrt{-g}} \ \frac{\delta \ \ }{\delta g^{\mu\nu}}\log Z\right]_{g_{\mu\nu}=\eta_{\mu\nu};\,\Lambda\to\infty}\,.
\label{gamma2}
\eea
Again we have explicitly shown all of the possible counterterms in eq.~\reef{ctact} which would be needed to regulate these two expectation values up to $d=9$.

Now several comments are in order: First we have introduced a background curved space metric in the partition function \reef{part}, even though we are evaluating the final expectation values in flat space. This is, of course, because the metric serves as the source of the stress tensor as in eq.~\reef{gamma2}. Further, in this vein, we have included counterterms linear in Ricci scalar in eq.~\reef{ctact} since even though these terms vanish in flat space, their variation still contributes to regulating the expectation value of the stress tensor in eq.~\reef{gamma2}. Of course, these terms are not needed to evaluate $\langle \phi^2 \rangle_{ren}$ in eq.~\reef{gamma1}. We have ignored terms involving higher powers of the Ricci Scalar since they do not contribute to the two one-point functions in eqs.~\reef{gamma1} and \reef{gamma2}. Further we have dropped any total derivative terms in the counterterm action, as well as terms that can be related to those appearing in eq.~\reef{ctact} by using integration by parts and the identity
$\nabla^\mu R_{\mu\nu}=\frac12\,\nabla_\nu R$. As a result, we were able to eliminate any counterterms linear in the Ricci tensor.

Implicitly above, we are treating the mass-squared as a background scalar field which is a function of all of the spacetime coordinates, \ie $m^2=m^2(x^\mu)$. For example, this assumption is evident in eq.~\reef{gamma2} where the variation yields the expectation value of the local operator operator $\phi^2(x^\mu)$. Now if the path integral \reef{part} is performed with a covariant action, the counterterm action, as well as the entire partition function, will be diffeomorphism invariant, as assumed with the presentation in eq.~\reef{ctact}. In particular, the derivatives of the mass only appear there as powers of the covariant d'Alembertian operator.\footnote{Again, integration by parts was used to reduce certain covariant counterterms to this form, \eg $(\partial m^2)^2=g^{\mu\nu}\partial_\mu m^2\,\partial_\nu m^2 \sim - m^2\Box m^2$.} Of course, in applying this counterterm action to study (global) mass quenches, we only consider the mass to be a function of time but the structure revealed here readily explains why all of the counterterm contributions in eq.~\reef{ict} have an even number of time derivatives. We might also comment that in the curved background geometry we have 
\beq
\Box m^2 =\frac1{\sqrt{-g}}\,\frac{\partial\ }{\partial x^\mu}\!\left(\sqrt{-g}\, g^{\mu\nu}\,\frac{\partial m^2}{\partial x^\nu}\right)
\label{dall}
\eeq
and hence these derivative terms also contribute nontrivially to regulating the stress tensor in eq.~\reef{gamma2}.

Let us observe that there are four terms at order $k^{d-9}$ in eq.~\reef{ict} but only three corresponding counterterms at order $\Lambda^{d-8}$ in eq.~\reef{ctact}. Hence the four counterterm contributions are not all independent. In fact, it is straightforward to show that for a time-dependent mass, the variation of the counterterm with $s_{41}\, m^4 \Box m^2$ is proportional to  $s_{41}\,(2 m^2\partial_t^2m^2+\partial_tm^2\partial_tm^2)$, which has precisely the ratio of coefficients with which these two terms appear in eq.~\reef{ict}. In fact, by carefully comparing eqs.~\reef{ict} and \reef{ctact}, we can identify the coefficients:
\beqa
&& s_{10}= - \frac{1}{2(d-2)\sigma_s} \,,\qquad\ \, s_{20}=  \frac{1}{8 (d-4)\sigma_s}\,,\nonumber\\
&& s_{30}= - \frac{1}{16(d-6)\sigma_s} \,,\qquad s_{31}= - \frac{1}{32(d-6)\sigma_s}\,,
\label{coeffs}\\
&& s_{40}= \frac{5}{128(d-8)\sigma_s} \,,\qquad\  s_{41}= \frac{5}{128(d-8)\sigma_s}\,,
\qquad s_{42}= \frac{1}{128(d-8)\sigma_s}\,.\nonumber
\eeqa
where 
\beq
\sigma_s\equiv\frac{2(2\pi)^{d-1}}{\Omega_{d-2}}\,.
\label{sigs}
\eeq
In principle, 
the adiabatic expansion in the last subsection could also be used to find the remaining coefficients in eq.~\reef{ctact}, which would be needed to regulate the expectation value of the stress tensor \reef{gamma2} --- in dimensions up to $d=9$. However, as we will explain in section \ref{stress_energy}, we can avoid this calculation, at least in evaluating the expectation value of the energy density (the $tt$ component of the stress energy tensor). The latter can be related to $\langle \phi^2 \rangle_{ren}$
using a diffeomorphism Ward identity.  We will explicitly apply this approach for $d=5$ in section \ref{stress_energy} and for $d=3$ in section \ref{sec_3d_stress}.

\subsection{Response to the mass quench}
\label{responseq}

In this subsection we calculate renormalized quantities which measure the response to a mass quench of the form (\ref{massprofile}) with $A = -B = 1/2$.

\subsubsection{Numerical results}
\label{numerical}

\begin{figure}[H]
        \centering
        \subfigure[$d=3$]{
                \includegraphics[scale=0.8]{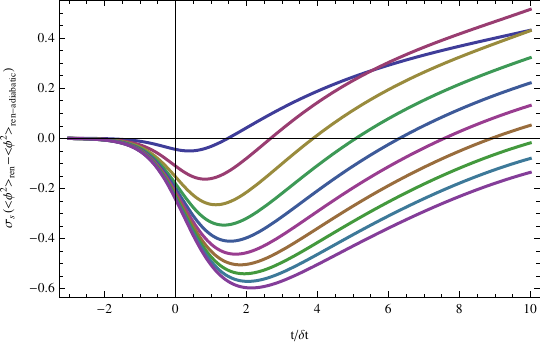}}
   		 \subfigure[$d=4$]{
                \includegraphics[scale=0.8]{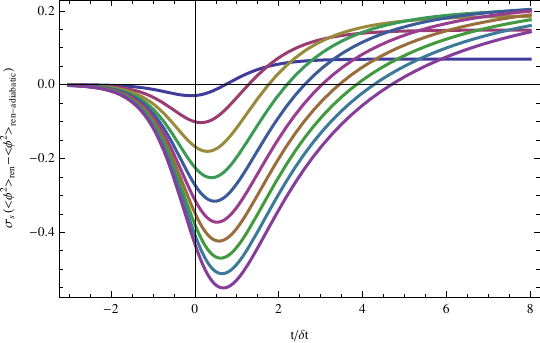}}
        \caption{(Colour online) Renormalized expectation values $\langle \phi^2\rangle_{ren}$ as a function of time $t/\dt$, for $d=3$ and $4$. In each plot, the different curves correspond to different quench rates: $\dt = 1/1,1/2, \cdots, 1/10$ where the curves exhibiting higher peaks (in absolute value) correspond to smaller values of $\dt$. Note that the expectation value is multiplied by the numerical constant $\sigma_s=\frac{2(2\pi)^{d-1}}{\Omega_{d-2}}$. Further, at each time, the expectation value for an `adiabatic' quench is subtracted.}\label{fig_scalars1a}
\end{figure}

\begin{figure}[H]
        \centering
         \subfigure[$d=5$]{
                \includegraphics[scale=0.8]{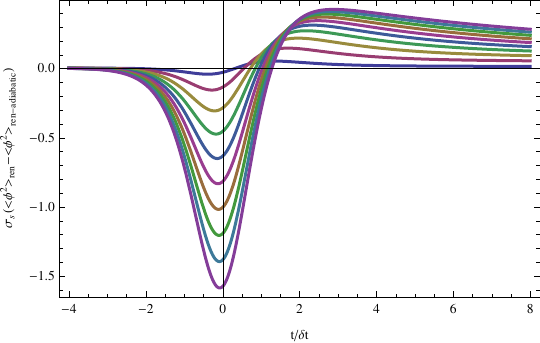}}
   		 \subfigure[$d=6$]{
                \includegraphics[scale=0.8]{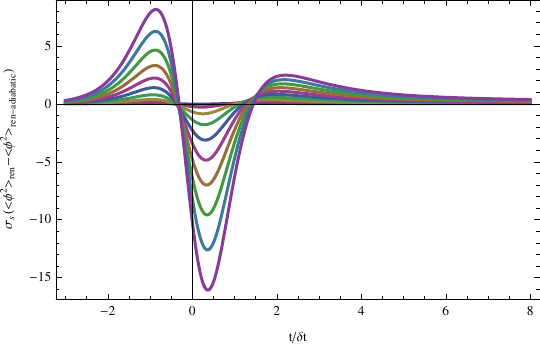}}
         \subfigure[$d=7$]{
                \includegraphics[scale=0.8]{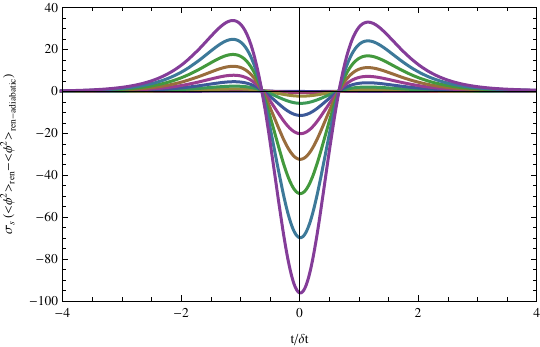}}
        \caption{(Colour online) Renormalized expectation values $\langle \phi^2\rangle_{ren}$ as a function of time $t/\dt$, for $d=5$, 6 and $7$. In each plot, the different curves correspond to different quench rates: $\dt = 1/1,1/2, \cdots, 1/10$ where the curves exhibiting higher peaks (in absolute value) correspond to smaller values of $\dt$. Note that the expectation value is multiplied by the numerical constant $\sigma_s=\frac{2(2\pi)^{d-1}}{\Omega_{d-2}}$. Further, at each time, the expectation value for an `adiabatic' quench is subtracted.}\label{fig_scalars1b}
\end{figure}

\begin{figure}[H]
        \centering
   		 \subfigure[$d=8$]{
                \includegraphics[scale=0.8]{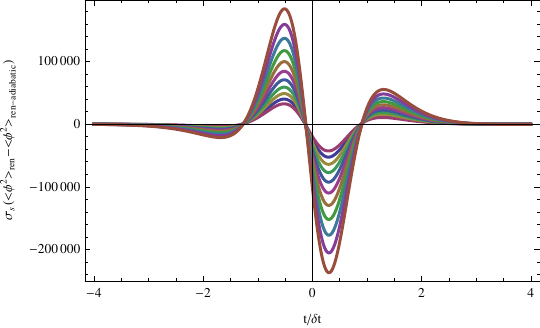}}
         \subfigure[$d=9$]{
                \includegraphics[scale=0.8]{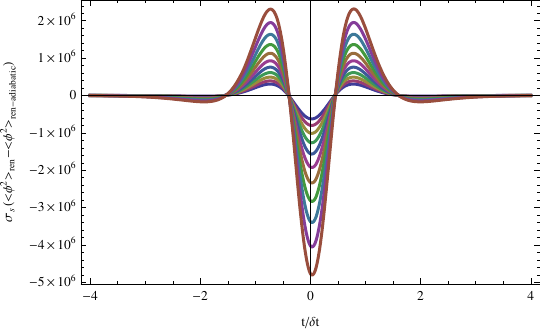}}
        \caption{(Colour online) Renormalized expectation values $\langle \phi^2\rangle_{ren}$ as a function of time $t/\dt$, for $d=8$ and $9$. In each plot, the different curves correspond to different quench rates: $\dt=1/20, 1/21, \cdots, 1/30$ where the curves exhibiting higher peaks (in absolute value) correspond to smaller values of $\dt$. As in the previous figure, the expectation value is multiplied by the numerical constant $\sigma_s=\frac{2(2\pi)^{d-1}}{\Omega_{d-2}}$. Further, at each time, the expectation value for an `adiabatic' quench is subtracted.}\label{fig_scalars2}
\end{figure}

Given eq.~\reef{renorm} for the renormalized expectation value and eq.~\reef{ict} for the necessary counterterm contributions, we are in position to compute $\langle \phi^2\rangle_{ren}$ for spacetime dimensions from $d=3$ to $9$. We first perform this computation numerically. The evolution of the resulting expectation value is shown in figs.~\ref{fig_scalars1a}, \ref{fig_scalars1b} and \ref{fig_scalars2} for different values of the quench rate $\dt$. In these plots, the expectation value for an `adiabatic' quench is subtracted, where the latter actually corresponds to $\dt=10$. We have verified that the expectation value is essentially independent of $\dt$ for larger values. Further, as discussed in section \ref{adiabat}, regulating the expectation value in even dimensions requires the introduction of additional renormalization scales. In the plots presented here, we have set all of these to one, \ie $\mu_i=1$.
Further we have also set $m=1$ in the mass profile \reef{massprofile}.

\begin{figure}[h!]
\setlength{\abovecaptionskip}{0 pt}
\centering
\includegraphics[scale=1]{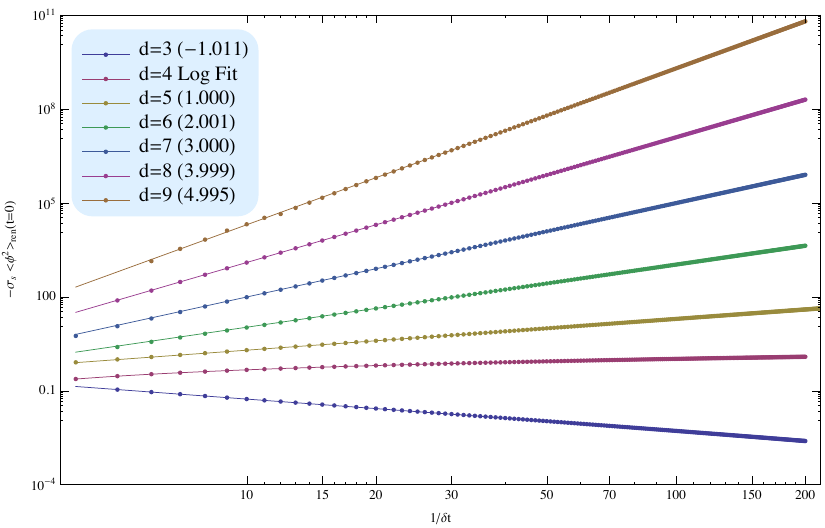}
\caption{(Colour online) Expectation value $\langle\phi^2\rangle_{ren}(t=0)$ as a function of the quench times $\dt$ for spacetime dimensions from $d=3$ to 9. Note that in the plot, the expectation values are multiplied by the numerical factor: $\sigma_s=\frac{2(2\pi)^{d-1}}{\Omega_{d-2}}$.
The slope of the linear fit in each case is shown in the brackets beside the labels.
The results  support the power law scaling $\langle\phi^2\rangle_{ren} \sim \dt^{4-d}$.} \label{fig_all_d_scalars}
\end{figure}

\begin{figure}[h!]
\setlength{\abovecaptionskip}{0 pt}
\centering
\includegraphics[scale=1]{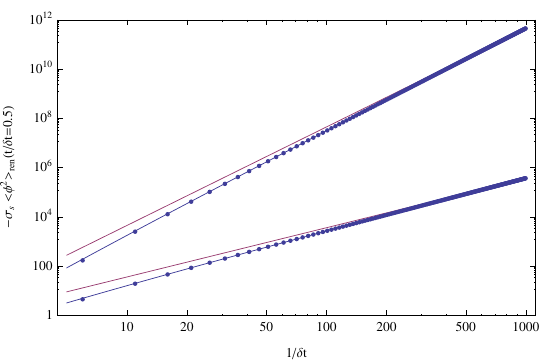}
\caption{(Colour online) Expectation value $\langle\phi^2\rangle_{ren}(t=\dt/2)$ as a function of $\dt$ for spacetime dimensions $d=6$ and $d=8$ --- the lower curve corresponds to $d=6$. As in previous plots, the expectation values are multiplied by $\sigma_s=\frac{2(2\pi)^{d-1}}{\Omega_{d-2}}$. 
We show in a blue solid curve the best fit by a function $f(\dt) = \dt^{-\alpha} (a \log \dt +b)$, where we get $\alpha = 1.9995$ for $d=6$ and $\alpha = 4.0097$ for $d=8$. The purple curve is the best fit for a function $f(\dt) = a \dt^{4-d}$. The plots clearly show that there is an extra logarithmic divergence in expectation values.
The results  support the scaling $\langle\phi^2\rangle_{ren} \propto \dt^{4-d} \log (\dt)$ for even $d$.} \label{fig_all_d_log}
\end{figure}

We can see in figs.~\ref{fig_scalars1a}, \ref{fig_scalars1b} and \ref{fig_scalars2} that the peaks in the expectation value grow (in absolute value) as $\dt$ becomes smaller, and that this growth becomes even faster when the spacetime dimension is increased. To quantify the growth more precisely, fig.~\ref{fig_all_d_scalars} shows $\langle \phi^2\rangle_{ren}(t=0)$ over a broad range of $\dt$, going from $\dt^{-1}=1$ to $\dt^{-1}=200$, in a log-log plot for $d=3$ to 9. Furthermore, for each value of $d$, the linear fits were made to the curve and the results indicate that the expectation value scales as $\langle \phi^2 \rangle_{ren} \sim \dt^{4-d}$ for small $\dt$.\footnote{Recall that we have set $m=1$ and hence $\dt\ll1$ should be interpreted as $m\,\dt\ll1$, in agreement with the fast quench condition \reef{accord}.} For the special case of $d=4$, where this formula seems to indicate no scaling, we found that there is actually a logarithmic scaling. Both of these facts match the scaling found in holographic analysis \cite{numer,fastQ}. 
In particular, the quenched operator is $\phi^2$ with $\Delta=d-2$ and hence the exponent in eq.~\reef{1-3} becomes $d-2\Delta=4-d$, precisely the scaling found with the linear fits. Further given that $\Delta$ is an integer, the holographic results also suggest that there should be an extra logarithmic enhancement for even dimensions \cite{fastQ}, \ie $\langle \phi^2 \rangle_{ren}\propto \dt^{4-d}\log (\dt)$. The logarithmic scaling found for $d=4$ certainly agrees with this expected enhancement, although there was no evidence of such an enhancement in $d=6$ or 8. In section \ref{analytical_cont}, we will see this occurs simply because for the particular tanh profile, the logarithmic contribution simply vanishes at $t=0$. Fig.~\ref{fig_all_d_log} shows similar plots of $\langle \phi^2\rangle_{ren}(t=\dt/2)$ over a broad range of $\dt$ for $d=6$ and 8. There the fit with the extra logarithmic enhancement is clearly preferred over the linear fit.\footnote{Note that as well as the usual fast quench condition \reef{accord}, we must also require that $\mu\,\dt\ll1$ for these logarithmic terms to dominate the scaling behaviour. \label{bang}} Hence we have found that effectively the mass quenches of a free scalar theory quench reproduces precisely the same early time scaling that was discovered with a holographic analysis \cite{numer,fastQ}.

Note that the holographic result is even valid for $d=3$, where there is no divergence but a linear relation to $\dt$. We leave the detailed analysis of this particular case after we discuss the analytical results in section \ref{analytical_cont}.

\subsubsection{Analytical leading contributions: $d \geq 5$}
\label{analytical_cont}

The numerical results above revealed that fast mass quenches in the free scalar theory have the same early time scaling \reef{1-3} as in the holographic quenches \cite{numer,fastQ}. However, looking at the curves of figs. \ref{fig_scalars1b} and \ref{fig_scalars2}, the entire time profile of the expectation value seems to take a relatively simple and possibly universal form.  In particular, for odd spacetime dimensions, one can easily verify that the response takes a form similar to a certain time-derivative of the mass profile. In particular, it seems that $\langle \phi^2 \rangle_{ren} \propto \partial_t^{d-4}m^2(t)$, where the power of the time derivative in this ansatz was chosen as it matches the power-law scalings already discussed. In this section, we will verify this universal form by developing an expansion of the hypergeometric functions which allows us to extract the leading behaviour of the expectation value in the limit in which $\dt \rightarrow 0$. In fact, we will show that this leading behaviour is in perfect agreement with the numerical response presented in previous subsection. In the case of even $d$, we perform a similar expansion to again extract the leading universal response for small $\dt$ and we will find an enhancement by a logarithmic factor.

\begin{figure}[h]
        \centering
        \subfigure[$d=5$]{
                \includegraphics[scale=0.8]{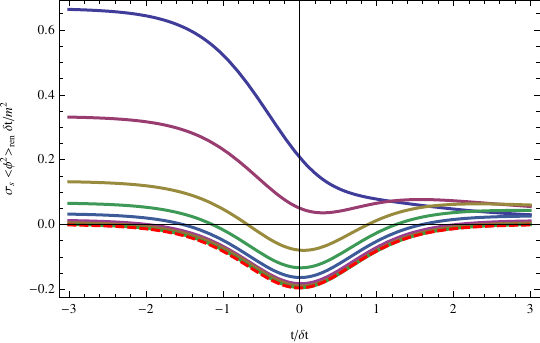}}
   		 \subfigure[$d=7$]{
                \includegraphics[scale=0.8]{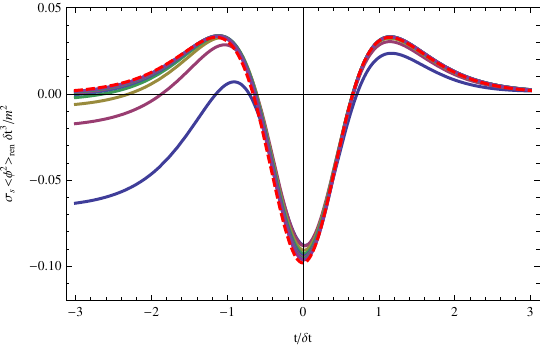}}
        \caption{(Colour online) $\dt^{d-4}\langle\phi^2\rangle_{ren}$ for different values of $\dt$ in odd spacetime dimensions. The curves approach the analytical leading order solution \reef{phi_odd}, shown as the dashed red line, as $\dt$ gets smaller. In panel (a) for $d=5$, running from top to bottom on the left hand side, the solid lines correspond to $\dt=\lbrace 1, 1/2, 1/5, 1/10, 1/20, 1/50, 1/100, 1/500\rbrace$. Similarly from bottom to top in panel (b) for $d=7$, the curves correspond to $\dt=\lbrace 1/2, 1/3, \cdots, 1/10 \rbrace$. }\label{fig_leading_s}
\end{figure}

We first define dimensionless parameters. The relevant physical variables in the quenches here are the initial mass $m$, the momentum $k$ and the quench rate $\dt$. With those, we define
\begin{eqnarray}
\kappa & = & m\, \dt\,, \label{demon}\\
q & = & k\, \dt \,. \nonumber
\end{eqnarray}
Now we want to expand the hypergeometric function for small $\kappa$ and fixed $q$. We will need to expand the hypergeometric series in eq.~(\ref{hyper}) to second order in $\kappa$, which gives
\beq
_2F_1 (a,b;c;z)   
=\sum_{n=0}^{\infty} \frac{(\frac{1-i \kappa^2}{4q})_n (\frac{-i \kappa^2}{4q})_n}{(1-iq -i\frac{\kappa^2}{2q})_n}\, \frac{z^n}{n!}\,, \label{array}
\eeq
where the notation $( \ )_n$ is as defined below eq.~(\ref{hyper}). Also note that given our definition \reef{demon}, terms with higher powers of $\kappa$ will contain extra factors of $\dt$ and so in the limit of $\dt\to0$, these contributions will be subleading, giving a slower scaling with $\dt$.
From eq.~\reef{array}, we see here that each term in the infinite series has an order $\kappa^2$ contribution. Indeed the contribution proportional to $\kappa^2$ is an infinite series in powers of $z$. However, we are only interested in computing $|_2F_1|^2$ and then integrating over all momenta. Remarkably it turns out that for a given $d$ these integrals which multiply factors of $z^p$ vanish for all $p \geq p_d$ where $p_d$ is an integer which depends on $d$. Therefore we can calculate the nonvanishing contributions to $\langle \phi^2 \rangle$ explicitly,
with only the first few terms. We also need to regulate the expectation value after making this expansion. So in the same way as before, we expand the series for large $q$ and subtract the divergent contributions. Of course, this procedure produces leading order expansion in $\kappa^2$ of the counterterm contributions that were found in eq.~\reef{ict}. We are able now to compute the leading contribution in $\kappa^2$ to the expectation value of $\phi^2$. This gives, for odd $d\geq 5$,\footnote{We are putting aside $d=3$ here --- that special case will be analyzed separately at the end of this subsection.}
\beq
\sigma_s \langle\phi^2\rangle_{ren} = (-1)^{\frac{d-1}{2}} \frac{\pi}{2^{d-2}}\, \partial^{d-4}_t m^2(t) + O(\dt^{6-d}).
\label{phi_odd}
\eeq
Note that to get this universal result, we need to use the same relations that were used in computing the counterterms in order to relate $z$ with $m^2(t)$. As we are considering the mass profile $m^2(t)=\frac{m^2}{2} (1-\tanh t/\dt)$, eq.~(\ref{phi_odd}) supports the early time scaling
\beq
\langle \phi^2\rangle_{ren} \sim m^2/\dt^{d-4}\,, \label{lazy}
\eeq
that was found numerically above. Here, it emerges from the leading term in an analytical expansion when $\dt\rightarrow0$. A nice way to visualize this behaviour is to replot the numerical results as $\dt^{d-4}\langle\phi^2\rangle_{ren}$ and compare the curves with the leading order contribution \reef{phi_odd}. This is shown for $d=5$ and $7$ in fig.~\ref{fig_leading_s}. As we see, the numerical curves collapse down onto the leading analytical profile as $\dt$ gets smaller and smaller. Further, the plots demonstrate that that the numerical curves converge to the leading behaviour \reef{phi_odd} more quickly in higher dimensions, as might be expected since the power law scaling is more pronounced.

\begin{figure}[H]
        \centering
        \subfigure[$\dt^2\langle\phi^2\rangle_{ren}$ for $d=6$]{
                \includegraphics[scale=0.8]{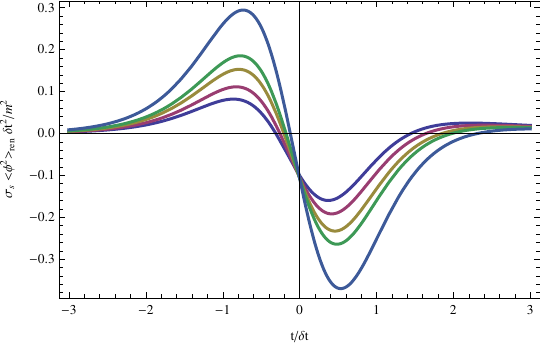}\label{no_log}}
   		 \subfigure[$\dt^2\langle\phi^2\rangle_{ren}/\log \dt$ for $d=6$]{
                \includegraphics[scale=0.8]{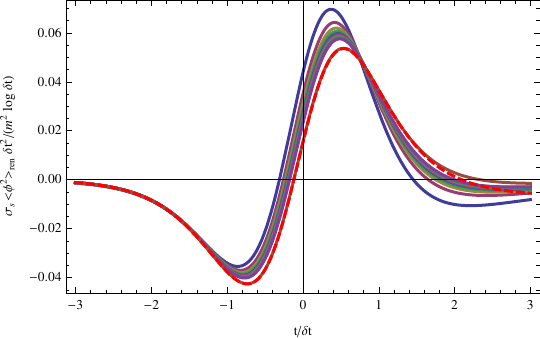}\label{log}}
        \caption{(Colour online) Renormalized expectation value for different values of $\dt$ in $d=6$. Panel (a) shows $\dt^2\langle\phi^2\rangle_{ren}$. As we reduce $\dt$ from $\dt=1/10$ to $\dt=1/1000$, the peaks in the response continue to grow. In particular, the curves correspond to $\dt=\lbrace 1/10, 1/20, 1/50, 1/100, 1/1000\rbrace$. Panel (b) shows $\dt^2\langle\phi^2\rangle_{ren}/\log \dt$. Here as $\dt$ decreases, the curves converge to the analytic expression (red dashed line). In this case, the amplitude of the left peak increases monotonically as $\dt$ shrinks and the various curves correspond to $\dt=\lbrace 1/10, 1/20, 1/30,\cdots, 1/100, 1/1000\rbrace$.}
        \label{fig_leading_s_even}
\end{figure}

Now let's turn to the case of even dimensions where the situation is more subtle. First, we have the IR regulator $\mu$ which we use to produce the dimensionless variable $\nu =\mu\dt$ along with $\kappa$ and $q$, as in eq.~\reef{demon}. Now we follow the same procedure as before: expanding to leading order in $\kappa^2$ and further expanding for large $q$ to find the counterterm contributions. The difference in this case is that in evaluating $\langle\phi^2\rangle_{ren}$, the integration over the momentum is divided into two regions, as described in subsection \ref{adiabat}, and this is where the $\nu$ dependence will appear. In fact, in a manner similar to that found above, we find that the entire $\nu$ contribution is encoded in the first few terms of the expansion of hypergeometric functions and after some manipulation, those terms simplify to yield
\beq
\sigma_s \langle \phi^2\rangle^{(d)}_{ren} = (-1)^{d/2} \log(\mu \dt)\, \frac{\partial^{d-4}_t m^2(t)}{2^{d-3}} + \cdots,
\label{even_log}
\eeq
where we already wrote the expectation value in terms of \textit{dimensionful} $\mu$ and the dots indicate terms independent of $\mu$. However, let us note that we will see that the latter contributions include terms that still scale as $\dt^{4-d}$. Further let us re-iterate the comment in footnote \ref{bang} that for the above behaviour contribution to become dominant, we need $\mu \dt \ll1 $ as well as $m\,\dt\ll1$ to be in the fast quench regime. Hence eq.~(\ref{even_log}) reveals a further logarithmic enhancement of the leading response over the power law scaling in eq.~\reef{1-3}. Rather for even $d$, we find
\beq
\langle \phi^2\rangle_{ren} \sim m^2 \frac{\log \dt}{\dt^{d-4}}\,,
\label{gonzo}
\eeq
where we have set $\mu=1$ above. In fact, this logarithmic enhancement is exactly the kind of behaviour found in the holographic studies \cite{numer,fastQ}. If we present the numerical response as $\dt^2\langle\phi^{d-4}\rangle_{ren}$, as is shown in fig.~\ref{no_log} for $d=6$, the peaks in the curves continue to grow as $\dt$ becomes smaller and smaller. This growth reflects the additional logarithmic factor appearing above in eq.~\reef{gonzo}.

In fig.~\ref{fig_leading_s_even},  we show instead $\dt^2\langle\phi^{d-4}\rangle_{ren}/\log \dt$ for $d=6$. There is also red dashed line that corresponds to the  leading order expression derived analytically and as expected, the numerical response collapses down onto this analytical profile as $\dt$ decreases. However, there is still a part of this analytic response that we need to describe. Basically, the hypergeometric function will give us a structure like
\beq
\langle \phi^2\rangle_{ren} = \phi_1(t) \dt^{4-d} \log (\mu \dt) + \phi_2(t) \dt^{4-d} + O(\dt^{6-d})\,,
\label{gonzo9}
\eeq
where $\phi_1(t)$ is given by eq.~(\ref{even_log}). Unfortunately, $\phi_2(t)$ cannot be expressed as neatly as in the case of $\phi_1(t)$, possibly indicating that the form of this contribution is not universal. In fact, all of the terms in the expansion \reef{array} of the hypergeometric functions contribute to this profile. The result for even dimensions $d\geq 4$ can be written
\beq
\phi_2(t) = \lim_{h\rightarrow\infty} (-1)^{d/2}\sum _{i=2}^h (-1)^{i} \frac{\log \left(i\right)\,i^{d-4}}{2\,(i-1)!} \sum _{j=1}^{h-1} \left(z^{j+1}\prod _{k=0}^{i-2} (j-k)\right).
\label{phi2}
\eeq
We have written the double sum in terms of a limit because we found that we can approximate the entire expression for $\phi_2(t)$ well with the expression above where $h$ is kept finite but large. In particular, the analytical profile shown in fig.~\ref{log} corresponds to eq.~\reef{gonzo9} evaluated with $\dt=10^{-3}$ and taking $h=25$ in eq.~\reef{phi2}, as well as $m=\mu=1$. Again, as shown in fig.~\ref{log}, there is essentially exact agreement between the numerical solution and this analytic profile. Note also that even for $\dt=10^{-3}$, $\log(\dt)\sim-6.9$ and so both terms in eq.~\reef{gonzo9} contribute significantly to the expectation value, \ie one must go to much smaller values of $\dt$ before $\phi_2(t)$ can be neglected.

Finally, let us turn to the question of why we did not see the logarithmic enhancement in the original numerical results, \ie in figs.~\ref{fig_scalars1a}, \ref{fig_scalars1b}, \ref{fig_scalars2} and \ref{fig_all_d_scalars}. Recall that in those plots, we were examining $\langle\phi^2\rangle_{ren}(t=0)$ as a function of the quench times $\dt$. The key point here is that we choose to evaluate the response at time $t=0$. Here we might note that in fig.~\ref{no_log}, all of the curves go through the same point at precisely $t=0$, \ie the entire scaling has been removed by multiplying by $\dt^{d-4}$ at this time. This effect arises because we are studying the specific mass profile $m^2(t)=m^2(1-\tanh(t/\dt))/2$. In this case, any even number derivatives of this profile precisely vanishes at $t=0$. Hence we were simply unlucky in our choice of the time at which to sample the response. As shown in figs.~\ref{fig_all_d_log} and \ref{no_log}, the logarithmic enhancement can be seen in the numerical results when we examine the response at any other value of $t$. 

\subsubsection{Analytical leading contributions: Low dimensional spacetimes} \label{low}

There are a number of reasons to treat $d=3$ and 4 separately. First,  eq.~(\ref{phi_odd}) does not make sense when $d=3$ since the latter would give a negative number of time derivatives in this formula. Moreover, for both $d=3$ and 4, all terms in the hypergeometric series expansion \reef{array} contribute. Finally, our expansion in powers of $\kappa$ has some problems in the IR related to simultaneously taking the limits $\kappa, q\rightarrow0$.

Let us illustrate the latter problem with $d=3$. In this case, the counterterm contribution \reef{ict} reduces to $f_{ct} (k, m(t)) = 1$ and hence in terms of dimensionless variables, eq.~\reef{renorm} can be written as
\beq
d=3\ :\ \ 
\langle\phi^2\rangle_{ren} = \frac{1}{4\pi\,\dt} \int dq \left( \frac{q}{\sqrt{q^2+\kappa^2}}\, |_2F_1|^2 - 1 \right)\,. \label{re3norm}
\eeq
However, if we now first expand the integrand in powers of $\kappa$ and then consider the limit $q\to0$, we find an extra divergent term: $-\kappa^2/(2q^2)$. Of course, this ill-behaved term arises because we are expanding $q/\w_{in}\dt=1/\sqrt{1+\kappa^2/q^2}$ for both $\kappa$ and $q$ around zero. For general dimensions, this term becomes $q^{d-2}/\w_{in}\dt=q^{d-3}/\sqrt{1+\kappa^2/q^2}$ and the order $\kappa^2$ term becomes $-\frac{\kappa^2}2 q^{d-5}$. Therefore a similar logarithmic divergence appears for $d=4$ but no extra divergence appears at order $\kappa^2$ for $d\ge5$. Furthermore, we observe that we do not encounter any IR divergence coming from the same expansion for the \textit{reverse} quench (\ie $A=B=1/2$) in any $d$. In the latter quenches, we have simply $\w_{in}\dt=q$. Finally, we note that no such IR divergence appeared in the numerical calculations for either $d=3$ or 4. Therefore we conclude that this is not a physical divergence of our system. Rather it is a spurious problem generated by our expansion in powers of $\kappa$. 

Hence we remove this divergence by simply subtracting the spurious term as an extra counterterm contribution, which yields for $d=3$,
\beq
\langle \phi^2\rangle_{ren} = \sigma_s^{-1} m^2 \dt\, \frac{\pi}{4} \sum_{i=1}^{\infty} \frac{z^i}{i}+ O(\dt^3) = -  \frac{\pi}{4\sigma_s}\,m^2 \dt  \log \left( \frac{1-\tanh t/\dt}{2} \right) + O(\dt^3)\,, \label{d3ad}
\eeq
where $\sigma_s$ was defined in eq.~\reef{sigs}. Above, the second expression is just the sum that appears when expanding the hypergeometric function and the third one is the result of summing all the terms in the sum. We might also mention that for the \textit{reverse} quench in $d=3$, we find $\langle \phi^2\rangle_{ren} = \frac{\pi}{4\sigma_s}m^2 \dt  \log \left( \frac{1-\tanh t/\dt}{2} \right) + O(\dt^3)$, which is just the negative of the above result.

\begin{figure}[H]
        \centering     
   		 \subfigure[$\dt=10^{-3}$]{
                \includegraphics[scale=0.8]{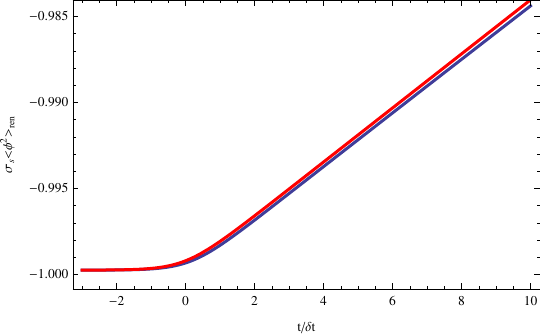} \label{d3-1}} 
         \subfigure[$\dt=10^{-4}$]{
                \includegraphics[scale=0.8]{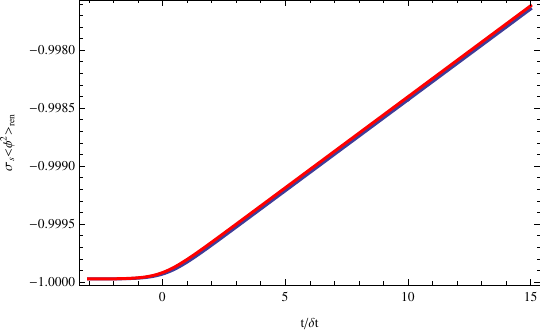} \label{d3-2}} 
         \subfigure[Reverse quench with $\dt=10^{-4}$]{
                \includegraphics[scale=0.8]{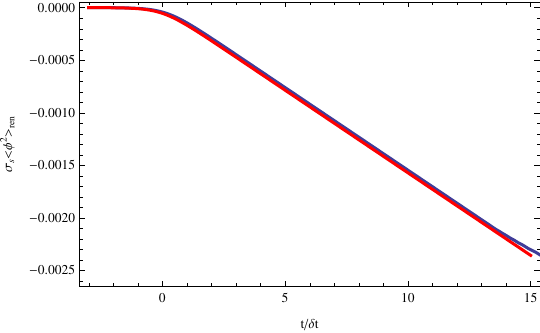} \label{d3-3}} 
        \caption{(Colour online) $\sigma_s \langle\phi^2\rangle_{ren}$ in three-dimensional space-time. The red curves correspond to the leading order analytic expression \reef{d3adv2} while the blue curves are the full numerical solution. By comparing panels (a) and (b), we can see that the difference between the two solutions is roughly of order $O(\dt)$. We can observe that apart from having an extra minus sign difference, the reverse quench in panel (c) starts from zero without needing to be shifted by the factor of $m$.  }\label{fig_d3} 
\end{figure}

Let us also say that subtracting that extra counterterm has its effect on the final expression for the expectation value. In fact, by carefully comparing the full numerical integration with the analytic answer we found that they are shifted by a factor $\sqrt{m^2}$. We write $m$ in this way to emphasize that this extra term is non-analytic in $m^2$, so in fact what we are finding is that
\beq
\langle \phi^2\rangle_{ren} = -\frac{m}{4\pi}  - \frac{m^2\dt}{16}\,  \log \left( \frac{1-\tanh t/\dt}{2} \right) + O(\dt^3) \,, \label{d3adv2}
\eeq
where we have substituted $\sigma_s=4\pi$ for $d=3$ using eq.~\reef{sigs}.
We can recognize, though, that this extra term is due to the $\kappa$-expansion because, for instance, it does not appear in the reverse quench where $\omega_{in}^2=k^2+m^2$ and there is no problem in taking both limits. This difference is illustrated for both types of quenches in figs.~\ref{d3-2} and \ref{d3-3}. In section \ref{late2}, we will see that this constant term simply corresponds to the renormalized expectation value for a fixed mass.

Now if the leading term in eq.~\reef{d3ad} is evaluated in the middle of the mass quench, we have $\langle \phi^2\rangle_{ren} (t=0)= \frac{\log2}{16}\,m^2 \dt$. Hence as observed in fig.~\ref{fig_all_d_scalars}, this result is linear in $\dt$ and so it actually approaches zero in the limit $\dt\to0$. This behaviour should be contrasted with the growing response found in higher dimensions, \eg as shown in eq.~\reef{lazy}. In fact, the same diminishing response will be found in $d=3$ when the expectation value is evaluated for any finite value of $t/\dt$. However, this scaling is deceptive as it may lead one to expect that the quench has a vanishing effect in the limit $\dt\to0$. Considering eq.~\reef{d3ad} but in the limit $t/\dt\gg1$ instead, we find 
\beq
\langle \phi^2\rangle_{ren}(t\gg\dt)\sim \frac18\,m^2\,t\,,
\label{lin3}
\eeq
which is independent of the quench rate!

We will analyze the late time behaviour of the quench in greater detail in section \ref{late2}. However, the above expression \reef{lin3} clearly indicates that the apparent scaling behaviour shown in fig.~\ref{fig_all_d_log} and eq.~\reef{d3ad} does not give an accurate characterization of the overall effect of the mass quenches in three dimensions. Pushing our numerical results to longer times, we were able to go as far as $t/\dt \sim 18$. At these `late' times, we found that the response is indeed linear and independent of $\dt$. For instance, a linear fit to the numerical results in fig. (\ref{fig_d3}) certainly respects the analytic limit.\footnote{We also observe that examining the curves in figs.~\ref{d3-1} and \ref{d3-2} shows that the analytic expression \reef{d3adv2} differs from the full numerical results only by terms that are roughly of order $\dt$.} However, in section \ref{late2}, we will show that for very late times, where $m^2 t^2 \gg 1$, the growth is no longer linear but rather logarithmic.

This result also highlights another key difference between eq.~\reef{d3ad} and the leading behaviour \reef{phi_odd} found in higher dimensions. In higher dimensions, the time profile of the leading analytic term approaches zero exponentially fast (with a `tanh' mass profile) for $t/\dt\gg1$, while in eq.~\reef{d3ad}, the corresponding time profile grows without bound at large times.

The situation for $d=4$ is quite similar, but now the $\kappa$ expansion generates an extra logarithmic divergence in the calculation of the response, as already commented above. However, the same discussion as in the case of $d=3$ still applies. Being in an even number of dimensions, the leading order response has two components as in eq.~\reef{gonzo9} and so for $d=4$, we have
\beq
\langle \phi^2\rangle_{ren} = \phi_1(t)\,\log (\mu \dt)  + \phi_2(t) + O(\dt^{2})\,.
\label{logphi}
\eeq
Here we find $\phi_1(t) = \frac{m^2}{4}(1+\tanh(t/\dt))$ and $\phi_2(t)$ is given by eq.~(\ref{phi2}) with $d=4$. We also note that for the \textit{reverse} quench in $d=4$, the only difference is that we find $\phi_1(t) = - \frac{m^2}{4}(1+\tanh(t/\dt))$. Evaluated at $t=0$, the leading contribution for small $\dt$ is just $\langle \phi^2\rangle_{ren} = \frac{m^2}{4}\log (\mu \dt)$. Hence as the numerical results in fig.~\ref{fig_all_d_scalars} showed, the leading contribution in $d=4$ scales logarithmically when $\dt\rightarrow 0$. Again, this logarithmic scaling was also agrees with the behaviour found in holographic quenches \cite{numer,fastQ}. We might also comment that for large times, \ie $t/\dt\gg1$, both $\phi_1(t)$ and $\phi_2(t)$ approach a constant in eq.~\reef{logphi}.

\subsubsection{The stress-energy tensor}
\label{stress_energy}
There is an elegant and independent consistency check of our results involving the energy density. In particular, we can consider the diffeomorphism Ward identity \cite{numer}
\beq
\partial_t \langle {\cal E}\rangle= - \langle {\cal{O}}_\Delta\rangle\ \del_t \lambda\,,
\labell{ward}
\eeq
where ${\cal E}$ is the (renormalized) energy density. In the case of a constant mass, this identity simply expresses the conservation of energy for the system, \ie the RHS vanishes identically. But in the case of a time-dependent coupling, eq.~\reef{ward} determines the work done by the quench. Following the conventions of \cite{numer}, in our case, $\lambda = m^2(t)$ and ${\cal{O}}_\Delta = -\frac{1}{2} \phi^2$, so with our previous analysis, we already have all of the information needed to compute the RHS of the identity. The independent consistency check will then consist of evaluating the time derivative of the energy density, \ie computing the LHS of eq. (\ref{ward}) directly.

The energy density, defined by the $T_{tt}$ component of the stress-energy tensor, is given by
\beq
{\cal{E}} = \frac{1}{2} \Big(\partial_t \phi\, \partial_t \phi + \partial_i \phi\, \partial_i \phi + m(t)^2\, \phi^2 \Big)\,,
\eeq
where the index $i$ is summed over the spatial dimensions. Given our mode expansion (\ref{fieldx}), this expression results the following expectation value,
\beq
\langle {\cal{E}} \rangle = \frac{\Omega_{d-2}}{2 (2\pi)^d} \int k^{d-2}dk\,\Big( |\partial_t u_\vk|^2+|\partial_i u_\vk|^2 + m(t)^2 |u_\vk|^2
\Big)\,.
\label{en_density}
\eeq
Now it is straightforward to check analytically that taking the time derivative of the above expression and simplifying the result with the equations of motion for the scalar field,
yields exactly $\partial_t \vev{{\cal{E}}} = \frac{1}{2} \partial_t m^2(t) \langle \phi^2 \rangle$, as required by eq.~\reef{ward}.

\begin{figure}[H]
        \centering
        \subfigure[ ]{
                \includegraphics[scale=0.81]{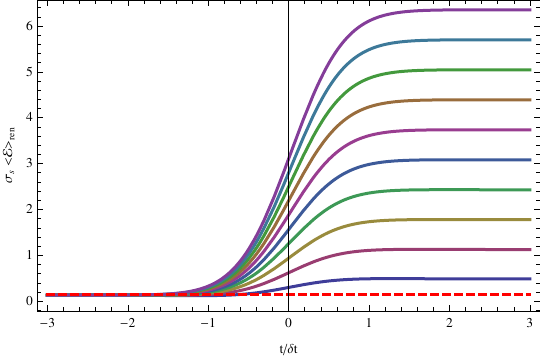}\label{energy1}}
   		 \subfigure[ ]{
                \includegraphics[scale=0.74]{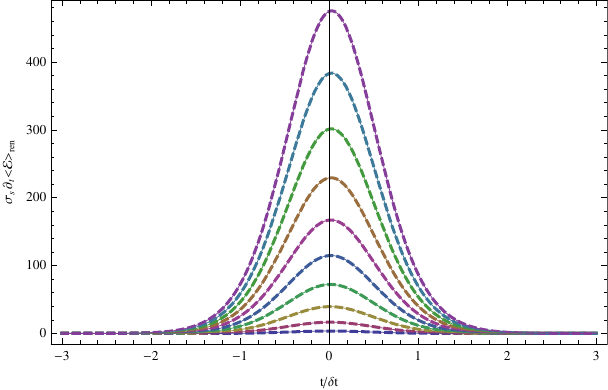}\label{energy2}}
        \caption{(Colour online) Numerical verification of the diffeomorphism Ward identity \reef{ward} for $d=5$. Panel (a) shows $\langle {\cal{E}} \rangle$ as a function of time. Panel (b) shows the corresponding $\partial_t \langle {\cal{E}} \rangle$ as a function of time (dashed) and the RHS of Ward identity (thin solid) evaluated using our previous results. In each case, the curves from top to bottom correspond to $\dt=1/10, 1/20, 1/30, \cdots,1/100$. The straight red dashed line in panel (a) shows $\langle {\cal{E}} \rangle$ for the constant mass case ($m^2=1$).}
        \label{fig_energy}
\end{figure}

We can also verify this agreement numerically. For simplicity, we will set $d=5$ and in this case, we know that all counterterms come from the zeroth order terms in the adiabatic expansion, which can be extracted from the constant mass expectation value --- see discussion below. In this case, eq.~\reef{en_density} reduces to
\beq
\langle {\cal{E}} \rangle_{m^2(t)=m^2} = \sigma_s^{-1} \int dk\, k^3 \sqrt{k^2+m^2} = \sigma_s^{-1} \int dk  \left( k^4 + \frac{m^2}{2} k^2 - \frac{m^4}{8} + O(k^{-2}) \right)\,. 
\label{ect}
\eeq
Hence we know the necessary counterterms in $d=5$ will to regulate the expectation value of the energy density by subtracting off these first three terms, with $m^2$ replaced by $m(t)^2$. With this subtraction, we can evaluate the finite part of eq.~(\ref{en_density}) to get fig. (\ref{energy1}). By numerically differentiating it with respect to time we should get exactly the RHS of eq. (\ref{ward}) and that is indeed the result, as shown in fig.~\ref{energy2}.

As a further check of our analysis, we can verify that the counterterm contributions have the expected form. That is, even though we are finding them separately and independently in eqs.~\reef{ict} and \reef{ect}, they should actually come from the same counterterm action, as discussed in section \ref{effaction}. In the present case of $d=5$, the action in eq.~\reef{ctact} reduces to five terms
\beq
S_{ct}(m^2,g_{\mu\nu},\Lambda) =-\int d^dx \sqrt{-g} \left[ s_{00} \Lambda^5 + s_{10} m^2 \Lambda^3  + s_{20} m^4\Lambda+
R \left(s_{50} \Lambda^3 + s_{51} m^2 \Lambda\right)\right]
\label{ctact22}
\eeq
The counterterm contributions to $\langle \phi^2 \rangle$ and $\langle {\cal{E}} \rangle$ are then determined from this action by eqs.~\reef{gamma1} and \reef{gamma2}, respectively. It is clear that the terms involving the Ricci scalar do not contribute to $\langle \phi^2 \rangle$ when the latter is evaluated in flat space. Similarly, the variation of the $s_{50}$ term to $\langle {\cal{E}} \rangle$, coming from the variation with respect to the metric, vanishes in flat space. Finally, the variation of the $s_{51}$ term yields a contribution of the form:\footnote{See the discussion related to eq.~\reef{blech3} below. We also note that while the $tt$ component vanishes here, this contribution would still be essential to regulate the pressure in the present quenches.} $\langle T_{\mu\nu}\rangle \sim \left(\partial_\mu\partial_\nu- \eta_{\mu\nu}\Box\right)m^2$. However, since the mass only depends on time, one finds that this particular contribution vanishes for the energy density, $\langle {\cal{E}} \rangle=\langle T_{tt}\rangle $. Hence, in fact, only the first three counterterms in eq.~\reef{ctact22} will contribute in the present case. That is, we should find
\begin{eqnarray}
\langle {\cal{E}} \rangle & \sim &  \frac12\left(s_{00}\, \Lambda^5 + s_{10}\,m^2\, \Lambda^3  + s_{20}\,m^4 \, \Lambda \right)
\,, \nonumber\\
\langle \phi^2 \rangle & \sim & s_{10}\, \Lambda^3 + 2 s_{20}\, m^2\, \Lambda \,.
\label{ghost}
\end{eqnarray}
Now if we integrate eq.~(\ref{ect}) up to a momentum $k_{max}$ and compare to the analogous result in eq.~\reef{naive}, we find 
\begin{eqnarray}
\langle {\cal{E}} \rangle & \sim & \frac{1}{\sigma_s}\left(\frac{ k_{max}^5}{5} + \frac{m^2 k_{max}^3}{6}-\frac{m^4 k_{max}}{8}\right)\,, 
\nonumber\\
\langle \phi^2 \rangle & \sim & \frac{1}{\sigma_s}\left(\frac{k_{max}^3}{3}-\frac{m^2 k_{max}}{2}\right)\,. \label{ghost2}
\end{eqnarray}
Hence we find that the coefficients of the cubic and linear divergences match between the two expectation values, as desired . Further, we can supplement the list of coefficients in eq.~\reef{coeffs} with $s_{00}=-1/(d\,\sigma_s)$, after generalizing eq.~\reef{ect} to $d$ dimensions.

\subsubsection{The energy density in three dimensions}
\label{sec_3d_stress}

\begin{figure}[h!]
\setlength{\abovecaptionskip}{0 pt}
\centering
\includegraphics[scale=1]{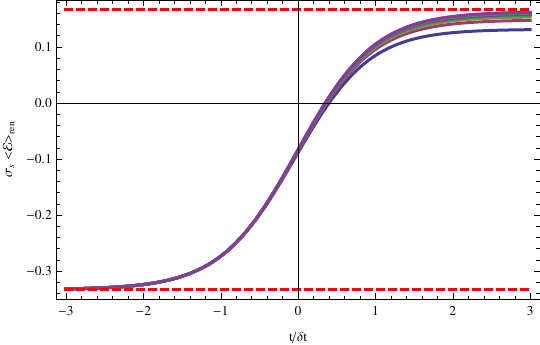}
\caption{(Colour online) Renormalized expectation value of the energy density as a function of time for different values of $\dt$. From bottom to top (on the right hand side), the different curves correspond to $\dt=1/10,1/20,1/30, \cdots, 1/100$. Hence with decreasing $\dt$, the curves accumulate towards the top red dashed line at late times. Note that all expectation values are multiplied by the constant $\sigma_s=4\pi$. The red dashed line at the bottom corresponds to the constant mass value (with $m^2=1$) while the one at top corresponds to 1/6 --- see main text for explanation of this value.} \label{fig_d3_en1}
\end{figure}

\begin{figure}[h!]
\setlength{\abovecaptionskip}{0 pt}
\centering
\includegraphics[scale=1]{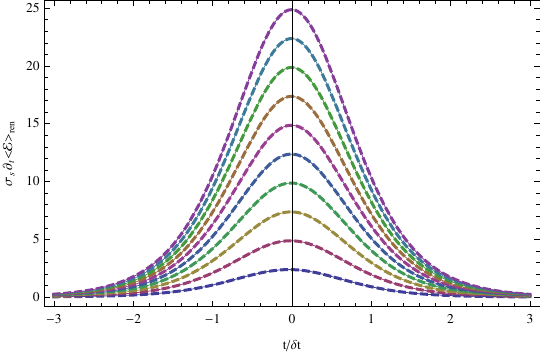}
\caption{(Colour online) Time derivative of the renormalized expectation value of the energy density as a function of time for different values of $\dt$. Different curves correspond to $\dt=1/10,1/20,1/30, \cdots, 1/100$, with curves with smaller $\dt$ correspond to higher peaks. Note that all expecation values are multiplied by a constant $\sigma_s=4\pi$. The dashed lines correspond to the time derivative of $\langle {\cal{E}} \rangle_{ren}$ while the thin solid lines correspond to evaluating $\frac{1}{2} \partial_t m^2(t) \langle \phi^2 \rangle_{ren}$. The agreement between both calculations shows that the diffeomorphism Ward identity is satisfied.} \label{fig_d3_en2}
\end{figure}

As in the case of section \ref{low}, it is interesting to repeat the above analysis but focusing on the $d=3$ case separately.
In this case, the scaling found for $\langle \phi^2 \rangle_{ren}$ in section \ref{responseq} is proportional to $\dt$. Hence on the RHS of the identity \reef{ward}, this is multiplied by $\partial_tm^2$ which gives a factor of $1/\dt$ and so one would find that $\partial_t\langle{\cal{E}}\rangle_{ren}$ does not scale at all with $\dt$. 
Since the quench essentially takes place over an interval $\dt$, this would then reproduce the na\"ive scaling $\delta\langle{\cal{E}}\rangle_{ren}\sim m^2\dt$ as suggested by eq.~\reef{1-2}, \ie no work is done in the limit $\dt\to0$. However, we will show below that this is not really the case and rather we find that $\partial_t \langle {\cal{E}} \rangle_{ren}$ scales as $1/\dt$ and that $\delta\langle{\cal{E}}\rangle_{ren}\sim m^3$ --- see figures \ref{fig_d3_en1} and \ref{fig_d3_en2}. Note that the latter result indicates that the work done is not analytic in the mass coupling, \ie $\delta\langle{\cal{E}}\rangle_{ren}\sim (m^2)^{3/2}$.

Let us start by computing the expectation value of the energy density for a constant mass. In this case and with $d=3$, eq.~\reef{en_density} yields
\beq
\langle {\cal{E}} \rangle_{m^2(t)=m^2} =   \int \frac{k\, dk}{4\pi}\, \sqrt{k^2+m^2} = \frac1{4\pi} \left( \frac{k_{max}^3}{3}+\frac{k_{max} m^2}{2}-\frac{1}{3} m^3 + O(1/k_{max}) \right)\,. \label{eq_constant_mass_d3}
\eeq
The first two divergent contributions would be removed by the counterterm contributions and hence the renormalized expectation value of the energy density would be $\langle {\cal{E}} \rangle_{ren} = -m^3/(12\pi)$ in the case of a constant mass. Again it is notable that this result is not analytic in the mass coupling. However, we can easily extract the counterterm contributions to regulate the expectation value in the case of a time-varying mass from eq.~(\ref{eq_constant_mass_d3}) to find $f_{ct}^{d=3} = k^2 + m^2(t)/2$. Subtracting these terms in the integral in eq.~(\ref{en_density}) with $d=3$ then yields the renormalized expectation value of the energy density. Then we computed this expectation value numerically for different values of $\dt$ ranging from $\dt=1/10$ to $\dt=1/100$, as shown in fig. (\ref{fig_d3_en1}). We observe that the energy density grows from its corresponding value at minus infinity --- as we set $m=1$, this means $4\pi\langle {\cal{E}} \rangle_{ren}(t\ll0) = -1/3$ --- to a certain constant value at late times. In particular, as $\dt$ becomes smaller, the latter constant seems to be independent of $\dt$. Hence, from this figure, we can see that the na\"ive power counting does not work, because as described above, it suggests that the change in energy density would be proportional to $\dt$.

Further, we can also compute the time derivative of this profile and compare it with the RHS of the Ward identity \reef{ward}, using our previous results for the expectation value of $\phi^2$. Again, we get perfect agreement, as shown in fig.~\ref{fig_d3_en2}. There we also see that $\partial_t \langle {\cal{E}} \rangle$ scales as $1/\dt$.

How can we understand this scaling? The key point is that the change in the expectation value of $\phi^2$ has a scaling proportional to $\dt$ but the full expectation value does not start from zero. Recall from eq.~(\ref{d3adv2}) that the expectation for $d=3$ is given by
\beq
 \langle \phi^2\rangle_{ren} = - \frac{m}{4\pi} - \frac{m^2\, \dt}{16}\,  \log \left( \frac{1-\tanh t/\dt}{2} \right) + O(\dt^3)\,.
\eeq
Now, as $\dt \to0$, the second term and all the subleading will go to zero and then $\langle \phi^2 \rangle_{ren} \simeq -m/(4\pi)$. So if we integrate the Ward identity \reef{ward} in this limit, we find
\beqa
\delta\langle {\cal{E}} \rangle_{ren}&=&\langle {\cal{E}} \rangle_{ren}(t=\infty) - \langle {\cal{E}} \rangle_{ren}(t=-\infty)\nonumber\\
& = &-\frac{m}{8\pi} (m^2(t=\infty) - m^2(t=-\infty))=\frac{m^3}{8\pi}\,. \label{blahblah}
\eeqa
Further at very early times (\ie $t\ll0$), the energy density will match that found in the case of a constant mass. Hence given the results in eq.~\reef{eq_constant_mass_d3}, we have $\langle {\cal{E}} \rangle_{t=-\infty}= -m^3/(12\pi)$. Hence for late times (and small $\dt$), we should find the energy density to approach $\langle {\cal{E}} \rangle_{t=\infty} = m^3/(24\pi)$, which is exactly what is shown for the long time behaviour in fig.~\ref{fig_d3_en1}.

To close this section, we reiterate that eq.~\reef{1-2} suggests the scaling of the energy should be $\delta\langle{\cal{E}}\rangle_{ren}\sim m^2\dt$ for $d=3$. This scaling was not realized here in eq.~\reef{blahblah} but this result depended on the fact that $\langle\phi^2\rangle=-m/(4\pi)$ in the past, \ie at the start of the quench. On the other hand, if we considered a `reverse' quench, where the mass starts at zero and rises to some finite $m$, this initial expectation value would vanish and hence the expected scaling would be fulfilled. That is, zero work is done by the reverse quench in the limit $\dt\to0$.

\subsubsection{Universal scaling of higher spin currents}
\label{higher_scaling}

It is known that free scalar field theory has an infinite set of higher spin conserved currents $j_{i_1\cdots i_s}$ \cite{Vasiliev:1999ba, Mikhailov:2002bp}. Apart from being conserved, these currents are symmetric in their indices and, in the case of massless theory, traceless. It is interesting, then, to analyze how these currents behave in the present quenches. In particular, we will be interested in determining how the higher spin currents scale in the fast quench limit.

Higher spin currents for a massless complex scalar field are given by \cite{Mikhailov:2002bp}
\begin{equation}
j_{i_1\cdots i_s} \propto
\sum\limits_{k=0}^s
{(-1)^k\over k!\left(k+{d-4\over 2}\right)!
 (s-k)!\left(s-k+{d-4\over 2}\right)!}
\partial_{i_1}\cdots\partial_{i_k}\phi^*\;
\partial_{i_{k+1}}\cdots\partial_{i_s}\phi-\mbox{traces},
\label{general_current}
\end{equation}
where indices $i_1,\cdots, i_s$ should be symmetrized above. In case of a complex scalar field, the even spin currents are symmetric under the interchange $\phi \leftrightarrow \phi^*$, while odd spin currents are antisymmetric. In our calculations, we are dealing with real fields and so the odd spin currents trivially vanish. Hence we will only consider the even spin currents.

Let us start by revisiting the spin-2 current, \ie the stress-energy tensor of the conformally coupled scalar. Hence we can obtain this current by varying the scalar field action with respect to the metric,
\begin{eqnarray}
j_{a b}^{(2)} = - \frac{2}{\sqrt{-g}} \frac{\delta S}{\delta g^{a b}},
\end{eqnarray}
where
\beq
S = -\frac{1}{2}\int d^d x \sqrt{-g} \left( \partial_\mu \phi^* \partial^\mu \phi + m^2 \phi^*\phi - \xi R\, \phi^* \phi \right) \label{actop}
\eeq
and $\xi$ takes the usual value for the conformal coupling: $\xi= \frac{1}{4} \frac{d-2}{d-1}$. Upon varying, we obtain
\begin{eqnarray}
j_{a b}^{(2)} &=& - \frac{1}{4} \frac{d-2}{d-1} (\phi \partial_{ab} \phi^* + \phi^* \partial_{ab} \phi) + \frac{d}{4 (d-1)} (\partial_a \phi \partial_b \phi^* + \partial_a \phi^* \partial_b \phi)\nonumber\\
&&\qquad\qquad - \frac{\eta_{ab}}{2 (d-1)} (\partial_c \phi \partial^c \phi^* + m^2 \phi^*\phi)\,.
\label{spin-2}
\end{eqnarray}
We note that the equation of motion, $\Box\phi=m^2\phi$, was used to simplify the above expression.
Further, we can verify that if we set $m^2=0$, the above result reproduces the $s=2$ current in eq.~(\ref{general_current}), up to an overall numerical factor. 
It will be convenient for the following to split the current into two parts: the minimally coupled current (obtained by setting $\xi=0$) and the remaining contribution coming from the conformal coupling term proportional to $R$ in eq.~\reef{actop}. Then we have
\beq
j_{a b}^{(2)} = j_{a b}^{(2) \text{min}} + j_{a b}^{(2) \text{conf}},
\eeq
where 
\begin{eqnarray}
j_{a b}^{(2) \text{min}} & = & \frac{\partial_a \phi \partial_b \phi^* + \partial_a \phi^* \partial_b \phi}{2} - \frac{1}{2} \eta_{a b} (\partial^c \phi^*\partial_c \phi  + m^2 \phi^*\phi)\,, \label{blech1} \\
j_{a b}^{(2) \text{conf}} & = & \frac{1}{4} \frac{d-2}{d-1}  \Big( - (\phi \partial_{ab} \phi^* + \phi^* \partial_{ab} \phi) - (\partial_a \phi \partial_b \phi^* + \partial_a \phi^* \partial_b \phi)  \label{blech2} \\
&&\qquad\qquad\qquad
+ \eta_{a b} (\phi^*\Box\phi+\phi\,\Box\phi^*+2\partial^c \phi^* \partial_c \phi ) \Big) \,. 
\end{eqnarray}
Of course, we have restored the terms involving $\Box\phi$ using the equations of motion in $j_{a b}^{(2) \text{conf}}$.
The reason for doing so is that it makes apparent that $j_{a b}^{(2) \text{conf}}$ is a total derivative, \ie
\beq
j_{a b}^{(2) \text{conf}} = \xi \left( \partial_{ab} (\phi^* \phi) - \eta_{ab} \Box(\phi^* \phi)\right)\,.
\labell{blech3}
\eeq 
Then, in our case (where $\langle \phi^2 \rangle$ only depends on time), we find that the $a=b=t$ component of this part vanishes and we are only left with the minimally coupled current. Therefore the energy density calculated with the full stress tensor \reef{spin-2} agrees with that found with the minimal stress tensor \reef{blech1}, as was done in the previous sections. 

Of course, for a constant mass, the spin-two current \reef{spin-2} is conserved. However, if we allow for a time-varying mass, the divergence of this current yields
\beq
\partial^a j_{a t}^{(2)} = \partial^t j_{t t}^{(2)} = \partial^t j_{t t}^{(2) \text{min}} = -\frac{1}{2} \partial_t m^2(t) \langle \phi^* \phi \rangle\,.
\eeq
Of course, we have reproduced the diffeomorphism Ward identity \reef{ward}, from which we can determine the energy which the quench injects into the system if we are given the expectation value $\langle \phi^* \phi \rangle$. The reason for revisiting this result for the spin-2 current is that we will now apply the analogous analysis with the spin-4 current and we will find the scaling of this higher spin current in the limit of fast quenches. Further, we will use this approach to argue for the scaling of all of the higher even spin currents.

First we must build the spin-4 current for the massive theory as follows: Take eq.~(\ref{general_current}) and explicitly symmetrize the indices. Then introduce all the necessary trace terms with the necessary coefficients to ensure that the result is traceless in the massless case. The next step is to generalize this current for a massive field. Here, we take the divergence of the massless expression and add all the necessary terms proportional to the mass to ensure that the divergence vanishes upon evaluation on the massive equation of motion. This procedure is explicitly carried out for the spin-4 current in Appendix \ref{appendix}. The final result is
\begin{eqnarray}
j^{(4)}_{i_1 i_2 i_3 i_4} = j^{(4) m^2=0}_{i_1 i_2 i_3 i_4} + \frac{m^2}{2 \left(\frac{d}{2}+2\right)! \left(\frac{d}{2}\right)!}\, \eta_{i_1 i_2} \left(  (d+1)\, j^{(2)}_{i_3 i_4} +\frac{2}{d-2}\, j^{(2) \text{conf}}_{i_3 i_4} \right)\,,
\end{eqnarray}
where again the indices in last term should be symmetrized. Now we are interested in obtaining the analogous Ward identity for the spin-4 current. In particular, we make the mass time-dependent and evaluate the time-derivative of the $j^{(4)}_{tttt}$ component. Note that the part proportional to the conformally coupled spin-2 current will vanish and hence we find
\beq
\partial_t \langle j^{(4)}_{t t t t} \rangle = \frac{d+1}{2 \left(\frac{d}{2}+2\right)! \left(\frac{d}{2}\right)!}\, \partial_t m^2(t)\, \langle j^{(2) }_{t t} \rangle = \frac{d+1}{2 \left(\frac{d}{2}+2\right)! \left(\frac{d}{2}\right)!}\, \partial_t m^2(t)\, \langle {\cal {E}} \rangle\,. \label{gabi4}
\eeq
To determine the scaling of this spin-4 `charge density' in the limit of fast quenches, we can use the scaling of the energy density $\langle {\cal E} \rangle \sim m^4/\dt^{d-4}$ to find:
\beq
\langle j^{(4)}_{t t t t} \rangle \sim \frac{(m^2)^3}{\dt^{d-4}}\,.
\eeq
Hence in the fast quench limit, the spin-4 charge diverges with precisely the same power of $\dt$ as the spin-2 charge and the spin-0 charge (\ie $\phi^2$), while an extra power of $m^2$ appears to make up the necessary dimension of the new operator.

Extending the construction of the spin-4 current, described above, to obtain higher spin currents in the massive theory is straightforward, though tedious. 
We expect that the massive terms for the spin-$s$ current can decomposed, as in the spin-4 case, in terms of the spin-($s$--2) current and a total derivative term. Then, it is easy to see that for a time-varying mass, we will get a hierarchy of generalized Ward identities,
\beq
\partial_t \langle j^{(s)}_{t \cdots t} \rangle \sim \partial_t m^2(t)\, \langle j_{t \cdots t}^{(s-2)} \rangle\, . \label{power2}
\eeq
Now integrating these identities will similarly yield a hierarchy of scalings for the final currents in the fast quench limit, \ie $\langle j^{(s)}_{t \cdots t} \rangle \sim m^2\, 
\langle j_{t \cdots t}^{(s-2)} \rangle$. Hence the scaling of all of the higher spin currents would be determined by that originally found from how $\langle \phi^2 \rangle$ scales. Then in general we should find that
\beq
\langle j^{(s)}_{t \cdots t} \rangle \sim \frac{(m^2)^{\frac{s}{2}+1}}{\dt^{d-4}}\,. \label{power}
\eeq
Of course it would be interesting to explicitly construct the currents in the massive theory and derive these scalings for the higher spin currents. However, our expectation is that after a quench, all of currents that will scale with precisely the same power of $\dt$. In particular then, for $d\ge4$, all of these currents will diverge as $\dt \rightarrow 0$.

\subsection{CFT to CFT quenches}
\label{cft_cft}

This subsection is devoted to study the response of the scalar field under a quench whose mass profile is asymptotically zero at both infinite past and future. We smoothly turn on the mass up to some $m^2$ and then go back to the critical point. The whole process is again characterized by a time length $\dt$. We may proceed analytically if we choose the following mass function
\beq
m^2(t) = \frac{m^2}{\cosh^2 (t/\dt)}\,.
\label{mass_pulse}
\eeq
%
%
Analysing this system is interesting because it provides a further check of our previous analysis. In particular, we should expect to have the same scaling behaviour for the renormalized expectation values in the limit of fast quenches. Moreover, the counterterms should be the same as in the previous case with the only difference that we should change the mass function (and its derivatives) to the new profile. Even though this is expected, it is not at all trivial : rather it provides a good confirmation of our results. Lastly, we will return to such CFT-to-CFT quenches later in section \ref{interacting} to give a general argument that should be valid for arbitrary CFTs and hence the present section provides an explicit example of these processes.
Finally, as in the case of the $\tanh$ profile, it is straightforward to extend the present analysis of these pulse-like quenches to include a constant mass, \ie
\beq
m^2(t) = m_0^2+\frac{m^2}{\cosh^2 t/\dt}\,.
\label{mass_pulse2}
\eeq
We will explicitly analyze quenches with this profile in section \ref{massive_massive}. However, our intuition suggests that universal scaling in eqs.~\reef{1-2} and \reef{1-3} should still hold if we satisfy both $m^2\dt^2\ll1$ and $m_0^2\dt^2\ll1$. Again, this emphasizes that what is important is that the theory has a UV fixed point, \ie, the UV description of the theory is a CFT. The IR details become unimportant in the fast quench limit, \ie when $1/\dt$ dominates all of the IR scales.

As with the $\tanh$ profile \reef{massprofile}, we can exactly solve this problem by decomposing the scalar field into momentum modes, as in eq. (\ref{fieldx}). This modes will satisfy the Klein-Gordon equation with mass given in eq. (\ref{mass_pulse}),
\beq
\frac{d^2 u_\vk}{dt^2} + \left( k^2 + \frac{m^2}{\cosh^2 t/\dt} \right) u_\vk =0\,.
\label{eom1}
\eeq
This equation can be written in hypergeometric form by expressing it in terms of variable $y=\cosh^2(t/\dt)$,
\beq
y (1-y) \frac{d^2 u_\vk}{dy^2} + \left(\frac{1}{2}-y \right) \frac{d u_\vk}{dy} - \left( \frac{k^2 \dt^2}{4} + \frac{m^2 \dt^2}{4 y}\right) u_\vk=0\,.
\label{eom pulse}
\eeq

We are interested in those solutions that behave purely as positive frequency waves in the infinite past, so we need to fix initial conditions so that $u_\vk (t\rightarrow -\infty) = \frac{1}{ \sqrt{4 \pi\omega_k}} \exp (-i \vk \cdot \vec{x} - i \omega_k t)$, where $\omega_k$ is just $\omega_k = k$ because in the infinite past we are in the massless theory.\footnote{The way to take this limit is to use identities that relate hypergeometric functions of argument $z$ with a linear combination of hypergeometric functions of argument $1/z$ --- see, for instance, \cite{abramowitz}. In our case, as $t\rightarrow -\infty$, $1-y\rightarrow -\infty$ and then, such identities are useful.} Then, the complete solution for the modes in terms of $k$ and $y$ is given by
\begin{eqnarray}
u_\vk = \frac{1}{\sqrt{4 \pi k}} & &\frac{2^{i k} y^\alpha}{E'_{1/2} E_{3/2} - E_{1/2} E'_{3/2}} \times \nnn \\ 
& & \times \left( E_{3/2} \ _2F_1 (a,b;\frac{1}{2};1-y) + E_{1/2} \sinh(t/\dt) _2F_1 (a+\frac{1}{2},b+\frac{1}{2};\frac{3}{2};1-y) \right), \nonumber \\
\end{eqnarray}
where
\begin{eqnarray}
E_{c} = \frac{\Gamma(c) \Gamma(b-a)}{\Gamma(b) \Gamma(c - a)} \  & , &  \ E'_{c} = E_{c} (a \leftrightarrow b)\,,
\nonumber \\
a = \alpha + \frac{ik \dt}{2} & , & \ \ \ b  =  \alpha - \frac{ik \dt}{2}, \label{a and b} \\
\alpha  & =  &  \frac{ 1 + \sqrt{1+4 m^2 \dt^2} }{4}\,.\nonumber
\end{eqnarray}

\begin{figure}[H]
        \centering
        \subfigure[$d=5$]{
                \includegraphics[scale=0.8]{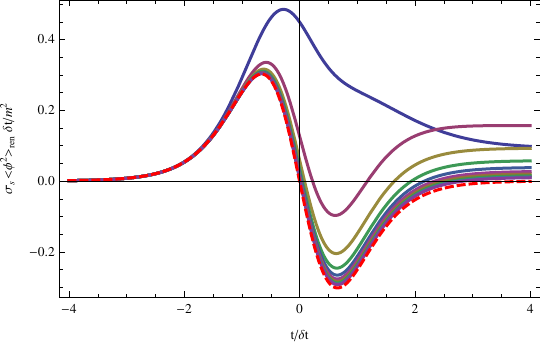}}
   		 \subfigure[$d=7$]{
                \includegraphics[scale=0.8]{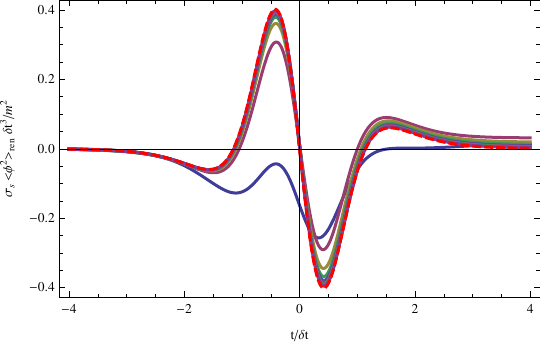}}
         \subfigure[$d=9$]{
                \includegraphics[scale=0.8]{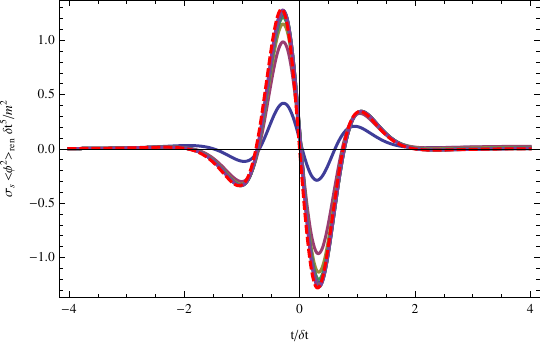}}
        \caption{(Colour online) $\langle\phi^2\rangle_{ren} \dt^{d-4}/m^2$ for different values of $\dt$ and different odd spacetime dimensions $d$. The solid curves correspond to $\dt=1,1/2,\cdots,1/10$ with $\dt$ decreasing as they converge to the analytical leading expression (\ref{phi_odd}), plotted with dashed red curve. This leading term has $\langle\phi^2\rangle_{ren}^{(d)} \sim (-1)^{\frac{d-1}{2}} \partial^{d-4}_t m^2(t)$. }\label{fig_leading_pulse}
\end{figure}

Now, as we did in the previous case, we integrate over momentum modes in evaluating the expectation value of $\phi^2$ and this integral is UV divergent. To get the finite renormalized expectation value, we must subtract the appropriate counterterm contributions. In section \ref{adiabat}, we used an adiabatic expansion to obtain the counterterms supposing only that the mass depends on time. Hence we can expect the counterterm contributions in eq.~(\ref{ict}) will regulate $\langle\phi^2\rangle$ for any mass profile. Hence, we use the \textit{same} expression here and only change the profile of $m^2(t)$ to the pulsed one \reef{mass_pulse}. In this way, we obtain 
\beq
\langle \phi^2 \rangle_{ren} = \int d^{d-1}k \left( |u_\vk|^2 - f_{ct} (k,m(t)) \right)\,, \label{carrot}
\eeq
which is UV-finite, as we will see below. A more nontrivial result is that these expectation values should yield the same leading order behaviour, as derived in section \ref{analytical_cont}, where the results were expressed in terms of derivatives of the mass profile. In fact, we found that this same universal behaviour indeed emerges for the pulsed profile and so eq.~(\ref{phi_odd}) also gives the correct result in this example. In particular, fig.~\ref{fig_leading_pulse} shows the renormalized expectation value of $\phi^2$ for odd dimensions $d=5,7,9$. As we did in the original quenches, here, we divide by the expected scaling $m^2/\dt^{d-4}$ and plot eq.~\reef{carrot} for different time intervals $\dt$. We see that the curves rapidly converge to the analytic expression given in eq.~(\ref{phi_odd}) as $\dt$ goes to zero. This clearly shows that both the expected scaling in eq.~\reef{1-2} and the leading analytical behaviour in eq.~\reef{phi_odd} are valid in the present example of a pulsed quench.

\begin{figure}[h!]
\setlength{\abovecaptionskip}{0 pt}
\centering
\includegraphics[scale=1]{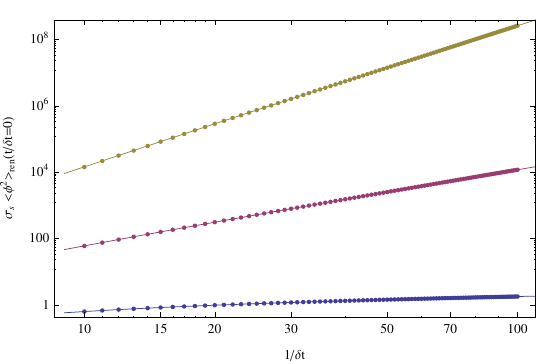}
\caption{(Colour online) Renormalized expectation value of $\phi^2$ at time $t=0$ as a function of $\dt$ for different even dimensions. The blue curve corresponds to $d=4$, where the fit by a function $\dt^{-\alpha} (a_1 - a_2 \log \dt)$ gives $\alpha=0.0028$, showing the expected logarithmic growth; the purple curve corresponds to $d=6$ and the same fit gives $\alpha=2.0006$; the yellow curve corresponds to $d=8$ and the fit results in $\alpha=4.0019$, just as expected by our power law scaling \reef{1-2}.} \label{fig_pulse_evend}
\end{figure}

\begin{figure}[h!]
\setlength{\abovecaptionskip}{0 pt}
\centering
\includegraphics[scale=1]{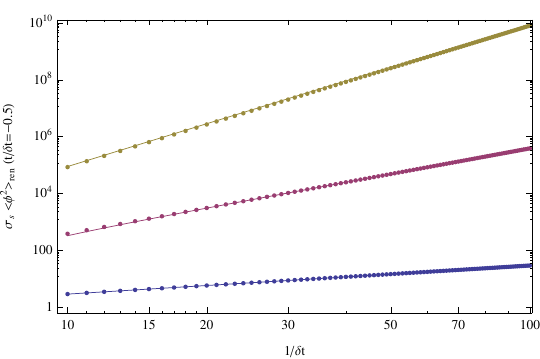}
\caption{(Colour online) Renormalized expectation value of $\phi^2$ at time $t/\dt=-0.5$ as a function of $\dt$ for different odd dimensions. The blue curve corresponds to $d=5$, where the fit by a function $\dt^{-\alpha} a_1 + a_2$ gives $\alpha=1.006$, showing the expected scaling; the purple curve corresponds to $d=7$ and the same fit gives $\alpha=2.991$; the yellow curve corresponds to $d=9$ and the fit results in $\alpha=4.977$, just as expected by our power law scaling \reef{1-2}.} \label{fig_pulse_oddd}
\end{figure}

For even $d$, we expect the scaling to be enhanced by a logarithmic factor, as discussed in section \ref{analytical_cont}. In the case of the previous case with the tanh profiles, we could not see this enhancement in our numerical results \cite{dgm} because the leading order term vanishes at $t=0$. However, in the present case, the even derivatives of the mass are not zero at $t=0$ and hence, we should be able to see the expected behaviour even at zero time. This can be seen exactly in fig.~\ref{fig_pulse_evend}, where the fits of the curves support the scaling $\langle\phi^2\rangle_{ren} \sim m^2/\dt^{d-4} \log \dt $. In contrast, for odd $d$, the corresponding derivatives of pulse profile \reef{mass_pulse} vanish at $t=0$. However, we can instead evaluate $\langle\phi^2\rangle_{ren}(t=-\dt/2)$ to reveal the same scaling applies in odd $d$, as shown in fig. (\ref{fig_pulse_oddd}). Of course, this scaling was already confirmed above by matching the leading analytic behaviour. 

\subsection{Universal scaling for arbitrary initial and final mass}
\label{massive_massive}

In this section, we would like to show that the universal scaling in eq.~\reef{1-2} is not  exclusive to quenches which involve a critical theory at the initial and/or final times, but are also found for arbitrary initial and final mass under certain assumptions. Basically what we need is $1/\dt$ to be the only relevant scale of the problem. So long the initial and final mass (and their difference) are much smaller than $1/\dt$, we will find the same scaling.

For the $\tanh$ profile \reef{massprofile}, this scaling can be explicitly seen by extending the analysis of the renormalized expectation value \reef{renorm} to general initial and final masses, \ie general $A$ and $B$ in eq.~\reef{massprofile}. Hence we have
\beq
\langle\phi^2\rangle_{ren} =  \int \frac{k^{d-2}dk}{\sigma_s\,\omega_{in}}\, \left[\left| _2F_1\left( 1+ i \omega_- \dt, i \omega_- \dt; 1 - i \omega_{in} \dt; \frac{1+\tanh(t/\dt)}{2} \right) \right|^2-f_{ct}(k,m(t))\right]\, ,
\label{phi squared app}
\eeq
where
\begin{eqnarray}
\omega_{in}^2 & = & k^2 + m^2 (A-B) \equiv k^2 + m_i^2 , \\ 
\omega_{out}^2 & = & k^2 + m^2 (A+B) \equiv k^2 + m_f^2 .
\end{eqnarray}
and the counterterm contributions $f_{ct}(k,m(t))$ are given by eq.~\reef{ict}.

Now let us redefine the integration variable in eq.~\reef{phi squared app}. We define 
\beq
\tilde{k}^2 \equiv k^2 + m_f^2
\label{arkan}
\eeq
and hence
\begin{eqnarray}
\omega_{in}^2 & = & \tilde{k}^2 + m_i^2 - m_f^2 \equiv \tilde{k}^2 + (\delta m^2) , \\ 
\omega_{out}^2 & = & \tilde{k}^2.
\end{eqnarray}
With this choice, eq.~\reef{phi squared app} starts to look like the expectation value for a quench from an initial mass-squared $(\delta m^2)$ to the massless case. In fact, the absolute value of the hypergeometric function in the integrand will look exactly like that. We have to take care about the rest of the integral. Applying the change of variables \reef{arkan}, eq.~\reef{phi squared app} becomes
\beq
\langle\phi^2\rangle_{ren} = \int_{m_f}^\infty \frac{\tilde{k}\,d\tilde{k}}{\sigma_s\,\omega_{in}}\, \left(\tilde{k}^2-m_f^2\right)^{\frac{d-3}{2}}\, \left[\left| _2F_1 \right|^2-f_{ct}(\sqrt{\tilde{k}^2-m_f^2},m(t))\right]\, .
\eeq
Further as in section \ref{analytical_cont}, we introduce a dimensionless momentum $\tilde{q} = \tilde{k} \dt$, which then yields
\beq
\langle\phi^2\rangle_{ren} = \frac{1}{\dt^{d-4}} \int_{m_f \dt}^\infty \frac{\tilde{q}^{d-2}d\tilde{q}}{\sigma_s\,\omega_{in}}\left(1-\frac{m_f^2\dt^2}{\tilde q^2}\right)^{\frac{d-3}{2}}\Big[\left| _2F_1 \right|^2-f_{ct}\left(\tilde q\,(1-m_f^2\dt^2/\tilde q^2)^{1/2},m(t)
\right)\Big] \, .
\eeq
In the limit of $m_f \dt \ll 1$, the expectation value becomes
\beq
\langle\phi^2\rangle_{ren} = \frac{1}{\sigma_s\,\dt^{d-4}} \int_{0}^\infty \frac{\tilde{q}^{d-2}d\tilde{q}}{\omega_{in}} \, \Big[\left| _2F_1 \right|^2-f_{ct}\left(\tilde q,m(t)\right)\Big]\, ,
\eeq
up to contributions suppressed by $m_f^2\dt^2$. Hence we have reproduced
exactly the expression computing the renormalized expectation value for a quench starting at $(\delta m^2)$ and ending at zero mass. Then, as shown previously, in the case where $\delta (m^2) \dt^2 \ll 1$, the expectation value of $\phi^2$ scales as $\delta m^2 \dt^{4-d}$. 

Hence to obtain the universal scaling \reef{1-2} in quenches with arbitrary masses, we need to satisfy two conditions:
\begin{eqnarray}
m_f \dt & \ll & 1\,, \\
\delta (m^2) \dt^2 = (m_i^2-m_f^2) \dt^2 & \ll & 1\,.
\end{eqnarray}
It is easy to check that these two conditions are equivalent to those in eqs. (\ref{accord}) and (\ref{extra}), \ie $(m_i^2-m_f^2) \dt^2 \ll 1$ and $(m_i^2 + m_f^2) \dt^2 \ll 1$.

Finally, we will comment on the case of the pulsed quench around any arbitrary mass. If our mass profile becomes $m(t)^2 = m_0^2 + \frac{m^2}{\cosh^2 t/\dt}$, then it is easy to verify that the only change in eq. (\ref{eom pulse}) is to add a term proportional to $m_0^2$ ending up with
\begin{eqnarray}
y (1-y) \frac{d^2 u_\vk}{dy^2} + \left(\frac{1}{2}-y \right) \frac{d u_\vk}{dy} - \left( \frac{(k^2+m_0^2) \dt^2}{4} + \frac{m^2 \dt^2}{4 y}\right) u_\vk=0.
\end{eqnarray} 
In analogy to eq.~\reef{arkan}, we define $\tilde{k}^2 = k^2 + m_0^2$, so that the equation becomes the same but with $k\rightarrow \tilde{k}$. Then the solution for the modes will be the same with the only difference that in eq.~\reef{a and b}, $k$ is replaced with $\tilde{k}$ (in $a$ and $b$). To obtain the expectation value, we will have to integrate over all momenta. In a way completely analogous to the previous case, we can perform a change of variables to integrate in $\tilde{k}$ and in the limit of $m_0^2 \dt^2 \ll 1$, we will get exactly the same integral as in section \ref{cft_cft}. Hence it is expected that the same scaling will appear. In conclusion, for the pulsed quench, the expectation value of $\phi^2$ will scale as $m^2 \dt^{4-d}$ provided that $m^2 \dt^2 \ll 1$ {and} $m_0^2 \dt^2 \ll 1$.

\subsection{Comparison to linear response} 

The results of sections \ref{sect21} and \ref{massive_massive} for the tanh profile are leading order in the dimensionless variable $ (m\delta t)^2 $. Therefore they should agree with a linear response calculation. In this section, we compute $\langle0|\phi^2|0\rangle_{in}$ in linear response theory for the quench starting from a CFT and ending with a massive theory and show that the result is in exact agreement with the expansion of the exact answer to $O(m^2)$, for each of the $k$ modes individually. This agreement should hold for the other kinds of protocols as well, such as the pulse profile in section \ref{cft_cft}. 

 The linear response result for the expectation value $\langle0|\phi^2|0\rangle_{in}$ is given by the expression
\ben
\langle0|\phi^2(x,t)|0\rangle_{in} - \langle0|\phi^2(x,t)|0\rangle_{in}|_{m^2=0}
= -\int d^{d-1} x^\prime \int dt^\prime\, m^2(t^\prime) \,
G_R(x,t;x^\prime,t^\prime)
\label{lr-1}
\een
where the retarded correlator is given by
\ben
G_R(x,t;x^\prime,t^\prime) = i\theta(t-t^\prime) _{in}\langle0|[ \phi^2 (x,t),  \phi^2 (x^\prime,t^\prime) ] |0\rangle_{in}
\label{lr-2}
\een
The correlation functions are to be evaluated in the initial theory, which is the massless free field theory. The right hand side can be computed exactly leading to 
\ben
\langle0|\phi^2(x,t)|0\rangle_{in} - \langle0|\phi^2(x,t)|0\rangle_{in}|_{m^2=0} = -\int \frac{d^{d-1}k}{(2\pi)^{d-1}}\frac{1}{2k^2}\int_{-\infty}^t dt^\prime \,m^2(t^\prime) \,\sin [2k(t-t^\prime)]
\label{lr-3}
\een
We will express the right hand side of eq.~(\ref{lr-4}) as a power series expansion in 
\ben
\eta = \exp (2t/\delta t)\,.
\label{lr-4}
\een
In eq.~\reef{lr-3}, we write $m^2(t^\prime) =\frac{ m^2}{2} (1+\tanh(t^\prime/\delta t)) = m^2 \frac{\eta^\prime}{1+\eta^\prime}$. Then expanding this expression as a series in $\eta^\prime$ and performing the intergral over $\eta^\prime$, we obtain
\ben
\langle0|\phi^2(x,t)|0\rangle_{in} - \langle0|\phi^2(x,t)|0\rangle_{in}|_{m^2=0} = m^2\int \frac{d^{d-1}k}{(2\pi)^{d-1}}\sum_{n=1}^\infty \frac{(-1)^n}{4|k| (k^2+n^2)}\eta^{n+1}
\label{lr-5}
\een
Let us now consider the $O(m^2)$ contribution to $\langle0|\phi^2(x,t)|0\rangle_{in}$ from the exact answer. This is given by
\ben
\langle0|\phi^2(x,t)|0\rangle_{in}= \int \frac{d^{d-1}k}{(2\pi)^{d-1}~(2|k|)}|_2 F_1 [1+i\omega_-\delta t,i\omega_-\delta t;1-i\omega_{in}\delta t; \frac{1}{2}(1+\tanh(t/\delta t))] |^2
\label{lr-6}
\een
where in this case
\ben
\omega_{in} = |k|,~~ \omega_{out}=\sqrt{k^2+m^2},~~\omega_\pm =\frac{1}{2}(\omega_{out}\pm\omega_{in})\,.
\label{lr-7}
\een
We need to expand the hypergeometric function to $O(m^2)$ and express the answer as a power series expansion in $\eta$. It turns out that 
\ben
|_2F_1|^2 = 1 + m^2 \sum_{n=1}^\infty \frac{(-1)^n}{2(k^2+n^2)}\eta^{n+1} +O(m^4)\,.
\label{lr-8}
\een
Substituting eq.~(\ref{lr-8}) into eq.~(\ref{lr-6}), it is easily seen that the $O(m^2)$ contribution to the exact answer matches the answer from linear response theory (\ref{lr-5}).

\subsection{Comparison with instantaneous quenches} \label{comparin}

The results of the previous sections appear to be at odds with the well studied examples of instantaneous (or abru
pt)  quenches in field theories, in particular \cite{cc2,cc3}. The behavior of \eg eq.~(\ref{1-3}) suggests that for $\Delta > d/2$, the expectation value of the operator ${\cal{O}}$ and hence the rate of energy production {\em diverges} in the limit $\delta t \rightarrow 0$. In contrast, the results of instantaneous quenches indicate that there is a smooth limit. In this section we resolve this apparent discrepancy.

The main point is that the fast quench limit, considered here, involves a quench rate, \ie  $1/\delta t$, which is {\em fast} compared to the scale set by the relevant coupling, but {\em slow} compared to the UV cutoff.  This is implicit in the above since we are working with renormalized quantities, where in fact the UV cutoff has been sent to infinity. However the abrupt quenches which are considered in the literature involve an instantaneous change of the Hamiltonian at some time, e.g. $t=0$. The wave function evolves from early times according to one time independent Hamiltonian $H_{in}$ up to time $t=0$. The resulting wavefunction at $t=0$ then acts as an initial condition for evolution with a different time independent Hamiltonian $H_{out}$. This process can be considered as a limit of a smooth time-dependent Hamiltonian provided the scale of variation is infintely fast compared to all scales in the problem. In a field theory, this means that $1/\delta t$ is large compared to all momentum scales including the UV cutoff scale $\Lambda$. This is clearly not the limit considered in our work.

To make this point explicit, we will now compute the two-point correlation function in position space in the free bosonic field theory with the time-dependent mass given by eq.~(\ref{massprofile}). We are interested in this at late times. For this purpose, it is convenient to work in terms of the ``out" modes,
\beq
\phi= \int\!\! \frac{d^{d-1}k}{(2\pi)^{d-1}}\ \left( b_\vk\, v_\vk + b^\dagger_\vk\, v^*_\vk\right)\,,
\labell{fieldx2}
\eeq
where
\begin{eqnarray}
v_\vk  & = & \frac{1}{\sqrt{2\omega_{out}}} \exp(i\vk\cdot\vec{x}-i\omega_+ t - i\omega_- \dt \log (2 \cosh t/\dt)) \times \nonumber \\
& & _2F_1 \left( 1+ i \omega_- \dt, i \omega_- \dt; 1 + i \omega_{out} \dt; \frac{1-\tanh(t/\dt)}{2} \right)
\label{modes2}
\end{eqnarray}
with the various frequencies defined in eq.~\reef{omegadef}.
These modes have the usual plane-wave behaviour at late times, $t \gg \delta t$,
\ben
v_\vk \rightarrow  \frac{1}{\sqrt{2 \omega_{out}}} \exp(i\vk\cdot\vec{x}-i\omega_{out} t).
\label{outplane}
\een
The $in$ and $out$ sets of modes ($u_\vk$ and $v_\vk$, respectively) are related by a Bogoliubov transformation
\bea
u_\vk & = &\alpha_\vk\  v_\vk + \beta_\vk\  v^\star_{-\vk}\,, \nnn \\
u^\star_\vk & = &\alpha^\star_\vk\ v^\star_\vk + \beta^\star_\vk\ v_{-\vk}\,.
\label{bogo}
\eea
The Bogoliubov coefficients have been evaluated in \cite{BD2},
\bea
\alpha_\vk & = & \sqrt{\frac{\omega_{out}}{\omega_{in}}} \, \frac{\Gamma (1-i\omega_{in}\delta t)\Gamma(-i\omega_{out}\delta t)}{\Gamma(-i\omega_+\delta t)\Gamma(1-i\omega_+\delta t)}\,, \nnn \\
\beta_\vk & = & \sqrt{\frac{\omega_{out}}{\omega_{in}}} \, \frac{\Gamma (1-i\omega_{in}\delta t)\Gamma(i\omega_{out}\delta t)}{\Gamma(i\omega_-\delta t)\Gamma(1+i\omega_-\delta t)}\,.
\label{bogocoeff}
\eea

The correlation function of the field is then given by
\bea
&&\langle in, 0|\phi (\vx,t) \phi(\vx^\prime,t^\prime)|in, 0\rangle  =  \int \frac{d^{d-1}k}{(2\pi)^{d-1}}\, u_\vk (\vx,t) \,u^\star_\vk (\vx^\prime,t^\prime) 
\label{correlator} \\
& &\qquad\qquad\qquad=  \int \frac{d^{d-1}k}{(2\pi)^{d-1}} \Big\{ |\alpha_\vk|^2 ~v_\vk(\vx,t) v^\star_\vk (\vx^\prime,t^\prime)+ \alpha_\vk \beta^\star_{\vk}~v_\vk(\vx,t) v_{-\vk} (\vx^\prime,t^\prime)+ \nnn \\
& &\qquad\qquad\qquad\qquad\qquad\qquad\quad
\alpha^\star_\vk \beta_\vk ~ v^\star_{-\vk}(\vx,t)v^\star_\vk(\vx^\prime,t^\prime) +|\beta_\vk|^2 ~ v^\star_{-\vk}(\vx,t)v_{-\vk}(\vx^\prime,t^\prime) \Big\}.
\nonumber
\eea
Using (\ref{bogocoeff}) one finds
\bea
|\alpha_\vk|^2& =& 1+ |\beta_\vk|^2  =  \frac{\sinh^2 (\pi \omega_+ \delta t)}{\sinh (\pi \omega_{in}\delta t) \sinh(\pi \omega_{out}\delta t)} \nnn \\
\beta_\vk \alpha^\star_\vk & = & \frac{\omega_{out}}{\omega_{in}} \frac{\pi \omega_{in}\delta t}{\sinh(\pi \omega_{in}\delta t)} \frac{[\Gamma(i\omega_{out}\delta t)]^2}{(-\omega_+\omega_- \delta t ^2)[\Gamma(i\omega_- \delta t)]^2 [\Gamma(i\omega_+\delta t)]^2}.
\label{alphasquare}
\eea

Consider now the limit 
\ben
\omega_{in}\delta t \ll 1~~~~~~~~\omega_{out}\delta t \ll 1\,,
\label{limits}
\een
in which the quantities appearing in eq.~(\ref{alphasquare}) become
\bea
|\beta_\vk|^2 & \rightarrow & \frac{(\omega_{out} - \omega_{in})^2}{4\omega_{out}\omega_{in}} \nnn \\
\beta_\vk \alpha^\star_\vk & \rightarrow & \frac{\omega_{out}^2 - \omega_{in}^2}{4\omega_{in}\omega_{out}}.
\label{limitvalues}
\eea
Let us now compute the correlation function (\ref{correlator}) taking both the limit (\ref{limits}) and considering late times
\ben
t/\delta t \gg 1\,, ~~~t^\prime/\delta t \gg 1\,. \label{latett}
\een
Using eqs.~(\ref{outplane}) and (\ref{limitvalues}), a short calculation yields
\begin{eqnarray}
&&\langle in, 0|\phi (\vx,t) \phi(\vx^\prime,t^\prime)|in,0 \rangle  \rightarrow 
\int \frac{d^{d-1}k}{(2\pi)^{d-1}}\, e^{i\vk\cdot (\vx -\vx^\prime)}  \label{xprime} \\
& & \qquad\qquad\times \left[
\frac{e^{-i\omega_{out}(t-t^\prime)}}{2\omega_{out}} + \frac{(\omega_{out}-\omega_{in})^2}{4 \omega_{out}^2\omega_{in}} \cos \omega_{out}(t-t^\prime) + \frac{(\omega_{out}^2 - \omega_{in}^2)}{4 \omega_{out}^2\omega_{in}} \cos \omega_{out}(t+t^\prime) \right]. \nnn
\end{eqnarray}
Note that $\delta t$ has disappeared from the result. In fact 
this reproduces the result for an instantaneous quench from a mass $m_{in}^2 = m^2(A-B)$ to a mass $m_{out}^2 = m^2(A+B)$, \eg see eq.~(8) of \cite{cc3}.

In this paper, we have concentrated on local quantities like $\langle \phi^2 \rangle$ or the energy density. These involve integrals over all momenta all the way to the cutoff, and clearly the limit (\ref{limits}) is not appropriate for large UV momenta in these integrals. In our analysis, we have worked with renormalized quantities which, as we noted above, implicitly involves taking the UV cutoff $\Lambda$ much larger than $1/\delta t$. This is why our limit of fast quenches is physically different from the instantaneous quenches, studies elsewhere, where the quench rate is necessarily fast compared to $\Lambda$. In fact in the continuum limit, it is unphysical to consider such an instantaneous quench. It would be interesting to investigate these issues in a theory with finite cutoff. In such a theory one would expect that the scaling discussed in this paper should hold in a protocol where $\Lambda^{-1} \ll \delta t \ll m^{-1}$. On the other hand, when $\delta t$ is the same order as $\Lambda^{-1}$, the answers should approach those for an instantaneous quench.

Nevertheless, for distances $|\vx -\vx^\prime| \gg \delta t$, only momenta much less than $\delta t^{-1}$ should be making
a substantial contribution to the correlation functions. For such quantities, the condition
(\ref{limits}) is effectively satisfied and so by the above analysis, one should expect only small
differences between a fast smooth quench and an instantaneous quench at late times, \ie when eq.~\reef{latett} is also satisfied. Details of this comparison will be discussed in \cite{dgm2} --- see also discussion in the following section and in section \ref{conclusions}.

\subsection{Late time behaviour}
\label{late2}

In section \ref{low}, we observed some interesting late time behaviour for the expectation value $\vev{\phi^2(x,t)}$ in three dimensions, \ie at late times, the expectation value is independent of $\delta t$.  This may lead us to suspect that this late time behavior agrees with the results of an instantaneous quench. In this section, we will show that in a suitable regime this is indeed true for $d=3$, but one finds that the same agreement does not generally occur in higher dimensions \cite{dgm2}. 

As we noted in section \ref{low}, the numerical analysis only allowed us to evaluate the expectation values out to times of order $t \sim 10\,\dt$. Hence given the values of $m$ and $\dt$ that we were using, we were always in a regime where $m^2t^2\ll 1$.  We therefore first compare the result obtained in section \ref{low} with the result of an instantaneous quench in this regime. We will show that with the results already obtained we can reproduce our previous results in this limit but also go beyond them and evaluate the proper long time behaviour of the scalar field for $m^2t^2\gg1$.

The starting point will be to consider the correlator for instantaneous quench, eq.~(\ref{xprime}) and evaluate this expression at coincident points in space and time, \ie $\vec{x}=\vec{x}'$, $t=t'$. This gives,
\beq
\langle \phi^2 (\vec{x},t) \rangle = \int \frac{d^{d-1}k}{(2\pi)^{d-1}} \frac{1}{4 \omega_{out}^2 \omega_{in}} \left( \omega_{in}^2 + \omega_{out}^2 - (\omega_{in}^2 - \omega_{out}^2) \cos (2 \omega_{out} t) \right) \, .
\label{coincident}
\eeq

Focusing on the quench to the critical point, \ie $A=-B=1/2$ in eq.~\reef{massprofile}, for which $\w_{out}^2 = k^2$ and $\w_{in}^2 = k^2 + m^2$, we find
\beq
\langle \phi^2 (\vec{x},t) \rangle = \frac{\Omega_{d-2}}{2 (2 \pi)^{d-1}} \int \frac{k^{d-4} dk}{\sqrt{k^2 + m^2}} \left( k^2 + m^2 \sin^2(kt) \right) \, . \label{longtime_full}
\eeq
Of course, this expectation value is divergent in the UV, so it must be regulated as described in section \ref{adiabat}.
While in general this is a somewhat involved procedure, we begin here by considering $d=3$ in which case there is a single mass-independent UV divergence --- see eqs.~\reef{naive} and \reef{naive2}. Hence the difference between the quenched expectation value and that for a fixed mass $m$ will produce a finite result.\footnote{Note that we are subtracting the expectation value with the mass fixed at the initial mass of the quench rather than the final mass (which would be zero). Either choice would leave a finite remainder but the expressions simplify somewhat here by using the initial mass.}  That is, we subtract
\beq
\langle \phi^2  \rangle_{fixed} = \frac{\Omega_{d-2}}{2 (2 \pi)^{d-1}} \int \Phi^2_{fixed}(k) dk= \frac{\Omega_{d-2}}{2 (2 \pi)^{d-1}} \int \frac{k^{d-2}\, dk}{\sqrt{k^2 + m^2}}  \label{adiab}
\eeq
from eq.~\reef{longtime_full} and then evaluate the finite difference
\begin{eqnarray}
\langle \phi^2 \rangle_{quench} - \langle \phi^2  \rangle_{fixed} & = & \frac{m^2}{4\pi}  \int \frac{dk}{k \sqrt{k^2+m^2}}\, \sin^2{kt}\nonumber \\
& = &  \frac{m^2 t}{4\pi} \int \frac{dp}{p\sqrt{p^2 + m^2 t^2}}\, \sin^2 p \label{all_t}
\end{eqnarray}
for $d=3$. In these expressions, we have substituted $\sigma_s=4\pi$ for $d=3$ using eq.~\reef{sigs}.  Above in the second line, we also introduced the dimensionless momentum $p=k\,t$. The first thing to verify is that we recover our previous results for $d=3$ in the regime where $m^2 t^2 \ll 1$ --- see discussion in section \ref{low}. In this limit, we can drop the  $m^2 t^2$ appearing in the denominator of the integrand to find
\beq
\langle \phi^2 \rangle_{quench} - \langle \phi^2  \rangle_{fixed} = \frac{m^2 t}{4\pi} \int_0^\infty \frac{dp}{p^2} \sin^2 p = \frac{1}{8}\, m^2 t \,. \label{lin33}
\eeq
This is exactly the same result we found in eq.~\reef{lin3}, showing a linear growth in the expectation value of $\phi^2$ with a slope that is independent of $\dt$. From these results, we can also identify the constant displacement in eq.~\reef{d3adv2} and in fig.~\ref{fig_d3} as the {\em renormalized} expectation value for a constant mass, \ie $\langle \phi^2  \rangle_{fixed,ren}=-m/(4\pi)$.

\begin{figure}[h!]
\setlength{\abovecaptionskip}{0 pt}
\centering
\includegraphics[scale=1]{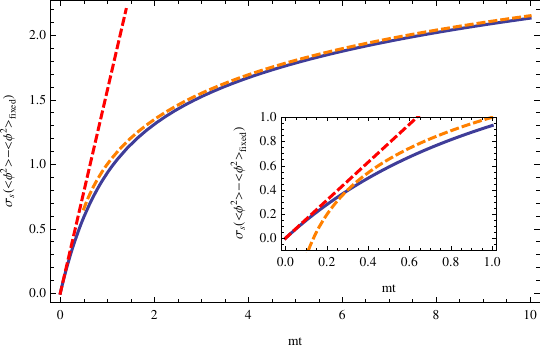}
\caption{(Colour online) Expectation value of $\phi^2$ as a function of time. We are in the limit of $t/\dt \gg 1$. The solid curve corresponds to the full solution for any value of $mt$. The red dashed line is the linear behaviour found in eq.~\reef{lin33} for $mt \ll 1$. The orange dashed line shows the logarithmic growth found in eq.~\reef{loglim} for $mt \gg 1$. Finally, the inset zooms in the region of small $mt$.} \label{fig_d3_long}
\end{figure}

However, given eq.~(\ref{all_t}), we can go further and analyze the behaviour of the expectation value for any value of $m^2 t^2$. In particular, this expression can be integrated exactly for any $m^2 t^2$ and evaluated in terms of generalized hypergeometric functions,
\beq
\langle \phi^2 \rangle_{quench} - \langle \phi^2  \rangle_{fixed} = \frac{m^2 t^2}{4\pi} \left(\frac{\pi}{2\,t}  \, _1F_2\left(\frac{1}{2};1,\frac{3}{2};m^2 t^2\right)- \, _2F_3\left(1,1;\frac{3}{2},\frac{3}{2},2; m^2 t^2\right)\right)\,. \label{d3exact}
\eeq
Fig.~\ref{fig_d3_long} shows a plot of this expectation value as a function of $mt$.  From the figure, we observe that the linear growth \reef{lin33} of the expectation value is only valid for $mt\ll1$. After that, the expectation value continues to grow but in a slower rate. In fact, one can take the limit $mt \to \infty$ in eq.~\reef{d3exact} to find
\beq
\lim_{mt \to \infty} \left(\langle \phi^2 \rangle_{quench} - \langle \phi^2  \rangle_{fixed}\right) = \frac{m}{8\pi} \,\log ({m t})\,.
\label{loglim}
\eeq

Hence we see that for very late times, \ie $mt\gg1$, the expectation value continues to grow but only logarithmically. In any event, if we look into infinite future time, the expectation value is divergent. At first sight, this unbounded growth may seem counterintuitive since, for example, it may seem that the physical work done by the quench will also diverge. However, if we recall that eq.~\reef{ward}, the time rate of change of the energy density is given by the product of this expectation value with the derivative of the mass coupling. For the mass profile \reef{massprofile}, the latter decays exponentially in time and hence the corresponding integral for the energy density remains finite and well-defined, despite the logarithmic growth of the expectation value \reef{loglim}. 

Let us now examine the question: why does the long time answer for smooth fast quench as defined in this paper agree with the instantaneous quench result for $mt \ll 1$.
We need to 
consider the validity of assumptions which were implicit in the above discussion. Our starting point was eq.~\reef{coincident} which was found by taking the limit of coincident points in eq.~\reef{xprime}. However, the latter correlator was simplified by assuming late times as in eq.~\reef{latett} but also small $\dt$ as in eq.~\reef{limits}. While the late time assumption is certainly valid here, it is not clear that the second assumption should hold. In particular, one expects that for sufficiently large momenta that the inequalities in eq.~\reef{limits} will be violated. However, if we examine the form of the integrand in eq.~\reef{all_t}, we see that it decays as roughly $1/p^2$ for large (dimensionless) momentum. Hence we can expect that the dominant contributions to the integral come from small and finite values of $p$. Further given that $p=kt$,  we will certainly satisfy $k\dt\ll1$ in the late time limit and hence eq.~\reef{limits} will be satisfied. For example, one can make a simple estimate of the error introduced in ignoring the very high momenta as follows: Certainly, eq.~\reef{limits} is satisfied for $k\sim m$ and hence the integrand in eq.~\reef{all_t} is accurate for dimensionless momenta at least up to $p=mt$. Then an upper bound on the error in our result is given by the integral from $p=mt$ to $\infty$ but removing the factor of $\sin^2p$. The final result of this integration is a fixed constant, \ie approximately $0.07\,m$. Hence at large $mt$, this upper bound on the error is small compared to the results given in eqs.~\reef{lin33} and \reef{loglim}. In fact, given that the numerical fits to the constant term were also good, this suggests that even this approximation is a gross over-estimate of the error.

\begin{figure}[H]
        \centering
        \subfigure[Integrands for $t=10$.]{
                \includegraphics[scale=0.8]{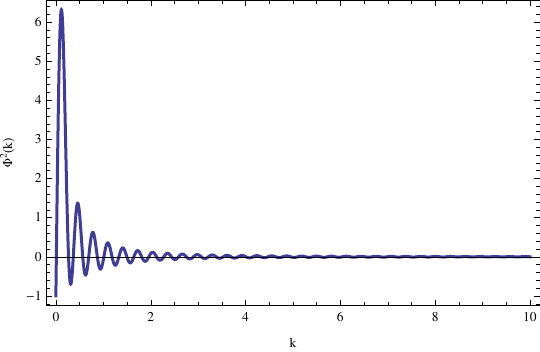} \label{approx_a}}
   		 \subfigure[$\vev{\phi^2}_{ren}$ as a function of time.]{
                \includegraphics[scale=0.8]{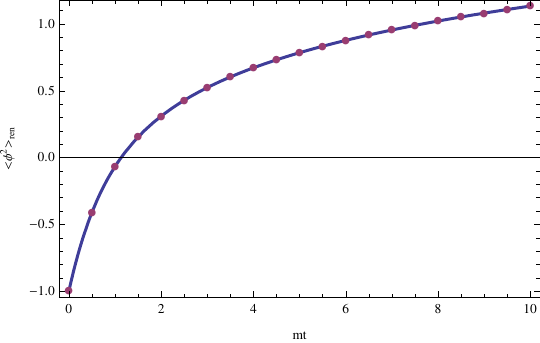} \label{approx_b}}
        \caption{(Colour online) Analysis of the approximation of low energies and late times in $d=3$, with $\dt=10^{-3}$ (where the units are set by $m$). Panel (a) shows the integrands in eqs.~(\ref{correlator}) and (\ref{longtime_full}) at $t=10$ but there is not visible difference between the  curves. Panel (b) shows $\vev{\phi^2}_{ren}$ as a function of $mt$. The solid blue curve corresponds to analytically integrating the expression for the instantaneous quench, eq.~(\ref{longtime_full2}) for $d=3$. The purple dots correspond to numerically evaluating the smooth quench expression of eq.~(\ref{full_late}). Again there is no visible difference between the two approaches.} \label{fig_approx_d3}
\end{figure}

In fact we can check the validity of our approximation by comparing the full integrand of eq. (\ref{correlator}) in the limit of late times, \ie by using eq. (\ref{outplane}) for the out-modes, with the approximate eq. (\ref{longtime_full}). Let's recall that eq. (\ref{correlator}) is not assuming any relation between the energies in the system and $\dt$, while eq. (\ref{longtime_full}) assumes $\omega \dt \ll 1$ for every $\omega$. Essentially, we want to compare the integrands of
\bea
\langle \phi^2 \rangle_{smooth} & = & \frac1{\sigma_s} \int \Phi^2(k)\, dk \label{full_late}\\
& = & \frac1{\sigma_s} \int dk \left(\frac{k^{d-2}}{\omega_{out}} \Big\{ |\alpha_\vk|^2 + \alpha_\vk \beta^\star_{\vk}~e^{2 i \w_{out} t}+
\alpha^\star_\vk \beta_\vk ~e^{-2 i \w_{out} t} +|\beta_\vk|^2 \Big\} - k^{d-3} \right), \nonumber
\eea
where $\alpha_\vk$ and $\beta_\vk$ are given by eq. (\ref{bogocoeff}), and
\beq
\langle \phi^2  \rangle_{instant} = \frac1{\sigma_s} \int \Phi^2(k)\, dk = \frac1{\sigma_s} \int dk \left( \frac{k^{d-4}}{\sqrt{k^2 + m^2}} \left( k^2 + m^2 \sin^2(kt) \right) - k^{d-3} \right) \, , \label{longtime_full2}
\eeq
for $d=3$. Fig.~\ref{approx_a} plots the integrands in these two expressions as a function of $k$ and in fact, there is no visible difference. The figure uses $m\dt=10^{-3}$ and $mt=10$ but similar results hold for different values of these parameters. It is clear that the integrand decays rapidly, \ie in fig.~\ref{approx_a}, it has become negligibly small around $mk\sim5$. Therefore the approximation $k \dt \ll 1$ is effectively satisfied since even though we are integrating over all momenta in the expressions above, the main contribution comes from very low momenta. The latter is explicitly verified in fig.~\ref{approx_b} which shows $\vev{\phi^2}$ after both the smooth and the instantaneous quench. As we show before, the instantaneous expression can be integrated analytically and the final result is given by the right-hand side of eq.~\reef{d3exact} plus $\langle \phi^2  \rangle_{fixed,ren}=-m/(4\pi)$. This result is shown in the figure with the solid blue curve. The purple points correspond to integrating numerically eq.~(\ref{full_late}). We see good agreement between both expectation values. In fact, if we compute the relative difference between them at late times, we see that it is of order $10^{-6}$ and hence we verify that both approaches give the same result at late times. 

\begin{figure}[h!]
\setlength{\abovecaptionskip}{0 pt}
\centering
\includegraphics[scale=1]{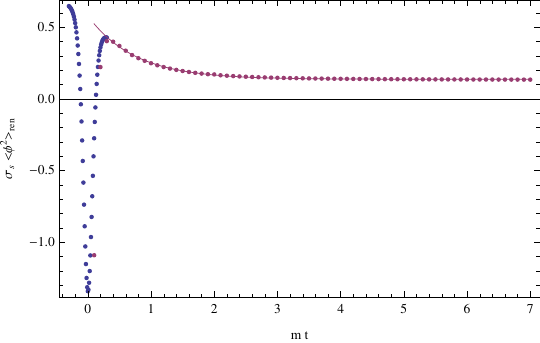}
\caption{(Colour online) Expectation value of $\phi^2$ as a function of time with $m\dt=10^{-1}$. The blue dots correspond to evaluating the expectation value as in section \ref{responseq}, while the purple dots are those coming from numerically integrating eq.~(\ref{full_late}). The solid line shows a fit by a function of the form $f(m t) = a + b \exp(-c\, m t)$, with parameters $a=0.136$, $b=0.442$ and $c=1.349$.} \label{fig_d5_phi_long}
\end{figure}

However, the same agreement does not hold in higher dimensions, as we will discuss in detail in \cite{dgm2}. However, let us present the late-time limit of the smooth quench in $d=5$ here. Recall that the desired expectation value is given by eq.~(\ref{full_late}) with $d=5$. Although this expression is quite complicated, we can integrate it numerically for different values of $m t$ and follow the evolution of the expectation value at late times, as shown in fig.~\ref{fig_d5_phi_long} for $m\dt=10^{-1}$.\footnote{We chose this value of $m\dt$ in order to compare with our previous results of section \ref{responseq}. Note that in that section, we were using units of time measured in units $\dt$ and so a very small $\dt$ would yield a plot that is very compressed around $t=0$ in $mt$ units.} In the figure, the blue dots are obtained by evaluating the absolute value of the hypergeometric, as we did in section \ref{responseq}. However, that analysis only allowed us to go relative short times, in units of $1/m$. The evaluation of eq.~(\ref{full_late}) is shown in purple dots for late times and we can see a nice continuity between the two approaches, showing its consistency. The exponential fit of the purple dots also shows that the expectation value is decaying as expected due to the exponential nature of the mass profile. Note that even though the decay is exponential, it does not decay to zero, but to a finite value. One can perform the analysis for different $\dt$'s and see that in the limit of $\dt \rightarrow 0$, that constant approaches to $\langle \phi^2 \rangle(t \to \infty) \simeq 0.168\, m^3/\sigma_s$. It would be interesting to have an analytical understanding of this asymptotic value and also to generalize these results to higher dimensions.

\section{Quenching a free fermionic field} \label{fermi}

Another way to test our universal scaling formulae in eqs.~\reef{1-2} and \reef{1-3} is to quench an operator with a different conformal dimension. In this section, we will be quenching the mass of a free Dirac fermion $\psi$ in $d$ dimensional spacetimes. Then, our operator of interest will be $\langle \calo_\Delta \rangle = \langle \bar{\psi}\psi \rangle$, whose conformal dimension is $\Delta=d-1$ and the corresponding coupling is the mass $\lambda (t) = m(t)$. It is interesting that in this case we should expect divergences to appear as $\langle\cO_\Delta\rangle_{ren}\sim\dg/\dt ^{2\Delta-d}=\dg/\dt ^{d-2}$ and hence, even for low dimensional spacetimes with $d=2,3$ we should be able to find divergent behaviours.

The calculations are analogous to those for the scalar field. In partiuclar, the situation can be related to one of fermions in curved space-times, where analytic solutions are known for specific mass profiles. Then one can compute the expectation values and find numerical solutions. We will also be able to find analytical leading order solutions in the fast quench limit when $\dt\rightarrow0$. The main conclusion is that the scaling relations, \reef{1-2} and \reef{1-3}, which were originally discovered by the holographic analysis also hold in this case. Further, the universal power-law scaling is enhanced by a logarithmic factor in the case of even $d$.

We will be following the conventions and notation of \cite{Duncan}, where the problem of fermions in flat FRW backgrounds is discussed. In this case, the equations of motion for a Dirac field $\Psi$ are not directly those that we are interested in, \ie the Dirac equation with a time-dependent mass. However, it is possible to do a confomal rescaling of the fields, $\Psi = C(t)^{\frac{1-d}{2}} \psi$, where $C(t)^2$ is the expansion factor, and then, one finds that $\psi$ satisfies the Dirac equation of motion,
\beq
\left( i \gamma^\mu \partial_\mu - m \, C(t) \right) \psi = 0, \label{eom_fermions}
\eeq
where we will define our time-dependent mass as $m(t) = m \,C(t)$. The exactly solvable model here requires $C(t) = A + B \tanh t/\dt$ \cite{Duncan}, and so
\beq
m(t) = m\, (A + B \tanh t/\dt)\,,
\label{mass_fermions}
\eeq
in contrast to the scalar case \reef{massprofile}, where it was the mass \textit{squared} that had the $\tanh$ profile. Solutions to eq.~(\ref{eom_fermions}) are given by
\beq
\psi = \left( \gamma^0 \partial_t + i k_j \gamma^j - m\, C(t) \right) e^{i \vec{k}\cdot\vec{x}} \phi_\vk(t),
\eeq
where $j$ denotes spatial coordinates and $\phi_\vk$ satisfies
\begin{eqnarray}
\ddot{\phi}_\vk + \left( \vk^2 + m^2 C^2 + m \gamma^0\, \dot{C}\right) \phi_\vk = 0.
\end{eqnarray}
For simplicity in the last expression we are not writing the time dependence on $C$ or the fields any more.

Now, the full solution for fermionic field $\psi$ can be written in terms of the $in$ modes as
\begin{eqnarray}
\psi = \frac{1}{(2\pi)^{(d-1)/2}} \int d^{d-1}k \sqrt{ \frac{m_{in}}{\omega_{in}}} \sum_{\lambda=1}^{2^{d/2-1}} \left( a_{in} (k,\lambda) U_{in} (k,\lambda,x,t) + b_{in}^\dagger (k,\lambda) V_{in} (k,\lambda,x,t) \right), \nonumber \\
\label{psi_full}
\end{eqnarray}
where
\begin{eqnarray}
U_{in} (k,\lambda,x,t)  & = & -\frac{1}{k} \sqrt{\frac{\omega_{in} + m_{in}}{2 m_{in}}} \left( -i \partial_t + i k_j \gamma^j - m \,C \right) \phi_k^{in (-)} (t) e^{i \vec{k}\cdot\vec{x}} u(0,\lambda), \nonumber \\
V_{in} (k,\lambda,x,t)  & = & -\frac{1}{k} \sqrt{\frac{\omega_{in} + m_{in}}{2 m_{in}}} \left( i \partial_t - i k_j \gamma^j - m \,C \right) \phi_k^{in (+) *} (t) e^{-i \vec{k}\cdot\vec{x}} v(0,\lambda), 
\end{eqnarray}
and the sum over the spinor index $\lambda$ runs up to $2^{(d-3)/2}$ if $d$ is odd. Here  $u(0,\lambda)$ and $v(0,\lambda)$ are constant basis spinors, the $\omega$'s here and below are defined as in eq.~\reef{omegadef} and $m_{in} = m(t=-\infty)$. Further, $a_{in}$ and $b_{in}$ are operators that annihilate the $in$-vacuum.

It can be shown that this solution reproduces the corresponding solutions for flat space at infinite past and infinite future --- see \cite{Duncan}. For the tanh mass profile  (\ref{mass_fermions}), there exist analytic solutions for $\phi_\vk$ that are of the form
\begin{eqnarray}
\phi_k^{in (\pm)} (t) & = &  \exp \left( -i \omega_+ t -i \omega_- \dt \log (2 \cosh t/\dt)\right) \times \label{sovlele} \\
& & _2F_1 \left(1+ i \omega_- \dt \pm i m B \dt, i \omega_- \dt \mp i m B \dt; 1- i \omega_{in}\dt; \frac{1+\tanh t/\dt}{2} \right), \nonumber
\label{phi_fermions}
\end{eqnarray}
where $_2F_1$ is the usual hypergeometric function. Note that this solution is similar but not equal to that appearing for the scalar field modes \reef{modes}. Further, we will again focus on quenches to the critical point, where $A=-B=1/2$.

Now we are interesting in finding the time evolution of the mass operator $\bar{\psi} \psi$ through the quench. This is given by
\begin{eqnarray}
\langle \bar{\psi}\psi \rangle \equiv \langle 0, in| \bar{\psi} \psi |0,in \rangle = \int \frac{d^{d-1}k}{(2\pi)^{\frac{d-1}{2}}} \left( \frac{m_{in}}{\omega_{in}} \right) \sum_{\lambda=1}^{2^{d/2-1}} \bar{V}_{in} V_{in},
\end{eqnarray}
that after some algebra it turns to
\begin{eqnarray}
\langle \bar{\psi}\psi \rangle & = & \sigma_f^{-1} \int \psi_{div}(k) dk  = - \sigma_f^{-1} \int k^{d-4} dk \left( \frac{m_{in}}{\omega_{in}} \right) \left( \frac{\omega_{in}+m_{in}}{2 m_{in}}\right) \nonumber \\
& & \times \left( \left( m^2(t) - k^2\right) |\phi_\vk|^2  + |\partial_t \phi_\vk|^2 - 2 m(t) \text{Im} \left( \phi_\vk \partial_t \phi_\vk^*\right) \right),
\label{barpsipsi}
\end{eqnarray}
where $\phi_\vk$ is actually $\phi_\vk^{in (+)}$ and $\sigma_f$ is a numerical factor that depends on the spacetime dimension as 
\beq
\sigma_f = \left \{
\begin{array}{ll}
2^{1-d/2}(2\pi)^{\frac{d-1}{2}} / \Omega_{d-2} & \ \ {\rm for\ even}\ d\,, \\
(2^{(3-d)/2}) (2\pi)^{\frac{d-1}{2}} / \Omega_{d-2}& \ \  {\rm for\ odd}\ d\,.\ 
\end{array} \right.
\label{sigmaff}
\eeq

As in the case of scalar fields, this expectation value is in general UV divergent so we need to regulate the result by subtracting the appropriate counterterm contributions
\beq
\langle \bar{\psi}\psi \rangle_{ren} \equiv \sigma_f^{-1} \int dk (\psi_{div}(k) - f_{ct} (m(t),k) )\,.\label{pppp}
\eeq
These counterterm contributions can again be found as for the scalar field in section \ref{adiabat}. In this way, we find that
\begin{eqnarray}
f_{ct}(m(t),k) & = & - m(t) k^{d-3} + \frac{m(t)^3}{2} k^{d-5} -  \frac{3 m(t)^5}{8} k^{d-7} + \label{ct_f} \\
& & +  \frac{1}{4} \partial^2_t m(t) k^{d-5} - \left( \frac{1}{16} \partial^4_t m(t)  + \frac{5 m(t)}{8} \left( \partial_t m(t) \partial_t m(t)  + m(t) \partial^2_t m(t)\right) \right) k^{d-7} , \nonumber
\end{eqnarray} 
are all the necessary terms needed to regulate theories up to $d=7$. Note that again contributions with time derivatives of the mass profile appear, now from $d=4$ onwards. Further, the first line of eq.~(\ref{ct_f}) corresponds to the counterterms that would appear in order to regulate the expectation value for a constant mass.

\begin{figure}[H]
        \centering
        \subfigure[$d=2$]{
                \includegraphics[scale=0.8]{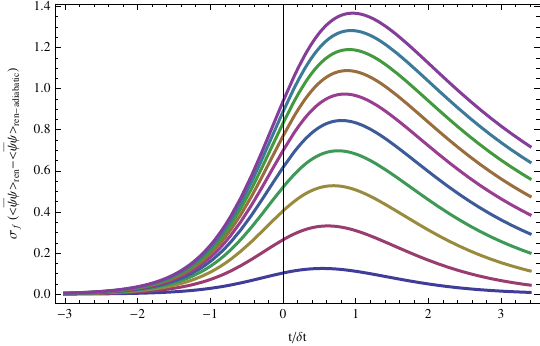}}
   		 \subfigure[$d=3$]{
                \includegraphics[scale=0.8]{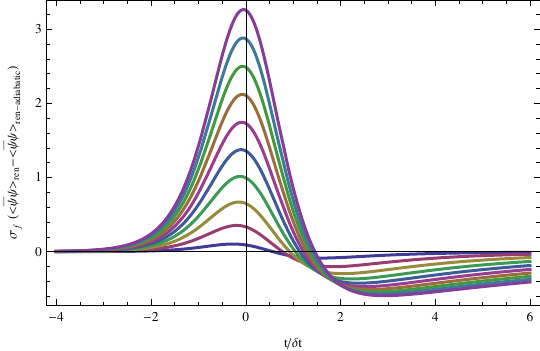}}
         \subfigure[$d=4$]{
                \includegraphics[scale=0.8]{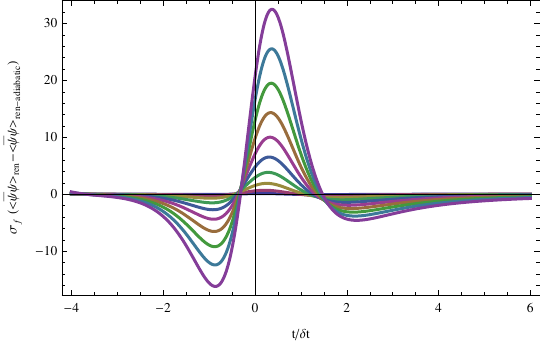}}
   		 \subfigure[$d=5$]{
                \includegraphics[scale=0.8]{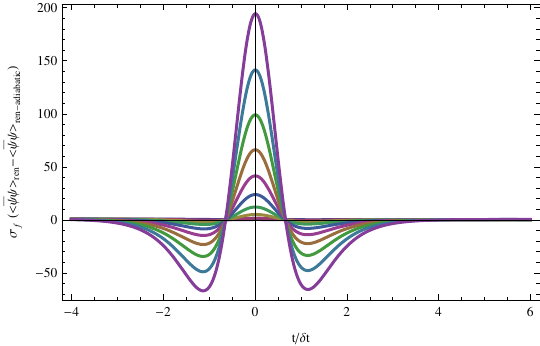}}
         \subfigure[$d=6$]{
                \includegraphics[scale=0.8]{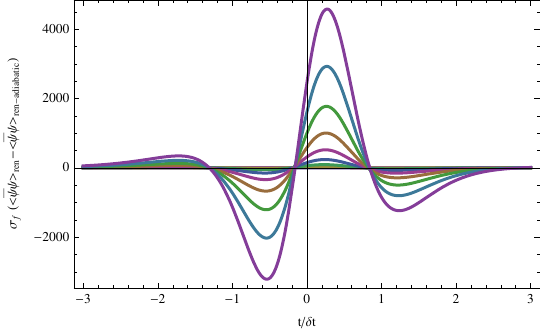}}
   		 \subfigure[$d=7$]{
                \includegraphics[scale=0.8]{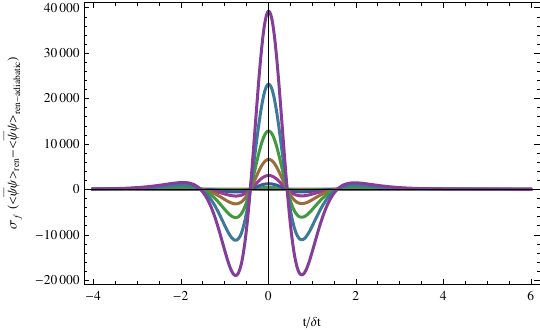}}
        \caption{(Colour online) Renormalized expectation values of $\bar{\psi}\psi$ as a function of time. The different curves correspond to $\dt = 1/1,1/2, \cdots, 1/10$. The curves are so that higher peaks (in absolute value) correspond to smaller $\dt$. Note also that we are plotting the expectation value multiplied by the numerical constant $\sigma_f$ that depends on the spacetime dimension. Also note that we are subtracting at each time the expectation value in the adiabatic case, for which we are using $\dt=10$. In even spacetime dimension $d$, the plots corresponds to having the renormalization scale set to $k_0=1$.}\label{fig_fermions}
\end{figure}

Given this finite expectation value \reef{pppp}, we are able to evaluate it numerically for all dimensions and different values of the quench rate $\dt$. The results are shown in fig.~\ref{fig_fermions}. Note that in these plots, we are subtracting  the expectation value in the adiabatic case, for which we are using $\dt=10$. We verified that the adiabatic expectation value is independent of $\dt$ as long as $\dt$ is large enough.  

As in the scalar case, we should distinguish between odd and even spacetime dimensions. In the case of even $d$, we also get logarithmic divergences (apart from the usual power-law divergences), that need extra renormalization scales in order to avoid infinities near $k=0$. Much in parallel to the scalar case, this will generate a logarithmic enhancement of the scaling behaviour in the expectation value.

\begin{figure}[h!]
\setlength{\abovecaptionskip}{0 pt}
\centering
\includegraphics[scale=1]{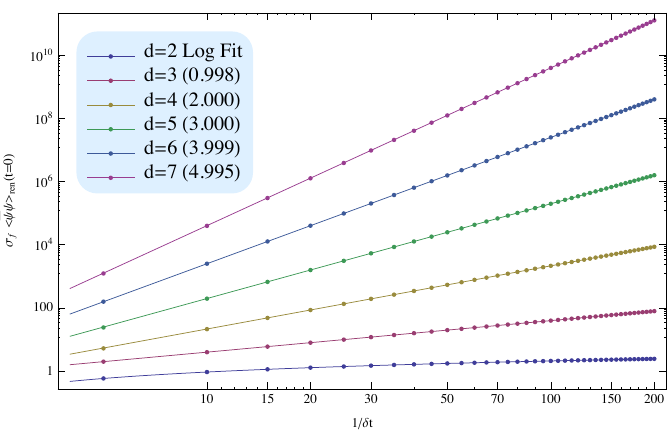}
\caption{(Colour online) Expectation value $\langle \bar{\psi} \psi \rangle_{ren}(t=0)$ as a function of the quench times $\dt$ for spacetime dimensions from $d=2$ to $d=7$. Note that in the plot, the expectation values are multiplied by a numerical factor $\sigma_f$ depending on the dimension.
The slope of the linear fit in each case is shown in the brackets beside the labels.
The results  support the power law scaling $\langle \bar{\psi} \psi \rangle_{ren} \sim \dt^{-(d-2)}$.} \label{fig_all_d_fermions}
\end{figure}

We can appreciate how the expectation values grow as we decrease $\dt$ in different dimensions in fig.~\ref{fig_fermions}. Note that in contrast with the scalar case, now we can see large growth appears as low as $d=2$. In order to quantify the precise nature of this growth, we compute the expectation value at a fixed time $t=0$ for a larger range of $\dt$ and plot it in a log-log scale in fig.~\ref{fig_all_d_fermions}. Even though the choice $t=0$ appears not to be appropriate to find the expected logarithmic enhancement, the figure and the linear fits there support completely the expected power-law scaling $\langle \bar{\psi} \psi \rangle_{ren}\sim m/\dt^{d-2}$. Again, if we do the same exercise but at a slightly shifted time, we find that in even dimensions there is a logarithmic enhancement of the divergences. This behaviour will be supported soon by analytical results in computing the expectation value. For now, we can only say that there is a logarithmic growth in $d=2$ that is in agreement with previous holographic results.

As in the case of scalars, one can recognize certain relationship between the expectation values and time-derivatives of the mass by looking at the plots of fig.~\ref{fig_fermions}. What we will show next is that if we compute the leading contribution in the limit of $\dt\rightarrow0$, we'll find precisely those mass derivatives.


\begin{figure}[H]
        \centering
        \subfigure[$d=3$]{
                \includegraphics[scale=0.8]{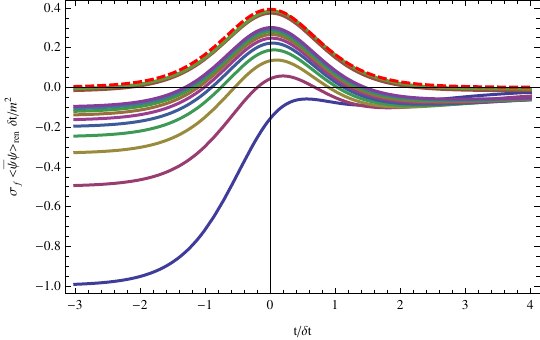}}
   		 \subfigure[$d=5$]{
                \includegraphics[scale=0.8]{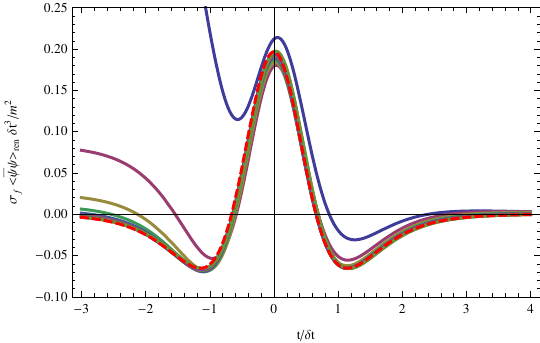}}
         \subfigure[$d=7$]{
                \includegraphics[scale=0.8]{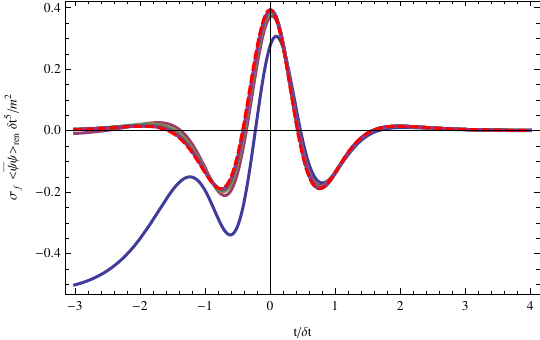}}
        \caption{(Colour online) $\langle\bar{\psi}\psi\rangle_{ren} \dt^{d-2}$ for different values of $\dt$ and different odd spacetime dimensions $d$. As the curves approach the  leading order analytic solution \reef{boat} shown with the dashed red line, $\dt$ gets smaller, with solid (numerical) curves going from $\dt=1$ to $\dt=1/10$. For $d=3$ we also included the curves with $\dt=1/50$ and $1/100$.}\label{fig_leading_f}
\end{figure}

The procedure is exactly the same as in the scalar case. We define dimensionless variables $q= k \dt$ and $\kappa = m \dt$ and then use the hypergeometric series expansion to get the leading terms in a $\kappa$-expansion. To get the counterterms to that order we can also expand for large $q$. The difference in this case is that the expectation value given in eq.~(\ref{barpsipsi}) also requires to compute the time-derivative of the hypergeometric function and in general, this can be an involved task. However, we should notice that the only time dependence in the hypergeometric function is in the last argument, \ie $z=(1+ \tanh t/\dt)/2$. The rest of the coefficients do not depend on time. Then,
\begin{eqnarray}
\partial_{t} \left( _2F_1 (a,b;c;z(t)) \right)  = \partial_t \left( \sum_{n=0}^{\infty} \frac{(a)_n (b)_n}{(c)_n} \frac{z(t)^n}{n!} \right) = \sum_{n=0}^{\infty} \frac{(a)_n (b)_n}{(c)_n} \frac{z(t)^{n-1}}{(n-1)!} (2z(t) - 2(z(t))^2), \nnn \\
\end{eqnarray}
where we use the usual trigonometric identities to express the time-derivative of $z(t)$ as a function of $z(t)$ itself. With this in mind, we can expand our hypergeometric series and note again that only the few first terms are needed in order to get the leading order $\kappa$ behaviour.
For odd $d\geq 3$, we obtain
\beq
\langle \bar{\psi}\psi \rangle_{ren} =  (-1)^{\frac{d-1}{2}} \frac{\pi}{2^{d-1}\sigma_f} \partial_t^{d-2} m(t) + O(\dt^{1-d})\,,
\label{boat}
\eeq
which correctly gives the expected scaling behaviour $m \dt^{2-d}$. Fig.~\ref{fig_leading_f} shows how the numerical solutions approximate this leading order analytic term as $\dt\to0$.

For even $d$ the situation is again a little bit different, since we have an extra logarithmic term. Then we can define
\beq
\langle \bar{\psi}\psi \rangle_{ren} = \sigma_f^{-1} \left( \psi_1 \log(k_0 \dt) + \psi_2 \dt^{2-d}  + \cdots  \right),
\eeq
where $k_0$ is the renormalization scale. Then, the universal term yields, for $d\geq4$,
\beq
\psi_1 = \frac{(-1)^{d/2+1}}{2^{d-2}} \partial_t^{d-2} m(t).
\label{universal_f}
\eeq
In contrast, $\psi_2$ is much more complicated and we expect that it is not universal. For the present tanh quenches, $\psi_2$ can be written as 
\begin{eqnarray}
\psi_2 & = & -\lim_{h\rightarrow\infty} \Bigg( \sum _{x=2}^h (-1)^{x+d/2+1} \log(x^2) \frac{x^{d-2}}{2} z(t)^x\sum _{j=1}^{h-1} \left(\frac{z(t)^{j-x+1}}{(x-1)!}\prod _{i=0}^{x-2} (j-i)\right) + \nonumber \\
& &+\sum _{x=2}^h z(t)^{h+1}(-1)^{x+d/2}\frac{x^{d-2}}{2} \log(x^2) \frac{1}{(x)!}\prod _{i=0}^{x-1} (h-i) \Bigg),
\label{sum_f}
\end{eqnarray}
where again $z(t)=0.5+0.5 \tanh(t/\dt)$.

\begin{figure}[H]
        \centering
        \subfigure[d=4 ]{
                \includegraphics[scale=0.8]{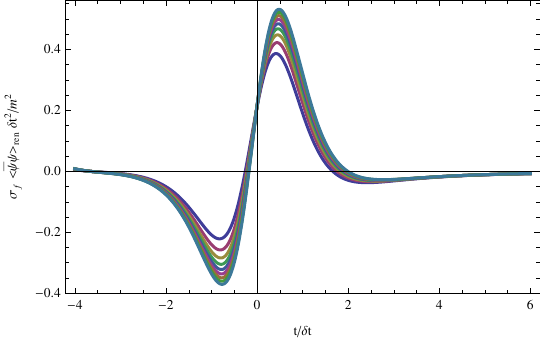}\label{no_log_f_4}}
         \subfigure[d=6 ]{
                \includegraphics[scale=0.8]{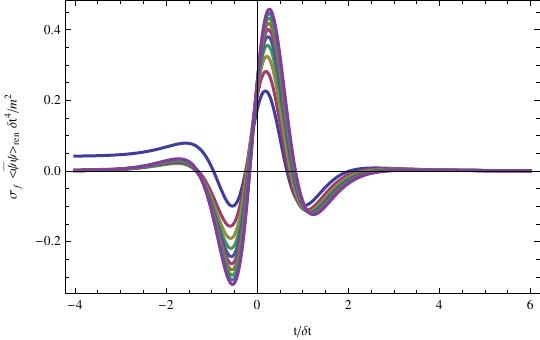}\label{no_log_f}}
   		 \subfigure[d=4 ]{
                \includegraphics[scale=0.8]{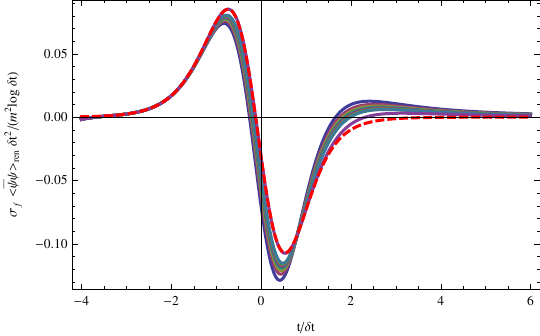}\label{log_f_4}}
   		 \subfigure[d=6 ]{
                \includegraphics[scale=0.8]{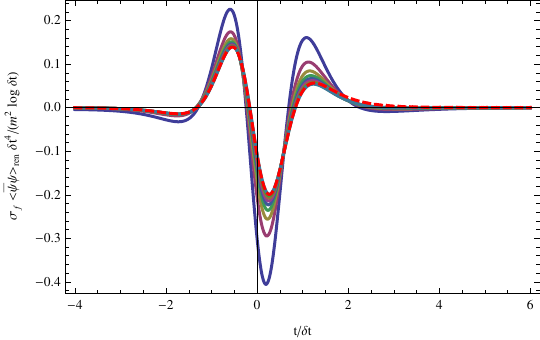}\label{log_f}}
        \caption{(Colour online) $\langle\bar{\psi}\psi\rangle_{ren}$ for different values of $\dt$ in $d=4$ and 6. In panels (a) and (b), we only divide by the expected power-law scaling. As we reduce $\dt$ from $\dt=1/10$ to $\dt=1/100$ for $d=4$ and from $\dt=1$ to $\dt=1/10$ for $d=6$, we see that the expectation value still grows, indicating the presence of an extra logarithmic factor. If we take the latter into account and divide by it as well, we find panels (c) and (d), where we see that the curves now converge towards the analytic expression (red dashed line).}
        \label{fig_leading_f_even_4}
\end{figure}

As in the case of the scalar field, these results support the holographic scaling where power-law growth is enhanced by a logarithmic factor in even $d$. We can appreciate these additional logarithmic factors by looking at figs.~\ref{no_log_f_4} and \ref{no_log_f}. There we divide out by the expected power-law scaling and we still see that the expectation value is growing as we decrease $\dt$. Finally, by using both eqs.~\ref{universal_f} and \ref{sum_f}, in figs.~\ref{log_f_4} and \ref{log_f}, we can see that the numerical solutions approach the analytical leading term for sufficiently small $\dt$'s.

\section{Quenches in general interacting theories}
\label{interacting}

Both the results in \cite{dgm} and in this paper show that different observables in free field theories after a smooth fast quench obey the same universal scaling relations as in quenches in holographic theories, as shown in \cite{numer, fastQ}. As the holographic CFT's are implicitly strongly coupled, we seem to have found the same scaling at two ends of the spectrum of possible interacting quantum field theories. Hence we should expect that the same result holds for a large variety of quenches in a wide range of interacting theories. In this section, we give arguments that the universal scaling in eq.~\reef{1-2} appears quite generally for fast quenches. The crucial assumption will be that the interacting theory which is being quenched approaches a UV fixed point, \ie its UV properties can be described by an appropriate CFT.

To motivate the general argument, we begin by considering quenches with a pulse profile in a CFT, as presented in \cite{dgm}.
In this case, we can use conformal perturbation theory.
The starting point is a CFT which is deformed as follows 
\ben
S = S_\mt{CFT} + \int d^d x\, \lambda(t)\, \calo_\Delta(x),
\een
where $\calo_\Delta$ is a relevant operator with dimension $\Delta < d$. We assume the profile for the corresponding coupling $\lambda(t)$ has the form
\beq
\lambda(t) = \delta \lambda\, h(t/\dt)\,,\labell{train8}
\eeq
where $\delta \lambda$ is the maximum coupling value and $h(y)$ is some smooth function that goes from 0 to 1 and back to 0 (at least roughly) in the interval $y =0$ to 1. Then, our coupling \reef{train8} has the form of a pulse in the time interval $t\in[0,\dt]$ with a maximum $\delta \lambda$. Note that this form essentially matches that of the profile \reef{mass_pulse} that we analyzed in section \ref{cft_cft}. There the mass profile was a pulse that goes from the critical point (\ie the massless theory) to the same critical point after passing through some maximum mass at $t=0$. Clearly this condition is not strictly necessary to obtain the universal scaling \reef{1-2}, as we have shown in quenches with a $\tanh$ profile for both the scalar and fermion masses yield the same scaling. However, the above framework will help to formulate our general argument.

Basically, since our theory is critical at both infinite past and infinite future (and anywhere outside the interval $t\in[0,\dt]$), we can calculate the expectation value of our operator using conformal perturbation theory, which yields
\bea
&&\langle \calo_\Delta(0)\rangle =
\langle \calo_\Delta(0) \rangle_\mt{CFT}- \delta \lambda \int d^d x\, h(t/\dt)\,  G_R (x,0) 
\label{gft}\\
&&\qquad\qquad\qquad+ \frac{\delta\lambda^2}{2} \int d^d x\,h(t/\dt)\,\int d^dx'\, h(t'/\dt) \ K(x,x',0)
+\cdots,
\nonumber
 \eea
where all expectation values are evaluated in the critical (conformal) field theory. Now, the first term in the RHS vanishes because $\calo_\Delta$ is a relevant operator, so its expectation value $\langle \calo_\Delta (0) \rangle_{\text{CFT}}$ must vanish. The second term is the linear response term where the retarded correlator is given by
\begin{eqnarray}
G_R (x,0) = i \theta(t)\ \langle\,  [\calo_\Delta (x) , \calo_\Delta (0) ]\, \rangle_\mt{CFT}\,.
\end{eqnarray}
The next-to-leading term is given by three-point correlator, 
\begin{eqnarray}
K(x,x^\prime,x^{\prime\prime}) & =  & \theta(t-t^\prime)\theta(t^\prime - t^{\prime\prime})
\langle|(\calo(x^\prime)\calo(x^{\prime\prime})\calo(x) + 
\calo(x)\calo(x^{\prime})\calo(x^{\prime\prime})|\rangle_\mt{CFT} \nonumber \\
&  &\qquad+ \theta(t-t^\prime)\theta(t - t^{\prime\prime})
\langle|\calo(x^\prime)\calo(x)\calo(x^{\prime\prime}) |\rangle_\mt{CFT} \label{threebus}
\end{eqnarray}

As in the free field cases analyzed in this paper, the expression in eq.~(\ref{gft}) is usually UV divergent so we need to regulate it by adding counterterms in order to get a finite expectation value. We will assume that renormalization can be done without problems to have a finite value that only depends on two renormalized parameters, $\dt$ and $\delta \lambda$. This assumption relies on the fact that, as mentioned in \cite{eva,timedeprg}, we do not expect that the quench protocol would lead to any unconventional RG flows. 

It is natural to expect that these counterterms are precisely given by the adiabatic expansion, as we have seen explicitly for the free field theory. Again, the reason is that the UV contributions to these quantities are insensitive to the quench rate so long as the rate is slow compared to the UV cutoff scale. Our protocol is chosen such that the rate is fast compared to the scale of the relevant coupling, as in eq.~\reef{1-1}, but always slow compared to the UV cutoff, \ie $\Lambda\dt\gg1$. For free field theories, it is easy to perform the adiabatic expansion since all we had to do is solve the wave equation in a WKB expansion, as described in section \ref{adiabat}. For interacting theories, this is no longer the case and we have to use the standard procedure in quantum mechanics starting with an expansion of the wave functional in terms of instantaneous eigenstates of the (time-dependent) Hamiltonian.

Now, as all correlators in eq. (\ref{gft}) are CFT correlators, they should be independent of the parameters $\delta \lambda$ and $\dt$. So, basically, $\dt$ will set the scale for the integrals and dimensional analysis will fix the form of all the possible terms in the expectation value. This means that
\ben
\langle \calo_\Delta(0)\rangle_{ren} =
 a_1\, \delta\lambda\,\dt^{d-2\Delta} 
+ a_2 \, \delta\lambda^2\,\dt^{2d-3\Delta} + \cdots,
\label{gft2}
\een
where the constants $a_n$ are finite numbers by assumption. Then, we can see that the first term, the linear response, is responsible of producing the universal scaling found, \ie $\langle \calo_\Delta \rangle \sim \delta \lambda / \dt^{2\Delta - d}$. However, we still have an infinite set of nonlinear contributions and so the next step is to show that the these become negligible once we take the limit of fast quenches \reef{1-1}. For that, it will be easier to define a dimensionless effective coupling, $g\equiv\delta \lambda\dt^{d-\Delta}$, so that eq.~(\ref{gft2}) becomes simply
\ben
\langle \calo_\Delta(0)\rangle_{ren}  =
 (\delta t)^{-\Delta} [\, a_1 g + a_2 g^2 + \cdots ]\,.
\label{stacker}
\een
That is, conformal perturbation theory \reef{gft} has expressed the expectation value in terms of a series expansion in terms of the dimensionless coupling.
The quenches we are considering correspond to keeping $\delta \lambda$ fixed, while taking $\dt\to0$. This means that we are taking our effective coupling small, \ie $g\rightarrow 0$, since we are quenching a relevant operator with $\Delta < d$. Hence with these protocols, the expansion \reef{stacker} is a very effective perturbation expansion and the leading behaviour is determined by just the first term, Of course, as we already noted, this term gives the desired scaling that was found in our previous calculations. We should note that \textit{any} pulsed quench will then give the desired scaling, independent, for instance, of the underlying CFT. Hence the present argument encompasses both the holographic CFTs of \cite{numer, fastQ} and the massless free fields studied here in previous sections.

In our discussion of free field quenches, we found that the universal scaling behavior \reef{1-2} is valid for profiles which are lot more general than the pulses considered above. Indeed, we now argue that the same scaling applies for general profiles subject to certain constraints and for general field theories subject to the assumption that the UV properties are described by a conformal fixed point.\footnote{In many respects, the following argument closely resembles the holographic analysis in \cite{fastQ}.} That is, we regard our original theory as emerging from an RG flow away from some perturbed CFT in the UV with the action
\ben
S_{init} = S_\mt{CFT} + \int d^d x\, \lambda_0\, \calo_{\Delta}(x)\,,\label{buss}
\een
where $\lambda_0$ is the coupling constant for some relevant operator $\calo_{\Delta}(x)$. Now consider a quench where the profile of the coupling $\lambda (t)$ only varies in the time interval $t\in[0,\delta t]$. At early times, $\lambda(t)$ will simply be fixed at $\lambda_0$ while after the quench it will take another constant value $\lambda_1$. For example, consider a coupling which interpolates between constant values $\lambda_0$ and $\lambda_1$ 
\[ \lambda(t) = \left \{
\begin{array}{ll}
\lambda_0 &\ \ \ {\rm for}\ t < 0\,,\\
\lambda_0 + \delta \lambda\, F(t/\delta t) &\ \ \  {\rm for}\ 0\le t\le \dt\,,\\ 
\lambda_1=\lambda_0 + \delta \lambda\, F(1) &\ \ \  {\rm for}\  t> \dt\,. 
\end{array} \right.
\]
We leave the details of the function $F(y)$ unspecified other than $F(y\le0)=0$ and $F(y\ge1)=1$ and the maximum is finite with $F_{max}\ge1$.
Further this profile may dip below zero by some finite amount and so we specify the minimum as $F_{min}\le0$.
Implicitly, we are also assuming that the profile is smooth.  Now we will work in the regime where
\ben
\lambda_0 \dt^{d-\Delta}\ll 1\,,\ 
\lambda_1 \dt^{d-\Delta} \ll 1\,,\quad
(\lambda_0+F_{max}\delta\lambda) \dt^{d-\Delta} \ll 1 \,,\quad
(\lambda_0+F_{min}\delta\lambda) \dt^{d-\Delta} \ll 1\,.
\label{lambdaregimes}
\een
We will calculate the expectation value of the operator at some time $t$ which is earlier than (or soon after) $t=\delta t $. Now motivated by the conformal perturbation expansion in eq.~\reef{gft}, we evaluate the change in $\langle\calo (t)\rangle$ relative to the initial theory \reef{buss} by expanding in $\delta \lambda$, \ie
\bea
&&\langle\calo(\vx,t)\rangle -\langle\calo(\vx,t)\rangle_{\lambda_0}  =
- \delta \lambda \int_0^t dt^\prime  F (t^\prime/\delta t) \int d^{d-1}\vx^\prime G_{R,\lambda_0}(\vx - \vx^\prime, t - t^\prime)
\label{lres1}\\
&&\qquad\qquad+ \frac{\delta\lambda^2}{2}  \int_0^t dt^\prime  F (t^\prime/\delta t) \int d^{d-1}\vx^\prime \int_0^t dt''  F (t''/\delta t) \int d^{d-1}\vx''
\ K_{\lambda_0}(t',\vec x';t'',\vec x'';t,\vec x)
+\cdots,
\nonumber
 \eea
where $G_{R,\lambda_0}$ denotes the retarded Green's function for the deformed CFT in eq.~\reef{buss} and similarly $K_{\lambda_0}$ denotes the analogous three-point correlator \reef{threebus} in this deformed theory. Of course, the first term in this expansion corresponds to the linear response. In writing the explicit range for the time integrals in eq.~\reef{lres1}, we have used the fact that the function $F(y)$ vanishes for $y \le 0$. Since all of the correlators in the above expansion are retarded, \ie only have support within the past light-cone, the spatial integrals are also limited to a range of order $t\le\dt$. That is, the integrals in eq.~\reef{lres1} only receive nonvanishing contributions from correlators where the operators are separated by a proper distance of less than $O(\dt)$. Now the fast quench regime defined by eq.~(\ref{lambdaregimes}) implies that these separations are all small compared to the inverse mass scales of the quenched theory. Hence the correlators will basically be the same as the CFT correlators, in eq.~\reef{gft} and up to small corrections, the integrals again all scale with the power of $\dt$ determined by dimensional analysis.
Therefore the change in the expectation value takes a general scaling form,
\ben
\langle \calo_\Delta(t)\rangle_{ren} - \langle \calo_\Delta(t)\rangle_{ren,\lambda_0} =
 (\delta t)^{-\Delta} [\, b_1 (t/\delta t)\, g + b_2 (t/\delta t)\,  g^2 + \cdots ]\,.
\label{stacker2}
\een
with none of the IR scales defining the deformed theory appearing in the problem. Again, the leading behaviour is determined by the linear response, \ie the term linear in the dimensionless coupling, and hence the change in the expectation value has the desired scaling, $\delta\lambda\,\dt^{d-2\Delta}$.

Further, the diffeomorphism Ward identity \reef{ward} still applies in the present context.
Hence energy is only injected into the system while $\partial_t\lambda(t)$ is non-vanishing, \ie only in
the interval $0 < t < \delta t$. But this is precisely the interval in which the change in the expectation value
was evaluated in eq.~\reef{stacker2} above. Now if we further assume that $\Delta>d/2$, then this change will be large compared to the unquenched
expectation value in the fast regime. Hence integrating the right hand side
of eq.~(\ref{ward}) will lead to the expected scaling of the energy, as given in eq.~\reef{1-3}.

Hence we have argued that the universal scaling in eqs.~\reef{1-2} and \reef{1-3} will emerge for a broad variety of quenches in a wide class of interacting field theories. 
Of course, the above framework could be made even more elaborate, \eg by introducing further deformations in the initial theory \reef{buss}. Again, the first essential ingredient in our argument was that the interacting field theory under study can be considered to emerge in the infrared from an RG flow away from a conformal fixed point in the UV. Further, we are considering fast quenches where the quench rate $1/\dt$ is much larger than any of the IR mass scales defining the initial theory or appearing in the quench protocol.  The upshot of this is that when $\delta t$ is the smallest physical
length scale in the problem, the early time response is entirely
governed by the conformal field theory at the UV fixed point, which
explains its universality. 
In particular, the scaling behavior is independent of the details of the protocol so long as eq.~(\ref{lambdaregimes}) is obeyed.

We can extend this discussion to make explicit the independence of the scaling behavior from the initial and final mass scales appearing in the quenches of the free field theory in section \ref{massive_massive}. As we have seen above, the leading result is given by the linear response. Hence we consider the linear response answer for  $\langle\phi^2\rangle$ in free scalar field with a mass profile similar to one considered there, \ie a profile which interpolates between $m_i^2$ and $m_f^2 =m_i^2+ \delta (m^2)$ with
\ben
m^2(t) = m_i^2 + \delta (m^2)\, F(t/\delta t)
\label{apd1}
\een
where the function $F(y)$ rises from zero around $y = 0$ and quickly settles to 1 soon after $ y=1$. The intitial and final masses, as well as $\delta m$  are small compared to the quench rate
\ben
m_i \delta t \ll 1\,,~~~~~~~~m_f \delta t \ll 1\,,~~~~~~~\delta m \delta t \ll 1\,.
\label{apd0}
\een
 The change in the expectation value is given by a generalization of eq.~(\ref{lr-3})
\begin{align}
\langle0|\phi^2(\vx,t)|0\rangle_{in} - & \langle0|\phi^2(\vx,t)|0\rangle_{in} |_{\delta m^2=0} \nonumber \\
& = -\delta( m^2)\,  \int \frac{d^{d-1}k}{(2\pi)^{d-1}}\,\frac{1}{2(k^2 + m_i^2)} \int_0^t dt^\prime F(t^\prime/\delta t) \sin [2(t-t^\prime)\sqrt{k^2+m_i^2} ]
\label{apd2}
\end{align}
where we have used the fact that the function $F(y)$ vanishes for $y<0$.

To estimate this, consider for example a function $F(x)$ which is piecewise constant
\[ F(y) = \left\{ 
\begin{array}{ll}
0 & \ \ {\rm for}\ y \le 0  \\
F_0 & \ \ {\rm for}\  0 < y < 1\\
1 & \ \ {\rm for}\  y \ge 1
\end{array}
\right.
\]
Then for any $t \leq \delta t$, eq.~(\ref{apd2}) becomes
\begin{align} 
\langle0|\phi^2(\vx,t)|0\rangle_{in} - & \langle0|\phi^2(\vx,t)|0\rangle_{in} |_{\delta m^2=0} \nonumber \\
& = - F_0\ \delta (m^2) \delta t ^{4-d} \int \frac{d^{d-1}q}{(2\pi)^{d-1}}\, \frac{\sin^2 [(\frac{t}{\delta t})\sqrt{q^2+(m_i\delta t)^2}]}{(q^2+(m_i \delta t)^2)^{3/2}}
\label{apd4}
\end{align}
where $q=k\dt$. Clearly $m_f$ has dropped out of this expression. Furthermore to the leading order in the limit (\ref{apd0}), the integral in (\ref{apd4}) becomes independent of $m_i$ as well. This leading answer is the same as in the case of a quench from a CFT.

Note that if we use the expression (\ref{lres1}) for times much longer
than $\delta t$, we will need to address issues of infrared divergences
associated with conformal perturbation theory for constant
deformations \cite{cpt}. For the question we are addressing here, we do not
need to do this. For a recent discussion of our scaling result in a
theory with an infrared regulator, see \cite{david}.

\section{Conclusions}
\label{conclusions}

In this paper, we have expanded on the results of fast but smooth quantum quenches that we previously presented in \cite{dgm}, and extended the results to more general quenches. We have given details of our calculations in free field theories, where both numerically and analytically we obtain the same scaling relations as in previous holographic studies of the same kind of quenches \cite{numer, fastQ}. This universal behavior in the early time response was found in a variety of quench protocols which interpolate between arbitrary constant masses so long as the quench rate $1/\dt$ is large compared to all other physical mass scales in the problem. In section \ref{interacting}, we provided a general argument that the universal scaling in eqs.~\reef{1-2} and \reef{1-3} will appear in fast quenches of any quantum field theory which flows from a conformal fixed point in the UV, \ie for any theory that can be described as a CFT deformed by some relevant operator(s). The scaling is purely a property of the UV conformal field theory, which emerges at early times as long as the duration of the quench is short compared to all other physical length scales in the problem, as in eq.~(\ref{regime}).

A key ingredient in our work, and in the corresponding holographic studies \cite{numer,fastQ}, is the renormalization of the underlying quantum field theory. Bare quantities, such as the expectation value $\langle \calo_\Delta \rangle$, are UV divergent and counterterms are needed in order to extract physically meaningful quantities. The problem here for quenches is quite similar in spirit to quantum field theories in curved space-times, \eg \cite{BD2,BD,Duncan}. In that case, the required counterterms involve operators made out of quantum fields, as well as curvature tensors of the background space-time. As discussed in section \ref{effaction}, for a global quench with a time-dependent mass, we need to add counterterms involving time derivatives of the mass function. In fact, to properly renormalize the expectation value of the stress tensor, we should also consider the theory in a curved background and include additional counterterms involving curvatures --- even if we are only considering these expectation values in a flat space background. 

However, we are still left with the problem of determining the precise coefficients of the counterterms which render the renormalized observables finite.  We argued that these coefficients can be determined by examining the quenches in an adiabatic limit and in section \ref{adiabat}, we demonstrated explicitly how to construct the necessary counterterms order by order in the expansion for slow quenches. Moreover, this procedure does not depend on any specific mass profile and so, the resulting counterterms should be universal. We verified that claim by correctly regulating quenches with a variety of different mass profiles using the same counterterms. Of course, it may appear surprising that an adiabatic expansion, which is an expansion in time derivatives, yields the correct counterterms for a {\em fast} quench. We argued that the physical reason behind this is that high momentum modes won't see whether the quench is fast or slow, so long as the quench rate is smaller than the cutoff scale $\Lambda$. In our cases we managed to take that cutoff to infinity while renormalizing the physical quantities, so we could expect that the counterterms would be the same in both slow and fast quenches. It would be interesting to test these assumptions in interacting field theories. Quenches in the large-$N$ vector model, for instance, have been studied previously in the literature \cite{misha}. This would be a good place to make explicit calculations and verify whether our intuition holds even when we have interacting theories.

\subsection*{Renormalized observables}

As emphasized above, our considerations refer to the renormalized quantities which require `removing' various UV divergences in our calculations. While this is, of course, the standard approach in quantum field theory, one may still ask how our renormalized observables would be related to measurements made in a physical experiment, where implicitly there is a finite UV cutoff? As a simple analogy, let us consider a quench which consists of suddenly applying external pressures to a crystal.  The phonons in the crystal would provide the analog of our quantum fields, \ie at least in a certain regime, they would have a QFT description. The quench will `excite' the final state of crystal in two ways. Naturally, the quench will generate phonon excitations in the crystal but the external pressure may also deform the crystal structure in the final configuration \eg modifying the dispersion relation for the phonons. The work done in deforming the crystal structure would then be the analog of the changes in the divergent `zero-point' energy that appears in the bare expectation value $\langle {\cal E}\rangle$ and which is subtracted by introducing mass-dependent counterterms to produce the renormalized energy density. Similarly, the energy available in the phonon excitations would correspond to the final $\langle {\cal E}\rangle_{ren}$. The latter is the energy that can be accessed and manipulated by probing the system with local operators. Let us add for the analogy of the crystal quench becomes more precise if we also insist that the quench time $\dt$ is larger than the lattice spacing, which provides the UV cutoff scale.
However, note that in this analogy, we have a cutoff which is itself time-dependent. This feature is quite different from the framework studied here where the cutoff is always fixed. Of course, more precise analogies without this defect could be developed, \eg by considering cold atoms trapped in a two-dimensional optical lattice  where the transverse potential is made to vary in a time-dependent but spatially homogeneous manner.

While the above analogy should make clearer the role of bare and renormalized quantities in a physical system with a finite cutoff, one may still ask what quantities would appear in experimental measurements. Answering this question becomes even more complicated for even dimensions, where in section \ref{adiabat} we found that logarithmic divergences introduced various renormalization ambiguities. Such ambiguities were also discussed in the holographic context in \cite{numer,fastQ}. The resolution there is that various fiducial experimental measurements would be made to fix these ambiguities. For example, examining eq.~\reef{ctact} for $d=4$, we find that there will be two such logarithmic terms.\footnote{These are the terms proportional to $\Lambda^{d-4}$, \ie with coefficients $s_{20}$ and $s_{51}$.} However, with some thought, we can see that the associated renormalization scales can be fixed by first measuring $\langle \phi^2\rangle_{ren}$ and $\langle {\cal E}\rangle_{ren}$ at some fixed finite mass.  

Implicitly in the previous discussion but more generally, we can work with quantities which are free of UV divergences by comparing expectation values at different times or in different quenches. For example, as discussed in section \ref{late2}, the difference $\langle \phi^2 \rangle_{quench} - \langle \phi^2  \rangle_{fixed}$ appearing in eq.~\reef{d3exact} is completely finite for quenches in $d=3$. Similarly, one can produce UV finite quantities by comparing the results for different quench protocols or by quenching different initial states with the same quench protocol, as in briefly discussed in appendix \ref{state9}. Of course, another family of UV finite observables would be correlators measured with finite separations, \eg as in eq.~\reef{correlator}. Further, one may be able to find evidence of universal scaling in the early time response with a strategic choice of the positions in the correlator. 
  
Of course, it would also be interesting to analyze cases where the cutoff remains finite, \eg in some lattice model. Though  the analysis would be more complex in such a case, we expect that our universal scaling properties should emerge in the regime where the scales are properly distinguished, \ie in a regime where $\Lambda \gg 1/\dt \gg m$. In fact, one might expect that as $1/\dt$ approaches the cutoff scale, one would recover the results of an instantaneous quench.

\subsection*{Comparison to instantaneous quenches}

Finally, we should comment on the relation between our smooth quenches and the instantaneous (or abrupt)  quenches that are usually studied in the literature \cite{cc2,cc3,gritsev}. Some preliminary discussion of the comparison between these two classes of protocols was given in section \ref{comparin} --- see also section \ref{late2} --- and a more detailed discussion will appear in \cite{dgm2}. Here, the universal scaling in eqs.~\reef{1-2} and \reef{1-3} suggests that divergences will appear as $\dt \rightarrow 0$ (whenever $\Delta\ge d/2$). This would seem to contradict instantaneous quench results.  However, as we already discussed in \cite{dgm}, these two types of quenches are different: while the present quenches evolve smoothly in a time-dependent scheme, the instantaneous quench approach can be thought as the evolution of a far-from-equilibrium initial state evolving under a fixed, time-independent, Hamiltonian. The scalings discussed in this paper hold for {\em renormalized} quantities and as emphasized above, the renormalization procedure demands that the quench rate is slow compared to the UV cutoff scale. On the other hand, instantaneous quenches necessarily involve quench rates which are fast compared to all scales, including the UV cutoff. Indeed we have explicitly shown that for free field theories, the momentum space correlator agrees with that for an instantaneous quench only when the momenta are small compared to the quench rate $1/\dt$.  {\em Local} quantities, like the one-point function of the mass operator or the energy density, involve an integral over all momenta and so this condition does not hold. 

However, this constraint above may still hold effectively if the contributions to the integral at high momentum are suppressed for other reasons. One case where the latter might apply is at late times after the quench. The intuition behind is that at late times, we expect only low energies (or momenta) contribute and hence the observables for fast smooth quenches and for instantaneous quenches may agree at late times. Section \ref{late2} presents some preliminary evidence for this conclusion.  In the free bosonic theory for $d=3$, we found that at sufficiently late times, the one point function becomes independent of $\dt$ and grows logarithmically in time. Further, this late time growth exactly agrees with the result from an instantaneous quench. For $d=5$, we showed that the late time result for a smooth fast again becomes independent of $\dt$ as $\dt \rightarrow 0$. However, as we will discuss in \cite{dgm2}, this answer only roughly agrees with the expectation value at late times after an instantaneous quench. More generally, the expectation values generated by the two different protocols fails to agree even at late times in higher dimensions \cite{dgm2} and hence the precise agreement in $d=3$ is quite exceptional. However, the disagreement found more generally should come as no surprise since the expectation value $\vev{\phi^2}$ involves a momentum integral up to the cutoff scale where, as we already argued, agreement should not be expected. However, we should add that a detailed analysis reveals agreement for the late time correlators at finite spatial separations which are large compared to $\dt$. Again a full discussion of these issue will be presented in \cite{dgm2}.

\subsection*{Higher spin currents} 

One interesting feature of the free field theories studied here is that they contain an infinite family of conserved higher spin currents. In section \ref{higher_scaling}, we began a study of the response of the higher spin currents in fast smooth quenches. In particular, we discussed the construction of the higher spin currents in the case of massive free fields --- see also appendix \ref{appendix}. This construction naturally leads directly to a generalization of the diffeomorphism Ward identity \reef{ward} for the higher spin currents. In general, there is a hierarchy of generalized Ward identities \reef{power2}, which can be used to understand how the `work' done in varying the mass parameter changes the various higher spin `charge densities.'  In the fast quench regime \reef{1-1}, the latter yields a simple universal scaling property \reef{power} for these higher spin densities, \ie $\langle j^{(s)}_{t \cdots t} \rangle \sim (m^2)^{\frac{s}{2}+1}/\dt^{d-4}$. Of course, for spin-2 and spin-0, these are scalings of the energy density and the mass operator, in accord with eqs.~\reef{1-2} and \reef{1-3}. We explicitly carried out this construction and demonstrated the corresponding scaling for the spin-4 current.
Hence it will be interesting to explicitly construct all of the massive higher spin currents and to explicitly find the corresponding generalized Ward identities. Of course, it would also be interesting to develop a better intuition for the physical meaning of this hierarchy of Ward identities and the resulting universal scaling.

Finally it would be interesting to make a connection to the physics of Kibble-Zurek scaling \cite{kibble,zurek}, which arises in the regime of slow quenches. One could, \eg consider a time-dependent mass which interpolates between finite values but vanishes at some intermediate time. In this case, one would expect Kibble-Zurek scaling to hold when the quench rate is slow compared to the initial mass. Further as the quench rate is increased, one should find a crossover to the scaling discussed in this paper. We leave this interesting problem for future study.

\section*{Acknowledgements} We would like to thank Joe Bhaseen, 
Anushya Chandran, Diptarka Das, Fabian Essler, Ganpathy Murthy, Erik Schnetter,
 Krishnendu Sengupta, Alfred Shapere, Tadashi Takayanagi and Larry Yaffe for discussions. 
S.R.D. would like to thank Yukawa Institute for Theoretical Physics, Kyoto University, for a Visiting Professorship.
Research at Perimeter Institute is supported by the
Government of Canada through Industry Canada and by the Province of Ontario
through the Ministry of Research \& Innovation. RCM and DAG are also supported
by an NSERC Discovery grant. RCM is also supported by research funding
from the Canadian Institute for Advanced Research. 
The work of SRD is partially supported by the National Science Foundation grants NSF-PHY-1214341 and NSF-PHY-0970069. 

\appendix

\section{Conserved higher spin currents for a massive scalar}
\labell{appendix}

In this Appendix, we will show explicitly how to construct the spin-4 current for a massive scalar field from the corresponding current for the massless conformally coupled scalar --- see eq.~\reef{general_current}.

The massless spin-4 current reads
\begin{eqnarray}
j_{abcd}^{(4)} & = & \frac{1}{576}\, \phi^* {\partial}_{a b c d} \phi + \frac{1}{576}\, \phi {\partial}_{a b c d} \phi^* - \frac{1}{36}\, {\partial}_{a} \phi^* {\partial}_{b c d} \phi - \frac{1}{36}\, {\partial}_{a} \phi {\partial}_{b c d} \phi^* + \frac{1}{16}\, {\partial}_{a b} \phi {\partial}_{c d} \phi^* + \nonumber \\
& & + \frac{1}{96}\, {\eta}_{a b} {\partial}_{e} \phi^* {\partial}_{c d e} \phi + \frac{1}{96}\, {\eta}_{a b} {\partial}_{e} \phi {\partial}_{c d e} \phi^* - \frac{1}{32}\, {\eta}_{a b} {\partial}_{c e} \phi {\partial}_{d e} \phi^* + \frac{1}{384}\, {\eta}_{a b} {\eta}_{c d} {\partial}_{e f} \phi {\partial}_{e f} \phi^*, \label{aa11}
\end{eqnarray}
where the traces coefficients have been chosen to make the current traceless and the whole expression should be symmetrized in all four indexes. When we take the divergence, it is straightforward to verify that this current is conserved:
\begin{eqnarray}
\partial^a j_{abcd}^{(4)} & = &  - \frac{1}{48}\, {\partial}_{b} \phi {\partial}\indices{^a_{a c d}} \phi^* - \frac{1}{48}\, {\partial}_{b} \phi^* {\partial}\indices{^a_{a c d}} \phi - \frac{1}{144}\, {\partial}\indices{_a^a} \phi {\partial}_{b c d} \phi^* - \frac{1}{144}\, {\partial}\indices{_a^a} \phi^* {\partial}_{b c d} \phi \nonumber \\
& & + \frac{1}{32}\, {\partial}_{c d} \phi^* {\partial}\indices{^a_{a b}} \phi + \frac{1}{32}\, {\partial}_{c d} \phi {\partial}\indices{^a_{a b}} \phi^* + \frac{1}{576}\, \phi^* {\partial}\indices{^a_{a b c d}} \phi + \frac{1}{576}\, \phi {\partial}\indices{^a_{a b c d}} \phi^*  \nnn \\
& & + \frac{1}{192}\, {\eta}_{b c} {\partial}^{a} \phi {\partial}\indices{^e_{e a d}} \phi^* + \frac{1}{192}\, {\eta}_{b c} {\partial}^{a} \phi^* {\partial}\indices{^e_{e a d}} \phi - \frac{1}{384}\, {\eta}_{b c} {\partial}_{d a} \phi {\partial}\indices{^{a e}_e} \phi^* \nnn\\
&& - \frac{1}{384}\, {\eta}_{b c} {\partial}_{d a} \phi^* {\partial}\indices{^{a e}_e} \phi = 0, \label{divergence_4d}
\end{eqnarray}
as all terms have $\partial\indices{^a_a} \phi$ or its conjugate, which vanishes because of the equations of motion in the massless case. To generalize eq.~\reef{aa11} to the massive case, we will need to add terms proportional to the mass squared, so that all these terms now are cancelled but upon evaluation in the massive equation of motion $\partial\indices{^a_a} \phi - m^2 \phi = 0$. So, for instance, the first term in the RHS of eq. (\ref{divergence_4d}) should be cancelled with one of the form $+\frac{m^2}{48} \partial_b \phi \partial_{c d} \phi^*$.

Now, all the possible $m^2$ terms we can add to the current are of the form
\begin{eqnarray}
j_{a b c d}^{(4)m^2}= m^2 \eta_{a b}  \Big[ A \left(\phi^* {\partial}_{c d}{\phi}+ \phi {\partial}_{c d}{\phi^*}\right)  - B\left( {\partial}_{c}{\phi}\,  {\partial}_{d}{\phi^*}\, + {\partial}_{d}{\phi}\,  {\partial}_{c}{\phi^*}\right)  + C\, \eta_{c d} {\partial}^{e}{\phi}\,  {\partial}_{e}{\phi^*}\, \Big]\,, \nnn 
\end{eqnarray}
where $A,B,C$ are constants to be determined (as always, the most general form of the current should be symmetrized). By taking the divergence we obtain,

\begin{eqnarray}
\partial^a j_{a b c d}^{(4)m^2} & = & 3 (A-2B) \left( \partial_b \phi \partial_{cd} \phi^* + \partial_b \phi^* \partial_{cd} \phi \right) + 3A \left( \phi^* \partial_{bcd} \phi + \phi  \partial_{bcd} \phi^* \right) + \nnn \\
& & 3(A-B+2C) \left( \eta_{bc} \partial^a \phi \partial_{ad} \phi^* + \eta_{bc} \partial^a \phi^* \partial_{ad} \phi \right) + \label{partial_j} \\
& & + 3A \left( \eta_{b c} \phi^* \partial\indices{^a_{a d}} \phi + \eta_{b c} \phi \partial\indices{^a_{a d}} \phi^* \right) - 3B \left( \eta_{b c} \partial\indices{^a_a} \phi^* \partial\indices{_d} \phi + \eta_{b c} \partial\indices{^a_a} \phi \partial\indices{_{d}} \phi^* \right)
\nnn
\end{eqnarray}
and so, we are left with a $3\times3$ system to solve for $A,B,C$ in order to have $j_{abcd}^{(4)} + j_{a b c d}^{(4)m^2}$ conserved. This gives $A=1/576, B=1/384, C=1/1152$. But we still need to add terms proportional to $m^4$ in order to cancel the terms that appear in the last line of eq. (\ref{partial_j}). For those we just need to add
\begin{eqnarray}
j_{a b c d}^{(4)m^4} = -3 m^4 (A-B) \eta_{ab} \eta_{cd} \phi^*\phi,
\end{eqnarray}
and then we have the full generalized 4-spin conserved current for the case of a massive scalar field.

We can do an analogue procedure in any spacetime dimension and we get
\begin{eqnarray}
j_{abcd}^{(4)} = j_{abcd}^{(4)m=0} + \frac{m^2 \eta_{ab}}{4 (d/2+2)! (d/2)!} \left( \frac{d}{2} j_1^{(2)} + \left( \frac{d}{2}+1 \right) j_2^{(2)} +j_3^{(2)} + m^2 j_0  \right),
\end{eqnarray}
where
\begin{eqnarray}
j_1^{(2)} & = & \phi^*\partial_{cd} \phi + \phi \partial_{cd} \phi^*, \\
j_2^{(2)} & = & - \partial_c \phi^* \partial_{d} \phi - \partial_c \phi \partial_{d} \phi^*, \\
j_3^{(2)} & = & \eta_{cd} \partial_e \phi^* \partial^e \phi, \\
j_0 & = & \eta_{cd} \phi^*\phi.
\end{eqnarray}

The crucial fact for our discussion in section \ref{higher_scaling} is that we can write the current as the sum of the minimally coupled and the conformally coupled spin-2 current and then it is direct to evaluate the generalized Ward identity.

Finally, we should say that this procedure is, in principle, easily generalized to any higher spin current. However, the procedure becomes tedious as the number of terms in the massless current grows quickly with the spin and so does the number of possible terms that should be canceled with mass terms. 

\section{Scaling of excited states in the scalar quench}
\label{state9}

It is interesting to also analyse the behaviour of excited states under a quench. This gives an extra observable to evaluate and it may be particularly useful in case one wants to make explicit contact with experiment. If we take the case of even dimensions, for instance, extra regulator ambiguities appear in the problem, as discussed in section \ref{adiabat}. So, if someone is performing an experiment, before looking at the scalings and so on, one should establish a way to fix these ambiguities. Interestingly, after being fixed, one should be able to compare different states using the same protocol, so excited states become useful observables to evaluate the behaviour of the system.

We can think of different possible excited states such as giving the system some excitations of in-modes or even think about a coherent state of in-modes. In any case, one interesting example is to compute $\langle \phi^2 \rangle_n\equiv\langle in,0| a^n_{\vk} \,\, \phi^2 \, a^{\dagger n}_{\vk}|in,0\rangle $, for any momentum $\vk$ and any number of excitations $n$.

However, we already know the exact solution to $\phi$ under the quenches we are considering, \ie see eqs. (\ref{fieldx}) and (\ref{modes}). So, in order to evaluate the excited expectation values we only need to use repeatedly the property of commutation of the $a_\vk$ modes --- see eq. (\ref{fieldx}). Explicitly, what we need to find are expressions of the type $\langle in,0| a_{\vk} \cdots a_{\vk} a_{\vec{k'}} a_{\vec{k''}}^\dagger a_{\vk}^\dagger \cdots a_{\vk}^\dagger |in,0\rangle$ and $\langle in,0| a_{\vk} \cdots a_{\vk} a_{\vec{k'}}^\dagger a_{\vec{k''}} a_{\vk}^\dagger \cdots a_{\vk}^\dagger |in,0\rangle$. After some algebra we get
\begin{eqnarray}
\langle \phi^2 \rangle_n = \langle \phi^2 \rangle_0 + 2 n^2 \frac{k^{d-2}}{\omega_{in}} |_2F_1|^2,
\end{eqnarray}
where $\langle \phi^2 \rangle_0$ means the vacuum expectation value and the hypergeometric function is evaluated at the same arguments as in the main body of this article but at a fixed momentum $k$. Now one can ask whether the difference between the excited states expectation value and the vacuum also scales as $\dt \rightarrow 0$. However, it is quite direct to show that as $\dt$ goes to zero with fixed momentum $k$, the hypergeometric function goes to 1, and so the difference $\langle \phi^2 \rangle_n - \langle \phi^2 \rangle_0$ would go to some constant depending on $k$ and $n$ but would not scale with some power of $\dt$, as found for the vacuum expectation value.

\end{document}